\documentclass[usenatbib]{mnras}

\usepackage[english]{babel}
\usepackage[utf8x]{inputenc}
\usepackage[T1]{fontenc}

\usepackage{amssymb}
\usepackage{amsmath}
\usepackage{graphicx}
\usepackage{booktabs}
\usepackage{ulem}
\usepackage{textcomp}
\usepackage{gensymb}
\usepackage{url}
\usepackage{multirow}
\usepackage[para,online,flushleft]{threeparttable}
\usepackage[table,xcdraw]{xcolor}
\usepackage{hyperref}
\usepackage{orcidlink}
\usepackage{siunitx}

\newcommand{\snia}{SN~Ia}
\newcommand{\sneia}{SNe~Ia}
\newcommand{\hsf}{Hawai`i Supernova Flows}
\renewcommand{\thefootnote}{\fnsymbol{footnote}}

\title[Hawai`i Supernova Flows]{Hawai`i Supernova Flows: A Peculiar Velocity Survey Using Over a Thousand Supernovae in the Near-Infrared}
\author[A.~Do et al.]{
Aaron~Do$^{1,2}$ \orcidlink{0000-0003-3429-7845}\thanks{E-mail: ajmd6@cam.ac.uk},
Benjamin~J.~Shappee$^2$ \orcidlink{0000-0003-4631-1149},
John~L.~Tonry$^2$ \orcidlink{0000-0003-2858-9657},
R.~Brent~Tully$^2$ \orcidlink{0000-0002-9291-1981},
Thomas~de~Jaeger$^{2,3}$ \orcidlink{0000-0001-6069-1139},
\newauthor
David~Rubin$^{2,4}$ \orcidlink{0000-0001-5402-4647},
Chris~Ashall$^{2,5}$ \orcidlink{0000-0002-5221-7557},
Christopher~R.~Burns$^6$ \orcidlink{0000-0003-4625-6629},
Dhvanil~D.~Desai$^2$ \orcidlink{0000-0002-2164-859X},
Jason~T.~Hinkle$^2$ \orcidlink{0000-0001-9668-2920},
\newauthor
Willem~B.~Hoogendam$^2$ \orcidlink{0000-0003-3953-9532},
Mark~E.~Huber$^2$ \orcidlink{0000-0003-1059-9603},
David~O.~Jones$^2$ \orcidlink{0000-0002-6230-0151},
Kaisey~S.~Mandel$^1$ \orcidlink{0000-0001-9846-4417},
\newauthor
Anna~V.~Payne$^7$ \orcidlink{0000-0003-3490-3243},
Erik~R.~Peterson$^8$ \orcidlink{0000-0001-8596-4746},
Dan~Scolnic$^8$ \orcidlink{0000-0002-4934-5849},
Michael~A.~Tucker$^{9,10,11}$ \orcidlink{0000-0002-2471-8442},
\\
$^1$Institute of Astronomy and Kavli Institute for Cosmology, Madingley Road, Cambridge, CB3 0HA, UK\\
$^2$Institute for Astronomy, University of Hawai`i, 2680 Woodlawn Dr., Honolulu, HI 96822, USA\\
$^3$CNRS/IN2P3, Sorbonne Universit\'{e}, Universit\'{e} Paris Cit\'{e}), Laboratoire de Physique Nucl\'{e}aire et de Hautes \'{E}nergies, 75005, Paris, France\\
$^4$Department of Physics and Astronomy, University of Hawai‘i, Honolulu, HI 96822, USA\\
$^5$Department of Physics, Virginia Tech, Blacksburg, VA 24061, USA\\
$^6$The Observatories of the Carnegie Institution for Science, 813 Santa Barbara Street, Pasadena, CA 91101, USA\\
$^7$Space Telescope Science Institute, 3700 San Martin Drive, Baltimore, MD 21218, USA\\
$^8$Department of Physics, Duke University, Durham, NC 27708, USA\\
$^9$ Center for Cosmology and Astroparticle Physics, The Ohio State University, 191 West Woodruff Ave, Columbus, OH 43210, USA\\
$^{10}$ Department of Astronomy, The Ohio State University, 140 West 18th Avenue, Columbus, OH 43210, USA\\
$^{11}$ Department of Physics, The Ohio State University, 191 West Woodruff Ave, Columbus, OH 43210, USA\\
}
\date{Accepted XXX. Received YYY; in original form ZZZ}
\begin{document}
\maketitle{}
\label{firstpage}
\pagerange{\pageref{firstpage}--\pageref{lastpage}}

\renewcommand{\thefootnote}{\arabic{footnote}}

\begin{abstract}
We introduce the \hsf{} project and present summary statistics of the first 1,217 astronomical transients observed, 668 of which are spectroscopically classified Type Ia Supernovae (\sneia{}).
Our project is designed to obtain systematics-limited distances to \sneia{} while consuming minimal dedicated observational resources.
To date, we have performed almost 5,000 near-infrared (NIR) observations of astronomical transients and have obtained spectra for over 200 host galaxies lacking published spectroscopic redshifts.
In this survey paper we describe the methodology used to select targets, collect/reduce data, calculate distances, and perform quality cuts.
We compare our methods to those used in similar studies, finding general agreement or mild improvement.
Our summary statistics include various parametrizations of dispersion in the Hubble diagrams produced using fits to several commonly used \snia{} models.
We find the lowest dispersions using the \texttt{SNooPy} package's EBV\_model2, with a root mean square (RMS) deviation of 0.165 mag and a normalized median absolute deviation (NMAD) of 0.123 mag.

The full utility of the \hsf{} data set far exceeds the analyses presented in this paper.
Our photometry will provide a valuable test bed for models of \snia{} incorporating NIR data.
Differential cosmological studies comparing optical samples and combined optical and NIR samples will have increased leverage for constraining chromatic effects like dust extinction.
We invite the community to explore our data by making the light curves, fits, and host galaxy redshifts publicly accessible.
\end{abstract}


\section{Introduction}
\label{sec:introduction}

\hsf{} is an ongoing effort to map the distribution of mass in the local universe (z < 0.1) using near-infrared (NIR) observations of Type Ia Supernovae (\sneia{}) in combination with untargeted optical surveys.
In this paper we provide an overview of the \hsf{} project to support future papers examining detailed science cases using data from \hsf{}.

The paper is organized as follows.
In this Section, we review the connection between Large Scale Structure and peculiar velocities, describe the largest contemporary peculiar velocity surveys, and motivate our use of \sneia{}.
Section \ref{sec:observations} describes the individual components of the project: including the target selection process, the observing facilities used, the photometric calibration and analysis, the identification of host galaxies, and the determination of their redshifts.
Section \ref{sec:distances} describes the three \snia{} fitting procedures we employ and how each set of fitting parameters is converted to distance moduli.
We validate our fitting methodology and photometry using data from the Dark Energy, $H_0$, and peculiar Velocities using Infrared Light from Supernovae survey \citep[DEHVILS;][]{peterson23} and the Carnegie Supernova Project's third data release \citep[CSP-I DR3;][]{krisciunas17} in Section \ref{sec:validation}.
Section \ref{sec:sample selection} lists the quality cuts used to define the various samples we describe, analyse, and discuss in Section \ref{sec:results}.

\subsection{Peculiar Velocities and the State-of-the-Art}

While the LSS contains some luminous, baryonic matter, the majority of its mass may only be studied through its gravitational effects \citep[for a review of dark matter, consider][]{bertone18}.
In comoving coordinates, objects accelerate towards denser regions of LSS and away from voids.
This motion is called peculiar velocity and its projection on our line of sight may be calculated as \citep{davis14}
\begin{equation}
    \label{eqn: pv_rad}
    v = c~\left(~\frac{z_\text{obs} - z_\text{cos}(d_L)}{1 + z_\text{cos}(d_L)}~\right)
\end{equation}
where $z_\text{obs}$ is the observed redshift and $z_\text{cos}(d_L)$ is the redshift at luminosity distance $d_L$ due to universal expansion in a given cosmological model with deceleration parameter $q_0$ and Hubble constant $H_0$ \citep{peebles93}
\begin{equation}
    z_\text{cos} \approx \frac{1}{1 - q_0}\left[-1 + \sqrt{1 + \frac{2H_0 d_L}{c}(1 - q_0)}\right].
\end{equation}

Peculiar velocities have been used to infer the distribution of LSS through a variety of approaches like the POTENT method \citep{bertschinger89, dekel90, dekel99}, the Wiener Filter and constrained realizations method \citep{ganon93, zaroubi95, zaroubi99, courtois12}, the unbiased minimal variance estimator \citep{zaroubi02}, and various Bayesian hierarchical approaches \citep{lavaux16, graziani19, valade22}.
These methods commonly assume the LSS formed through gravitational instabilities, and is thus irrotational on large scales ($\nabla \times \vec{v} = 0$) \citep{peebles80}.
Variations between the methods typically represent different approaches to minimizing the systematic effects of smoothing, uneven sky coverage, and biases in peculiar-velocity surveys.
Modern cosmographic surveys are not limited by analytical tools, but by the number and precision of distance measurements.

Many peculiar velocity surveys use either the Fundamental Plane \citep[FP;][]{djorgovski87, dressler87} or the Tully-Fisher relation \citep[TF;][]{tully77} to measure distances because these methods can be applied to a significant fraction of all galaxies, whereas other methods require relatively rare phenomena like a gravitational lens \citep{refsdal64}, a maser \citep{herrnstein99} or megamaser \citep{gao16}, a gravitational wave event \citep{holz05}, or a supernova (SN).
However, while the FP and TF methods have significant advantages in target availability, the resulting distance measurements are often five to ten times less precise than measurements from more narrowly applicable distance probes.
The FP and TF methods, along with most photometric measures of distance, produce errors in distance modulus, which causes error to increase with the distance.
This proportionality is directly passed on to the uncertainties in peculiar velocity.
While independent peculiar velocity measurements of $N$ neighboring galaxies can be combined to reduce the statistical uncertainty by a factor of $\sqrt{N}$, galaxies have a finite amount of neighbors.
A volume-limited peculiar velocity survey will always find a noise floor that scales with the uncertainty in the distance-measuring technique and inversely with the root of galaxy number density.
Put another way, a survey with an explicit precision requirement has a maximum effective range that cannot be extended without more precise measures of distance.

For this reason, two of the three largest homogeneous collections of peculiar velocities extend no farther than a cosmic microwave background (CMB) rest-frame redshift of $z_\text{CMB}=0.05$.
These are the Cosmicflows-IV TF catalogue \citep[CF4-TF, 9,792 galaxies;][]{kourkchi20} and the FP-based 6-degree Field Galaxy Survey peculiar velocity sample \citep[6dFGSv, 8,885 galaxies;][]{springob14, campbell14}.
The Sloan Digital Sky Survey (SDSS) peculiar velocity catalogue \citep[SDSS-PV, 34,059 galaxies;][]{howlett22} is the first FP- or TF-based survey to extend to $z_\text{CMB}=0.1$, but so far only covers the SDSS North Galactic Cap contiguous area (7,016 deg$^2$).
The largest compilation of extragalactic distances is the heterogeneous catalogue Cosmicflows-IV \citep[CF4, 55,877 galaxies;][]{tully23}, which consolidates these and other surveys and uses both FP and TF measurements, as well as surface brightness fluctuations \citep{tonry88}, core-collapse SNe \citep{hamuy02}, and \sneia{} \citep{phillips93}.

The SDSS-PV sample has not yet produced any detailed cosmographic studies, but the authors measured a bulk flow in mild excess ($p \sim 0.06 - 0.20$ depending on cuts) of what a fiducial dark energy and cold dark matter ($\Lambda$CDM) model would suggest.
This excess has been suggested before using various independent data sets \citep[e.g.][]{pike05, feldman08, kashlinsky08, lavaux10, feldman10}.
Contemporary analyses extend the scale of the issue, with \citet{watkins23} finding that CF4 data indicate excess bulk flows on scales of 200 h$^{-1}$ Mpc that have a $1.5 \times 10^{-6}$ chance of occurring in the standard cosmological model using CMB-derived parameters.
\citet{howlett22} theorize that the Shapley Supercluster as seen in the 2M++ redshift compilation \citep{carrick15} could be responsible, but because it is not in the SDSS-PV survey footprint it will be difficult to test.
A survey that trades depth for sky coverage will still struggle to constrain the effects of the Shapley Supercluster, as \citet{carrick15} find their bulk flow measurements prefer a contribution from sources at $z > 0.067$ at a 5.1$\sigma$ level.

Thus far, the \hsf{} project has obtained peculiar velocity measurements over three quarters of the sky to a depth of $z \sim 0.1$.
This encompasses the gravitational sources thought to dominate local dynamics, including the Shapley supercluster, the Dipole repeller \citep{hoffman17}, and the Cold Spot repeller \citep{courtois17}.
Equation \ref{eqn: pv_rad} shows that peculiar velocities require an assumed cosmology and two measurements: an observed redshift and a proper distance.
The redshift can be measured to high precision with a single spectrum, but measuring distances is more difficult.
Techniques have been developed and refined to excel in various niches of a parameter space spanning applicability, maximum range, and precision.
Our project uses optical and NIR observations of \sneia{} to measure distances.

\subsection{Type Ia Supernovae}

Following the discovery that \sneia{} could be used as standardizable candles \citep{pskovskii77, phillips93, tripp98} there have been continuous efforts to improve the accuracy and precision of \sneia{} distance inference.
These efforts include refining theoretical models of \sneia{} progenitors and explosions \citep[reviewed in][]{liu23}; increasing the sample of well-studied \sneia{} \citep[e.g.][]{amanullah10, scolnic18, phillips18, brout22}; and empirically identifying correlations between \sneia{} luminosities and observable parameters like
host-galaxy mass \citep{kelly10, lampeitl10, sullivan10},
host-galaxy specific star formation rate \citep{uddin17},
local H$_\alpha$ surface brightness \citep{rigault13},
host-galaxy metallicity \citep{moreno_raya16},
host-galaxy colours \citep{roman18},
ejecta velocity \citep{leget20},
and more.
Accompanying these efforts are improvements to fitting and modelling techniques
(\texttt{BayeSN}, \citet{mandel09, mandel11, thorp21, mandel22, grayling24};
\texttt{MLCS2k2}\footnote{Multicolor Light-Curve Shapes}, \citet{jha07};
\texttt{SALT}\footnote{Spectral Adaptive Light curve Template}, \citet{guy05, guy07, guy10, kenworthy21, pierel22};
\texttt{SiFTO}, \citet{conley08};
\texttt{SNEMO}\footnote{SuperNova Empirical MOdels}, \citet{saunders18};
\texttt{SNooPy}\footnote{SuperNovae in Object Oriented Python}, \citet{burns11, burns14};
\texttt{SUGAR}\footnote{SUpernova Generator And Reconstructor}, \citet{leget20}).
This body of work has established \sneia{} as excellent probes of distance.
We choose to use them over competing distance measuring techniques for three reasons.

Firstly, \sneia{} are abundant.
With modern surveys across the globe constantly scanning the sky, SNe are no longer rare targets of opportunity, but are discovered every night.
\citet{desai24} use data from the All-Sky Automated Survey for SuperNovae \citep[ASAS-SN;][]{shappee14, kochanek17, hart23} to find a \snia{} volumetric rate of $\sim 2.3 \times 10^4 \text{ yr}^{-1} \text{ Gpc}^{-3} h^3_{70}$, which amounts to about 20 each night within $z < 0.1$.
\citet{wiseman21} use results from the Dark Energy Survey (DES) to calculate a rate of \sneia{} per galaxy between 1 every 3,000 years to more than one every 100 years depending on host-galaxy properties.
This means that although the number of usable galaxies in a \sneia{}-based peculiar velocity survey is relatively low compared to TF or FP surveys, it scales with time and can exceed competing methods with enough observational support.

Secondly, \sneia{} are bright enough to be used at the distances we require.
The demonstration of accelerating expansion relied on measurements of \sneia{} at redshifts near unity \citep{riess98, perlmutter99}.
Our interests are more local, extending to redshifts $z < 0.1$.
The mean absolute magnitude of \sneia{} before correcting for host-galaxy extinction is about $-18.6$ mag in $B$ and $-18.7$ mag in $V$ \citep[e.g.][]{ashall16}.
At $z = 0.1$ this corresponds to an apparent magnitude of about 19.6 or 19.5 mag, within the limiting magnitude of two of the all-sky surveys described in Sec. \ref{sec:all-sky}.

Lastly, \sneia{}-based distance measurements are far more precise than those of competing methodologies.
This is not to say that \sneia{} are the most precise of all distance indicators.
Distances based on Cepheid period-luminosity relations \citep{leavitt1912} or the Tip of the Red Giant Branch \citep{freedman20, anand21} are typically more precise than those based on \sneia{}, but the objects of study for these probes are about 13-16 magnitudes fainter than \sneia{}.
This restricts them to $z < 0.023$ even with 22 HST orbits per galaxy (PI D. Jones; proposal 16269).
\hsf{} extends about 4 times farther.
The SDSS-PV sample has used the FP method to measure distances at $z \sim 0.1$, but these distances are only precise to around 20\%.
\sneia{}-based distances can be systematically corrected to a root mean square (RMS) scatter between 4-7\% \citep{burns18, scolnic18}.
This means that it would take several dozens of independent TF or FP measurements to reach the precision of a single \snia{} distance measurement.

Optical \sneia{} light-curves have been used as standardizable candles for several decades \citep[e.g.]{phillips93, hamuy95, riess98, perlmutter99}, but a growing body of evidence \citep[e.g.][]{kasen06, wood-vasey08, burns11, dhawan18, avelino19} suggests that the NIR may offer substantial advantages.

\subsubsection{\sneia{} in the Near-Infrared}
\label{sec:sneia_in_nir}
NIR bandpasses like $Y$, $J$, $H$, and $K$ are 5-11 times less affected by dust than the traditionally used B band \citep{cardelli89, odonnell94, fitzpatrick99}.
The total-to-selective extinction parameter $R_V$ is known to vary based on the properties of dust, even in our own galaxy \citep{draine03}.
\citet{brout21} and \citet{popovic23} proposed that the dispersion in Hubble residuals of red \sneia{} may be largely due to the uncertain properties of extragalactic dust, which varies as a function of position in the host galaxy.
The effects of dust correlate with the colour of any given SN Ia, making any added uncertainty a systematic issue that may not be resolved with a larger sample.
Studying \sneia{} in the NIR suppresses the systematic error associated with dust.

Additionally, \sneia{} have been claimed to be more uniform in the NIR \citep[e.g.][]{wood-vasey08, kattner12, barone-nugent12, stanishev18, avelino19, galbany22, jones22}.
\citet{avelino19} used NIR light curves to determine distances consistent with those determined using optical light curves.
Notably, \citet{avelino19} did not apply the typical standardizations to the NIR light curves, but did correct the optical light curves for decline rate, host-galaxy extinction, and host-galaxy mass.
The empirical regularity of \sneia{} peak magnitudes in the NIR is supported by theory \citep{kasen06}, with radiative transfer calculations showing how decreases in bolometric flux are balanced by increases in relative emission at longer wavelengths.
The remarkable uniformity of \sneia{} peak absolute magnitudes in the NIR makes any distance measurement much more robust against systematic uncertainties.

\section{Project Components and Observational Facilities}
\label{sec:observations}

Initial testing showed that \sneia{} observations spanning the NIR-peak produce RMS dispersions in Hubble residuals $\sim 10-30\%$ lower than values obtained for \sneia{} only observed after the peak.
Thus, to obtain distances to \sneia{} and recessional velocities for their host galaxies, we require three types of data: high-cadence photometry to find \sneia{} before they reach their NIR peaks, NIR photometry of each \snia{} near their peaks, and spectroscopically determined redshifts of their host-galaxies.

\begin{figure*}
    \centering
    \includegraphics[width=0.9\textwidth]{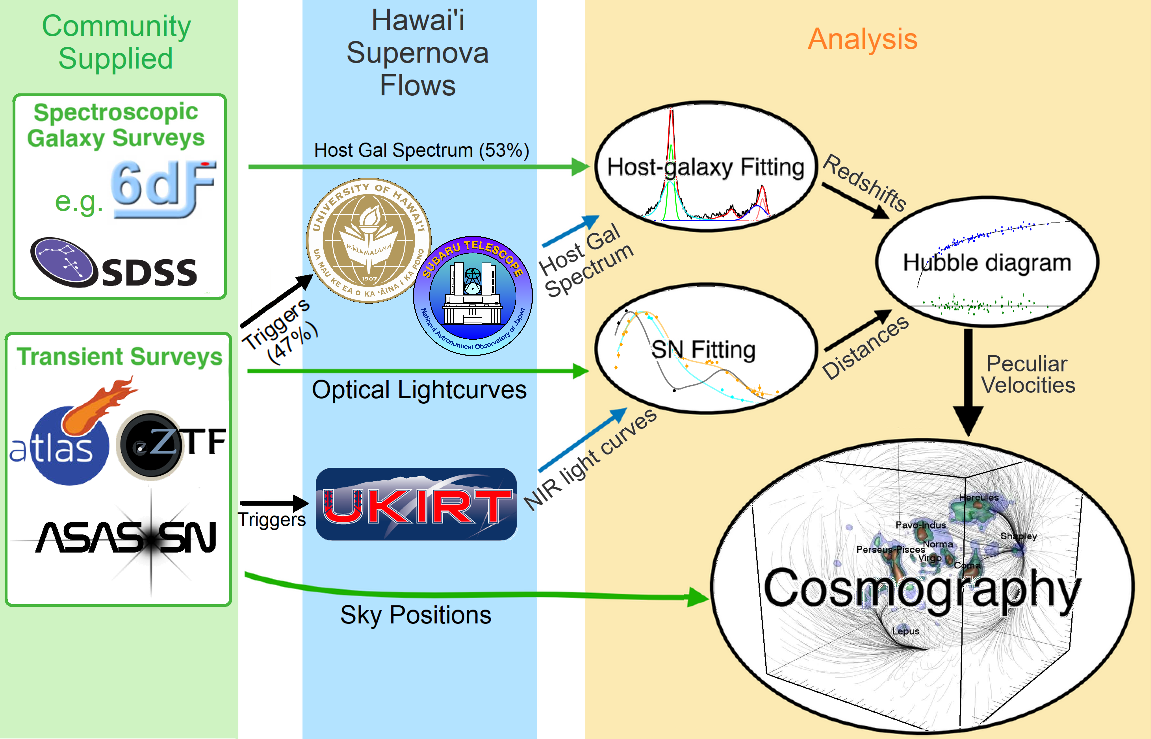}
    \caption{Our project uses archival and survey data as a foundation for supplementary observations.
    Whenever a new transient is reported, we collect optical light curves from ATLAS, ASAS-SN, and ZTF.
    We trigger NIR observations for targets that are either classified as \sneia{} or are unclassified and have a \snia{}-like light curve.
    About 53\% of targets we observe are associated with galaxies that have spectroscopic redshifts, and we pursue spectroscopic observations for the remaining 47\% with either the University of Hawai`i (UH) 2.2 m telescope or Subaru depending on their surface brightness profiles.
    The optical and NIR light curves allow us to infer luminosity distances, which we combine with host-galaxy redshifts to derive peculiar velocities.
    }
    \label{fig:workflow}
\end{figure*}

Figure \ref{fig:workflow} illustrates the various components of the program, delineating what is supplied from the community and what requires dedicated observing resources.

\subsection{Triggers from All-Sky Surveys}
\label{sec:all-sky}

The entire sky is imaged multiple times per night by All-Sky Surveys like the Asteroid Terrestrial-impact Last Alert System \citep[ATLAS;][]{tonry18}, the Zwicky Transient Facility \citep[ZTF;][]{bellm19}, and ASAS-SN.
These surveys operate with different cadences and depths to cover a range of science cases, but they all search the sky for objects that vary on timescales of hours, days, or months.
\sneia{} are in this class of astronomical objects, with light curves that increase in brightness for a few weeks before peaking, declining over a month, and then exponentially decaying.
Here we describe the archival and observational facilities used, and how we access, store, and process the data.

\subsubsection{The Transient Name Server}
The Transient Name Server (TNS)\footnote{\url{https://www.wis-tns.org/}} is the official International Astronomical Union repository for extra-galactic transients.
Large observational campaigns such as Pan-STARRS \citep{chambers16}, GaiaAlerts\footnote{\url{http://gsaweb.ast.cam.ac.uk/alerts}} \citep{gaia16, gaia18}, the surveys described in the following sections, and many more automatically generate reports within minutes to hours of exposure read-out.
Averaging overall reports from TNS, about 10\% of transients receive observational follow-up and spectroscopic classification, and of these, about 70\% are \sneia{}.\footnote{\url{https://www.wis-tns.org/stats-maps}}
The majority of transients fade and become unobservable without being classified.

\begin{figure}
    \centering
    \includegraphics[width=0.45\textwidth]{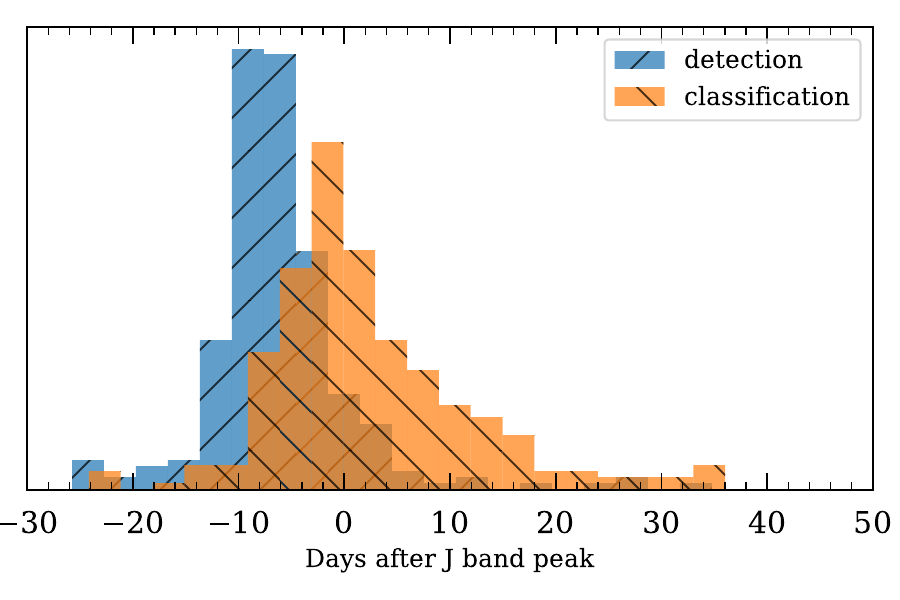}
    \caption{The distribution of detection dates and public classification dates for \sneia{} relative to the $J$-band maximum light. About 40\% of all \sneia{} are classified more than a day before the NIR peak.}
    \label{fig:tns_hist}
\end{figure}

The \hsf{} project uses the TNS-provided Python code\footnote{\url{https://www.wis-tns.org/sites/default/files/api/tns_api_search.py.zip}} to solicit new and recently updated reports every half hour, and uses these reports to generate a list of \sneia{} candidates.
We ignore transients that are classified as anything other than a \snia{} or non-peculiar sub-type, but still consider unclassified transients as potential \sneia{}.
This leads to some NIR observations of targets that are later classified as non-\snia{}, but we cannot afford to wait for spectroscopic classification of each target, which often occurs after the NIR-peak as seen in Fig. \ref{fig:tns_hist}.

The reduction in efficiency can be mitigated in several ways.
The \hsf{} team relays targets of interest to the Spectroscopic Classification of Astronomical Transients (SCAT) program \citep{tucker22}.
The SCAT team classifies astronomical transients using spectra primarily from the University of Hawai`i (UH) 2.2 m telescope (instrumentation described in more detail in section \ref{sec:snifs}), but has recently expanded to the Australian National University 2.3 m telescope through a collaboration with Melbourne University.
In a random sampling of TNS objects, one would expect 10\% to be classified, but by providing SCAT with a list of candidates to observe, we increase the fraction of classified transients in our observed sample to about 73\%.
Additionally, \citet{moller20} demonstrated that using whole light-curves, \sneia{} and non-\sneia{} can be identified with up to 95\% accuracy, or 98\% accuracy when including host-galaxy information.
Even when restricting the light curves to early times, the difference in light-curve shape between various SNe allows us to avoid observing unclassified targets that are unlikely to be \sneia{}.
The demographics of \hsf{} targets are presented in Table \ref{tab:demographics}.

\begin{table}
    \begin{tabular}{ |r|l| }
    \hline
    Type & Number \\ \hline
        SN Ia-norm & 637 \\ \hline
        SN Ia-91T-like & 25 \\ \hline
        SN Ia-91bg-like & 6 \\ \hline
        \hline
        Unclassified & 327 \\ \hline
        \hline
        SN Ia-pec & 3 \\ \hline
        SN Ia-CSM & 2 \\ \hline
        SN Iax[02cx-like] & 2 \\ \hline
        SN Ia-SC & 1 \\ \hline
        SN II & 93 \\ \hline
        SN IIn & 15 \\ \hline
        SN Ic & 12 \\ \hline
        SN Ib & 11 \\ \hline
        CV & 7 \\ \hline
        SN IIP & 6 \\ \hline
        SN & 4 \\ \hline
        SN Ibn & 4 \\ \hline
        SN Ib/c & 3 \\ \hline
        SN I & 3 \\ \hline
        SN Ic-BL & 3 \\ \hline
        Nova & 2 \\ \hline
        SLSN-II & 2 \\ \hline
        LRN & 1 \\ \hline
        AGN & 1 \\ \hline
        SN Ib-Ca-rich & 1 \\ \hline
        Varstar & 1 \\ \hline
        SLSN-I & 1 \\ \hline
        Impostor-SN & 1 \\ \hline
        ILRT & 1 \\ \hline
    \end{tabular}
    \caption{The spectroscopic classifications of our observed targets show that a strong majority of our targets receive classification, and most of those classifications are some kind of \snia{}.
    We need to select targets before they are classified in order to observe the NIR peak, which precedes the optical peak by several days.
    This results in some observations of non-\snia{} objects, but these targets are promptly removed from the observing queue.}
    \label{tab:demographics}
\end{table}

The following sections describe three untargeted surveys with publicly available light-curve generation services that we use to improve our triggering process, and as later detailed in section \ref{sec:distances}, improve our distance determinations.

\subsubsection{ATLAS}
\label{sec:atlas}
ATLAS consists of four fully robotic, 0.5 m f/2 Wright Schmidt telescopes that image the entire night sky about once every two days \citep{tonry11, tonry18}.
This system was designed to identify potentially hazardous asteroids, and optimizations for that purpose affect the utility of ATLAS in studying astrophysical transients.

An object's orbital elements are fairly decoupled from its spectral properties, so to increase throughput ATLAS uses two non-standard broad filters, a ``cyan'' filter covering 420–650 nm and an ``orange'' filter covering 560–820 nm.
This aids its primary science mission by increasing ``survey speed'' \citep{tonry11}, but presents unique challenges for integrating observations with other filter systems which we describe in section \ref{sec:snpy}.

Additionally, to specialize in moving object detection, the telescope system observes each field of view with four 30-second exposures over a one-hour interval.
Under nominal conditions, each 30-second exposure reaches a median $5\sigma$ detection limit of $o \sim 19.1$ AB mag and $c \sim 19.6$ AB mag.
For stationary targets, these exposures can be co-added to improve depth by about 0.75 AB mag and increase the Signal-to-Noise-Ratio (SNR) at a given brightness by a factor of 2.
However, we found that inter-observational variation in point spread function (PSF), pointing, and atmospheric conditions made combining multiple exposures difficult.
Instead, we combine the four photometric measurements of each object using an inverse variance weighted median, excluding any measurement more than three times its uncertainty away from the median flux.
Additionally, we ignore measurements where the object is within 40 pixels of a chip edge or has an axis ratio greater than 1.5 and measurements where the sky brightness is under 16.

Although ATLAS specializes in astronomy at the Solar System scale, it is a leading source of high-cadence data for studying astrophysical transients.
\citet{smith20} describe the utility of ATLAS in this context and how to access data using the ATLAS Forced Photometry server.\footnote{\url{https://fallingstar-data.com/forcedphot/}}
\hsf{} continues to use the proprietary channel we developed to access light curves before the forced photometry server came online, but the data collected exactly match the publicly available data.

\subsubsection{ASAS-SN}
ASAS-SN is a globally distributed system of 20 fully robotic telescopes focused on discovering bright, nearby SNe \citep{shappee14, kochanek17, hart23}.
Each of the five ASAS-SN sites employs four 14 cm telescopes sharing a common mount.
The original two sites used the Johnson \textit{V}-bandpass, but since 2019 all observations use the Sloan \textit{g}-bandpass \citep{holoien20}.
Each pointing consists of three dithered 90 second exposures, reaching median $5\sigma$ detection limits of 17.8 AB mag each \citep{kochanek17}.
These exposures can be co-added to improve depth by about 0.6 AB mag and increase SNR by a factor of $\sqrt{3}$.
The system images the entire sky about once every 20 hours, with few losses due to weather because of the numerous sites.

The ASAS-SN light-curve server described in \citet{kochanek17} has grown into the ASAS-SN Sky Patrol,\footnote{\url{https://asas-sn.osu.edu/}} which serves light-curves for any position on the sky.
As with ATLAS, we access this publicly available data using a proprietary channel to minimize overheads.

\subsubsection{ZTF}
ZTF uses the Palomar 48-inch Schmidt telescope to pursue science objectives across a range of cadences, depths, and areas, with an emphasis on SNe \citep{bellm19, graham19}.
Through the public surveys, ZTF covered the night sky North of $\delta = -31\degree$ once every three days, increasing to once every two days with ZTF-II.

ZTF uses custom \textit{g-}, \textit{r-}, and \textit{i-} band filters designed to avoid prominent sky lines at the Palomar site.
These filters reach 30-second exposure $5\sigma$ limiting magnitudes of 20.8, 20.6, and 19.9 mag respectively.
Each field of view is typically imaged twice, once in ZTF-\textit{g} and once in ZTF-\textit{r} \citep{bellm19}.

The ZTF alert distribution system produces over a million alerts each night, which feed into brokers that parse the data and make it publicly available.
We access ZTF light-curves through the Automatic Learning for the Rapid Classification of Events (ALeRCE) broker's Python client\footnote{\url{https://alerce.readthedocs.io/en/latest/}} \citep{forster21}.

\subsubsection{Triggering Criteria}
When our half-hourly sync with TNS reveals a new target, we obtain light-curves from ATLAS and ZTF, and if the target is brighter than 18 mag in any filter we also obtain an ASAS-SN light-curve.
We then attempt to fit the data to a \snia{} model using \texttt{SNooPy} \citep{contreras10, burns11} and SALT3-NIR \citep{pierel22} (our fitting procedure is discussed further in Section \ref{sec:distances}).
We manually inspect the light-curves and fits to address two points: is the candidate consistent with a \snia{} and is it possible to obtain observations at or before the NIR peak?
If the candidate does not have spectroscopic classification, we assess the quality of successful fits.
If the residuals indicate a poor fit to the data, or if the reduced $\chi^2$ is greater than 2, we reject the candidate or defer judgment until more photometry becomes available.
We estimate the time of peak brightness in the NIR using the best-fitting SALT3-NIR parameters.
If the candidate is either classified as a \snia{} or is photometrically consistent with one, and if it has not yet reached its NIR peak, we pursue NIR observations as described in the following section.

\subsection{\hsf{} NIR Photometry}
\label{sec:NIR}

\subsubsection{UKIRT -- WFCAM}
For NIR observations, \hsf{} uses the Wide Field Camera (WFCAM) mounted on the UKIRT 3.8 m telescope owned and operated by the University of Hawai`i\footnote{\url{https://about.ifa.hawaii.edu/ukirt/}} \citep{hodapp18}.
UKIRT is a 3.8-m Cassegrain telescope on the summit of Maunakea.
It has a declination limit of $-40\degree < \delta < 60\degree 07'$, granting access to about 3/4 of the sky.
The Cambridge Astronomical Survey Unit (CASU) continues to provide data processing services and the Wide Field Astronomy Unit at the University of Edinburgh maintains the WFCAM Science Archive \citep{hambly08} through which data are distributed.

WFCAM is a NIR imager developed specifically for large-scale surveys \citep{casali07}.
Its four detectors are Rockwell Hawaii-II (HgCdTe 2,048$\times$2,048) arrays \citep{hodapp04} each covering $13.65'\times13.65'$ at a scale of about $0\farcs4$ per pixel.
With its $0.9\degree$ diameter focal plane, WFCAM enabled the UKIRT Infrared Deep Sky Survey \citep{lawrence07} and the UKIRT Hemisphere Survey \citep{dye18}.
\citet{hodgkin09} explain that an astrometric distortion causes the pixel scale to vary radially, with percent level differences in pixel area between the centre and edge of the focal plane.
This changes the flux from the sky in each pixel, but their Equation 1 provides a method for correcting this effect.
We confirm this spatial variation and its resolution through the provided correction.

WFCAM uses a set of five broad-band filters, ZYJHK, and two narrow-band filters, H2 1-0 S1 and 1.644 FeII.
Each detector is equipped with its own set of filters, with inter-detector filter variations leading to photometric differences of no more than 0.01 mag \citep{hewett06}.
The performance of WFCAM in the above filters were analysed in \citet{hodgkin09}, who compared instrumental magnitudes against the Two Micron All Sky Survey (2MASS) Point Source Catalog \citep{skrutskie06}.
We use the $J$-band colour equation they derive to convert 2MASS $J$ and $H$ magnitudes to WFCAM $J$ magnitudes, which we use to calculate zero-points for each image.

\citet{hodgkin09} also identified spatially correlated photometric variability, even when accounting for the astrometric distortion mentioned previously.
The exact cause of the issue is unknown, but CASU provides an empirically derived table of corrections on a monthly basis.
We address this spatial correlation independently by treating each image's zero-point as a 2nd-order two-dimensional polynomial centred on the SN candidate, inferred with the probabilistic programming language Stan (implemented through PyStan \citet{pystan}) for each image \citep{carpenter17, stan}.
Stan provides a framework for specifying fully Bayesian statistical models and conditioning them on data using a no-U-turn sampler \citep[NUTS;][]{hoffman11, hoffman14, betancourt13}, an adaptive variant of Hamiltonian Monte Carlo sampling \citep[HMC;][]{duane87, mackay03, neal11}.
The scale of the effect is $\sim$0.021 mag from the centre to the edges of the image, comparable to the tables provided by CASU.

\subsubsection{Source Characterization and Galaxy Subtraction}
The data distributed through the WFCAM Science Archive include catalogues of photometric parameters for sources extracted with the program \texttt{imcore}.\footnote{\url{http://casu.ast.cam.ac.uk/surveys-projects/software-release/imcore}}
Initial testing highlighted issues in the catalogues when point sources coincided with extended sources.
This compromised the photometry of most \sneia{} that were not exceptionally well separated from their host-galaxy.

\begin{figure*}
    \centering
    \includegraphics[width=0.9\textwidth]{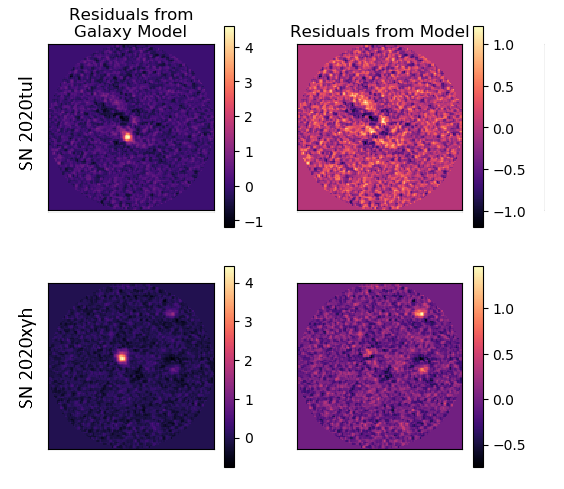}
    \caption{The forward-modelling code used to make photometric measurements produces diagnostic images showing the observed flux, the galaxy model, the residuals after subtracting the galaxy model (``Residuals from Galaxy Model''), and the residuals after subtracting both the galaxy and supernova models (``Residuals from Model''). We present a few examples representative of bad subtractions. SN 2020tul shows spatially correlated structure after galaxy subtraction, indicating the galaxy was not accurately modeled. Additionally, the supernova appears oversubtracted. This effect is more clearly seen in SN 2020xyh, which occurred near the nucleus of its host galaxy. The images on the left seem to show the galaxy has been subtracted, leaving only point sources at the location of the supernova and two nearby galaxies, but the images on the right show that SN 2020xyh appears to leave a small hole in some unmodelled structure.}
    \label{fig:1D3_vs_2D_examples}
\end{figure*}

Leveraging the multiplicity of our observations, we analysed each supernova and host-galaxy image series as an ensemble using the forward-model (or scene-model) code from \citet{rubin21}.
In short, this procedure assumes a series of images contains a time-independent two-dimensional surface (modelled with splines) and a time-varying point-source.
This allows for degeneracies when `sharp' features in the galaxy (such as the nucleus) coincide with the \snia{}, but late-time observations of the galaxy taken after the \snia{} has faded resolve this issue by essentially providing a traditional reference image for subtraction.
We manually determine which host-galaxies require late-time observations using diagnostic images such as those in Figure \ref{fig:1D3_vs_2D_examples}.
We pursue late-time observations if the galaxy model exhibits sharp features at the site of the SN, or if the residuals after subtracting either the galaxy or the galaxy and SN appear to have spatial structure.

\begin{figure*}
    \centering
    \includegraphics[width=0.9\textwidth]{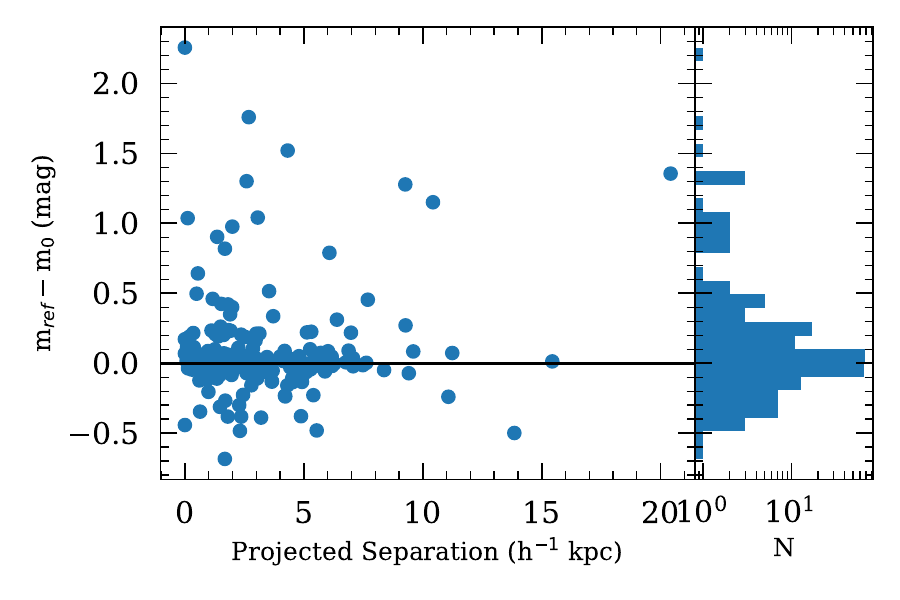}
    \caption{The differences between measurements made with and without late-time observations are minimal for a large number of targets, indicating accurate reconstruction of the galaxy surface profile.
    However, there are many targets where a late-time observation is crucial for decoupling the SN and host galaxy.
    The abundance of targets needing late-time observation is biased high in the plotted data because all targets were manually determined to potentially benefit from late-time observations.
    }
    \label{fig:1D3_vs_2D}
\end{figure*}

We use the subsample of targets with late-time observations to validate our methodology against an independent data reduction process using traditional image subtraction performed with \texttt{ISIS} \citep{alard98, alard99} and source characterization using \texttt{tphot} \citep{sonnett13}.
The differences between the forward-modelled and image-subtracted photometry have a median of 0.008 mag and a standard deviation of 0.07 mag.
We also examine how the forward-modelling code performs without late-time observations, and find the median difference remains low at 0.02 mag, but the standard deviation increases to 0.826 mag.
This increase is driven by a few cases where the forward-modelling code struggled to separate the galaxy and the transient.
Figure \ref{fig:1D3_vs_2D} shows the average difference in a galaxy's forward-modelled photometry with and without late-time observations as a function of projected separation between the supernova and host-galaxy nucleus.
The histogram shows that in the majority of cases, late-time observations do not result in significantly different photometry.
In a few cases, the observations break degeneracies in the forward-modelling process, resulting in photometry up to a few magnitudes different.
These cases are visually conspicuous, as seen in Figure \ref{fig:1D3_vs_2D_examples}.
In Appendix \ref{appendix:1D3_vs_2D}, we fit a Gaussian mixture-model to the photometric differences ($\Delta m$) using Stan \citep{carpenter17, stan} and find 74.0\% of the differences appear tightly dispersed ($\Delta m \sim \mathcal{N}(0.01 \pm 0.004 \text{ mag}, (0.08 \pm 0.005 \text{ mag})^2)$), and the remaining 26.0\% vary much more dramatically ($\Delta m \sim \mathcal{N}(0.33 \pm 0.050 \text{ mag}, (0.68 \pm 0.037 \text{ mag})^2)$).
The fraction of targets reliant upon late-time observations for accurate photometry is vastly exaggerated in this analysis because the subsample comprises only targets manually determined to potentially benefit from late-time observations.
Forward-modelled photometry is thus as accurate as traditional image subtraction, and more economical in that it often does not require a late-time observation.

\subsection{Host Galaxy Redshifts}
\label{sec:host z}
Although dozens of surveys have collectively measured redshifts for millions of galaxies, about half of the \sneia{} in our sample have host galaxies with no publicly available redshifts.
Furthermore, the redshift measurements that are publicly available come from heterogeneous methodologies and at times are inconsistent with other measurements of the same galaxy.
Here we describe how we identify host-galaxies for each \snia{}, incorporate data from extant surveys, and obtain redshifts for galaxies that do not have publicly available spectroscopic redshifts.

\subsubsection{Identifying Host Galaxies}
All SN host galaxies in our survey were identified manually.
This decision introduces an unquantified systematic error in our final peculiar velocity measurements due to the possibility of inaccurate host galaxy identification.
Without a detailed simulation, it is unclear how often we misidentify host galaxies.
However, the error rate is definitively lower than an algorithmic approach we tested, which produced obvious misidentifications.
This alternative approach is detailed in Appendix \ref{appendix:DLR}.

The \snia{}-galaxy associations produced manually were flagged if the host galaxy was ambiguous or otherwise problematic.
These manual flags allow us to exclude these \sneia{} in our analyses, but introduce a hard-to-quantify bias \citep{gupta16}, and will not scale well if operations significantly expand.
Recent work \citep[e.g.][]{aggarwal21, qin22} has formalized various methods of associating transient events with their host-galaxies using objective parameters, but still critically depends on the completeness and accuracy of galaxy catalogues.
Automatic association will become necessary when our sample expands, but we will continue to associate \sneia{} and their host galaxies manually while it remains accurate and practical.


\subsubsection{Incorporating Redshifts from Literature}
Before we pursue spectroscopic observations to find each host-galaxy's redshift, we search for existing measurements in the HyperLEDA database \citep{paturel03, paturel03b, makarov14}, which is based on the Lyon-Meudon Extragalactic Database \citep[LEDA;][]{paturel88} and Hypercat \citep{prugniel96}.
This significantly reduces our observational needs, but the variety of measurement techniques necessitates the careful handling of systematic differences.
HyperLEDA uses a system of quality flags\footnote{\url{http://leda.univ-lyon1.fr/a110/}} to hierarchically combine optical and radio redshift measurements, and applies corrections on a reference by reference level to minimize systematic offsets between data sources \citep{paturel97}.
If a host galaxy does not have a radial velocity in HyperLEDA, we pursue spectroscopic observations.

\subsubsection{UH 2.2 m -- SNIFS}
\label{sec:snifs}
The primary instrument we use for measuring host-galaxy redshifts is the Supernova Integral Field Spectrograph \citep[SNIFS;][]{lantz04} on the UH 2.2 m Telescope.
SNIFS samples a $6'' \times 6''$ field with $0\farcs4 \times 0\farcs4$ spaxels, each of which produces two spectra, one blue (320 -- 560 nm, R(430 nm) $\sim$ 1,000) and one red (520 -- 1,000 nm, R(760 nm) $\sim$ 1,300).
Our exposure times are manually chosen based on galaxy surface brightness, atmospheric conditions, and galaxy spectral type, with late-type galaxies typically featuring emission lines and thus requiring less integration.
The average exposure time was 1,800 s
We use the data reduction pipeline described in \citet{tucker22} to produce one-dimensional spectra.
Absolute wavelength calibration is provided by arc-lamp exposures taken immediately after each science exposure.
We include the average discrepancies between the arc spectra and their models when calculating redshift uncertainties, though the contribution is typically sub-dominant at $\sim$1 km s$^{-1}$.
All galaxy spectra are converted to the heliocentric rest frame.

\subsubsection{Subaru -- FOCAS}
When a galaxy is too faint for SNIFS, we use the 8.2 m Subaru telescope's Faint Object Camera and Spectrograph \citep[FOCAS;][]{kashikawa02} with its 300B grating with no filter (365 -- 830 nm, R(550 nm) $\sim 700$) and a 0\farcs6 or 0\farcs8 wide slit depending on the atmospheric conditions \citep{ebizuka11}.
Subaru's mirror has over 13 times more light-gathering power than the UH 2.2 m mirror.
This allows us to increase our limiting magnitude from $r < 19.1$ mag to $r < 22.9$ mag using comparable exposure times.

In addition to the increased light-gathering power, FOCAS's slit spectroscopy has proven necessary for very diffuse galaxies.
Our reduction pipeline for SNIFS spectra struggles with sky subtraction if the entire $6'' \times 6''$ microlens array is filled.
In such a case, we would need to obtain a sky observation for proper subtraction, doubling the exposure time required per object.
For each galaxy, we perform a 900 s exposure and examine the summit-pipeline-reduced spectrum.
If the galaxy has no strong emission lines, we pursue one or two additional 900 s exposures as deemed necessary by the observer.
We perform bias subtraction and flat-fielding data using the routines described in the FOCAS Cookbook.\footnote{\url{https://subarutelescope.org/Observing/DataReduction/Cookbooks/FOCAS_cookbook_2010jan05.pdf}}
We use skylines for relative wavelength calibration, and use Subaru's location, the time of each exposure, and the position of each target to transform all spectra to a heliocentric rest frame.

\subsubsection{Redshift Determination and Uncertainties}
Once we have spectra from either SNIFS or FOCAS, we compare them with spectral templates from SDSS DR5\footnote{\url{https://classic.sdss.org/dr5/algorithms/spectemplates/spectemplatesDR2.tar.gz}} \citep{adelman-mccarthy07} using the weighted cross-correlation routine in the SeeChange Tools\footnote{\url{https://zenodo.org/record/4064139\#.YHkLvC1h2X0}} \citep{hayden21}.
We tested the accuracy of this method by calculating redshifts for 158 galaxies using spectra from SDSS DR12, removing cross-correlations with an $r$-value less than 5 \citep[as defined in][]{tonry79}, and comparing our recession velocities with those in HyperLEDA. 
The differences averaged to $\sim$7 km s$^{-1}$ with a standard deviation of $\sim$45 km s$^{-1}$.
Thus we include a 45 km s$^{-1}$ uncertainty when inferring host-galaxy redshifts using this cross-correlation technique.

\begin{figure}
    \centering
    \includegraphics[width=0.45\textwidth]{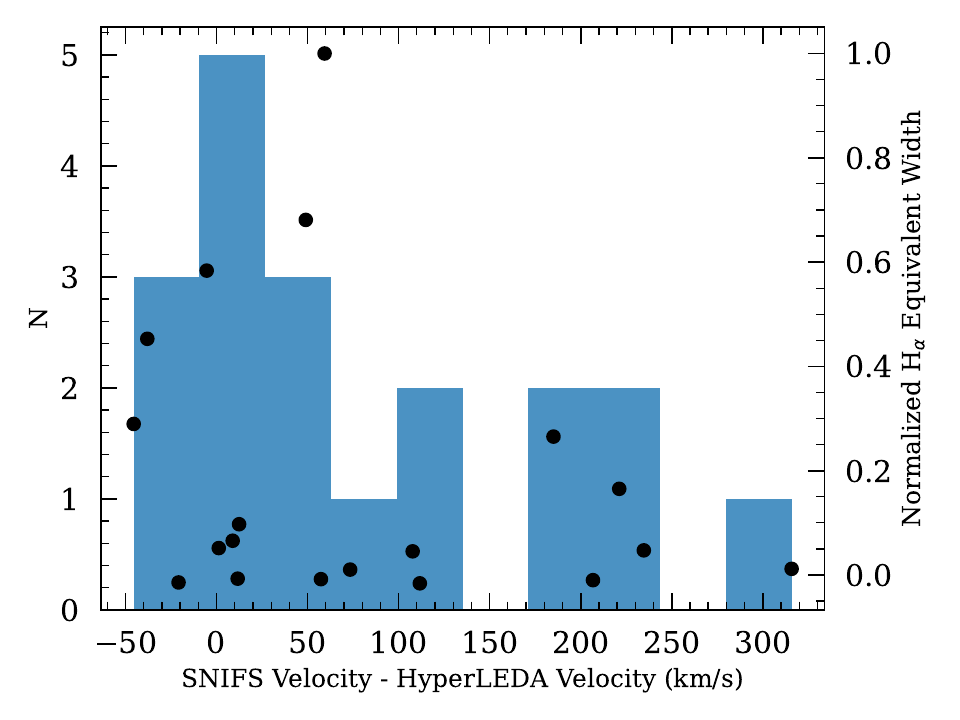}
    \caption{We compare the differences between our measured SNIFS velocities and HyperLEDA's aggregated velocities, finding two distinct groups.
    The 12 galaxies on the left side of the histogram have an average difference of 27 km s$^{-1}$ and a standard deviation of 48 km s$^{-1}$.
    The 4 galaxies on the right are offset by several hundred km s$^{-1}$.
    Three of the four HyperLEDA velocities come from relatively older sources, and could be the result of inaccurate methodologies.
    Each galaxy in the histogram also has a marker with a y-value corresponding to its rescaled H$_\alpha$ equivalent width.
    The presence of weak equivalent widths in the sample with low velocity differences suggests that the four discrepancies are not due to weak spectral features.}
    \label{fig:snifs_abs_cal}
\end{figure}

Additionally, we looked for systematic differences in absolute wavelength calibration between redshifts from literature and redshifts from our SNIFS and FOCAS spectra.
We observed 24 galaxies with redshifts available in HyperLEDA using SNIFS, and 4 using FOCAS.
Five of our SNIFS spectra had insufficient SNR and are not included in this analysis.
The 19 remaining spectra yielded redshifts within about 100 km s$^{-1}$ of their HyperLEDA values, with a few exceptions.
We measure five galaxies to have redshifts several hundred km s$^{-1}$ greater than their literature values.
In descending order of discrepancy, these galaxies are PGC 40363, 4579, 29889, 13428, and 1033041, shown in the right side of Fig. \ref{fig:snifs_abs_cal}.
These galaxies include early and late-type morphologies, emission and absorption spectra, and their colours are not at the extremes of the 19 galaxy sample.
The only unifying theme is that HyperLEDA sources the PGC 40363, 4579, and 13428 from relatively older sources \citep{eastmond78, sakai94, thoraval99}, whereas PGC 29889 and 1033041 have more recent measurements, such as those from SDSS or 6dF.
HyperLEDA aggregates and weights various sources, which should privilege more accurate observations, but these galaxies have only been spectroscopically observed once or twice before our observations with SNIFS.
It is unclear why our measured redshifts are uniformly greater than their literature values.
Disregarding these five exceptions, the average difference between the SNIFS-derived and HyperLEDA redshifts is $\sim$27 km s$^{-1}$ with a standard deviation of $\sim$48 km s$^{-1}$.
Including them, the average and standard deviation rise to $\sim$81 and $\sim$102 km s$^{-1}$ respectively.
We subtract $\sim$27 km s$^{-1}$ from our SNIFS-derived redshifts and interpret the $\sim$48 km s$^{-1}$ standard deviation as a rough confirmation of the previously identified $\sim$45 km s$^{-1}$ uncertainty.
We also note that redshifts in HyperLEDA that have not been verified through repeated observations could benefit from additional measurements.

\begin{figure*}
    \centering
    \includegraphics[width=0.9\textwidth]{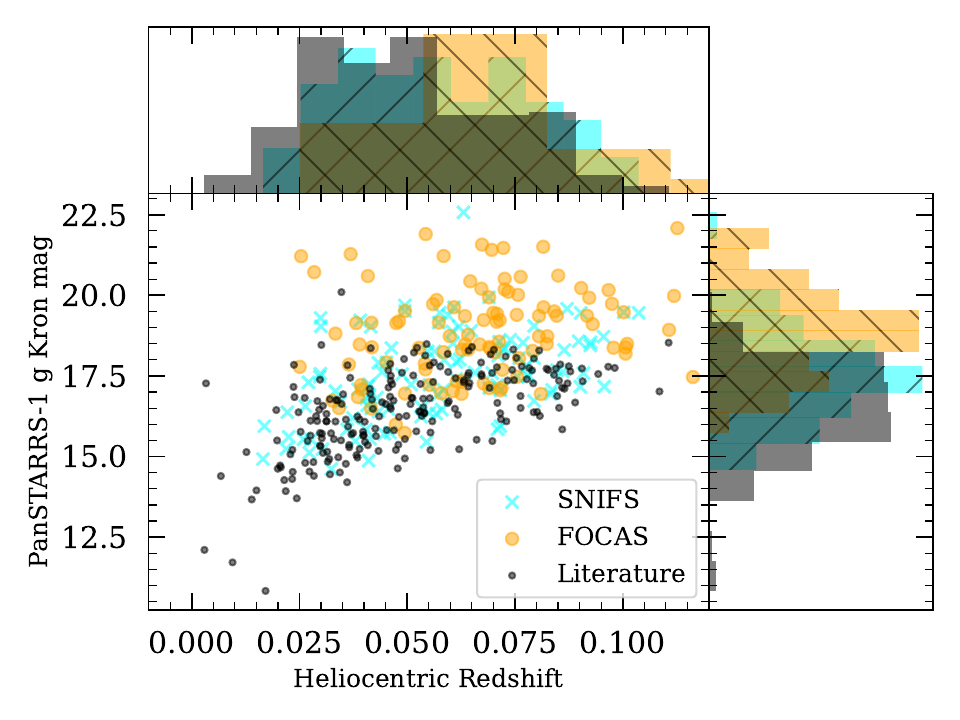}
    \caption{The normalized histograms of targets from SNIFS, FOCAS, and literature in $z$ show distinct redshift distributions.
    Each galaxy's Pan-STARRS $g$ Kron mag is plotted against redshift to show that the distribution in magnitudes are also distinct.
    Triggering only on transients in galaxies with redshifts in the literature biases the sample towards lower redshifts and brighter galaxies.
    By triggering on galaxies regardless of redshift availability, we mitigate this issue.
    }
    \label{fig:z_distribution}
\end{figure*}

We note that galaxies in larger groups will have an additional velocity term due to intracluster dynamics, and that using the group redshift would likely probe large-scale flows more robustly, as done in \citet{peterson22}.
However, pursuing spectroscopic observations for all members of an associated group would reduce the number of \sneia{} host galaxies we could observe.
We note that our analysis will benefit from future large spectroscopic surveys such as the Multi-Object Spectroscopy of Transient Hosts survey \citep[MOST Hosts;][]{soumagnac24} Dark Energy Spectroscopic Instrument \citep[DESI;][]{desi22}.

All redshift uncertainties are converted to uncertainties in distance modulus via the distance-redshift relation for an empty universe presented in \citet{kessler09sdss}:
\begin{equation}
    \sigma^z_\mu = \sigma_z \left(\frac{5}{\ln{10}}\right)\frac{1+z}{z(1+z/2)}.
    \label{eqn:redshift_uncertainty}
\end{equation}
Different cosmological models produce negligible differences in $\sigma^z_\mu$, which is already subdominant compared to other sources of uncertainty in the distance modulus.

\section{Distance Determination}
\label{sec:distances}

In this section, we describe the specific methodology used to convert our data into distance moduli using \texttt{SNooPy} and \texttt{SALT3-NIR} as they were the only publicly available fitting programs that can utilize optical and NIR observations when our analyses began.
We only intend to describe our fitting procedures to contextualize the results presented in section \ref{sec:results}, and as such we will not be claiming one program is more accurate or more appropriate for our use case.
We leave such an analysis for future work, where we will also incorporate fits from BayeSN, which was made public with \citet{mandel22}, and has been updated with \citet{grayling24}.

\subsection{SNooPy}
\label{sec:snpy}
\texttt{SNooPy} is a Python package designed for fitting light-curves of \sneia{} from the Carnegie Supernova Project \citep[CSP;][]{contreras10, burns11}.
It estimates luminosity distances by comparing data spanning flux, phase, and a shape parameter to filter-specific three-dimensional models \citep{burns11}.
These models were produced using high-cadence observations of \sneia{} in the CSP photometric system \citep{hamuy06}.
We use version 2.6.0, which does not yet include the improved models of \citet{lu23}.
We look forward to reprocessing our sample when \texttt{SNooPy} incorporates these templates.
Decreased systematic uncertainties in the NIR SED could increase the weight of $J$-band photometry in a multi-band fit.


\texttt{SNooPy} is described by the CSP\footnote{\url{https://csp.obs.carnegiescience.edu/data/snpy}} as ``not a fixed algorithm for fitting lightcurves, but rather a collection of tools that are useful for building your own fitter (or fitting interactively)''.
As such, there are a variety of non-trivial decisions that influence the distance moduli inferred using \texttt{SNooPy}.
In version 2.6.0, there are three primary decisions:
\begin{itemize}
    \item Parametrizing shape with $\Delta m_{15}$ or $s_{BV}$.
    \item Selecting one of the available models: EBV\_model, EBV\_model2, max\_model, max\_model2, Rv\_model, color\_model, SALT\_model, and MLCS\_model.
    \item Selecting a ``calibration'' to describe the correlation between SN parameters and absolute magnitude.
    \item Selecting a reddening law.
\end{itemize}
We describe and qualitatively justify our choices here, but refer the reader to Appendix \ref{appendix:snpy_calibration} for a quantitative analysis exploring alternative decisions.

\subsubsection{Choice of Shape Parameter}
\texttt{SNooPy} offers two distinct ways to characterize the shape of a \snia{} light-curve; one being the decline rate parameter \citep[$\Delta m_{15}$;][]{phillips99}, and the other being the colour-stretch parameter \citep[s$_{BV}$;][]{burns14}.
The latter is less sensitive to changes in reddening (varying $\sim1$\% across A$_V$ = 3 mag) and does not become degenerate for fast-declining \sneia{} ($s_{BV} < 0.7$), as seen with $\Delta m_{15}$ \citep{burns14}.
As such, we use $s_{BV}$ when characterizing light curves with \texttt{SNooPy}.

\subsubsection{Choice of Model}
The \texttt{SNooPy} models are described more comprehensively in the online documentation\footnote{\url{https://users.obs.carnegiescience.edu/cburns/SNooPyDocs/html/models.html}}, but we summarize them here to provide context for our decision.

The EBV\_model and EBV\_model2 use light curves in numerous filters to infer four parameters of each \snia{}: the shape, the time of B-band maximum, the colour excess of the host galaxy ($E(B-V)_\text{host}$), and distance modulus ($\mu_\text{cos}$).
The EBV\_model is restricted to using $\Delta m_{15}$ while the EBV\_model2 can use that or $s_{BV}$ to parametrize shape.
They also differ in that the former model approximates the luminosity-shape correlation as a linear function using the 6 calibrations presented in \citet{folatelli10}, whereas the latter model uses a quadratic function calibrated with additional CSP data.
In the EBV\_model2, the cosmological distance modulus $\mu_\text{cos}$ is related to the observed apparent magnitude in observer-frame bandpass $OF$ at time $t$ since $B$-band maximum ($m_{OF}(t)$) using a template in rest-frame bandpass $RF$ with shape factor $s_{BV}$ at de-redshifted time since $B$-band maximum $t'$ ($T_{RF}(t', s_{BV})$) with the following equation.
\begin{equation}
    \begin{split}
        m_{OF}(t) =& T_{RF}(t', s_{BV}) + P_0 + P_1(s_{BV} - 1) + P_2(s_{BV} - 1)^2 \\
        & + \mu_\text{cos} + R_{OF} E(B-V)_\text{MW} + R_{RF} E(B-V)_\text{host} \\
        & + K_{OF,RF}(T_{RF}(t', s_{BV}), z, R_{OF}, R_{RF})
    \end{split}
    \label{eqn:snpy_ebv_model2}
\end{equation}
where $P_0$, $P_1$, and $P_2$ are polynomial coefficients defined by the calibration, $R_{OF}$ and $R_{RF}$ are the total-to-selective absorptions of the Galactic and host galaxy dust, and $K_{OF, RF}$ is the cross-band k-correction (described in Section \ref{sec:K_S_corrections}).
$R_{OF}$ and $R_{RF}$ are calculated by using an $R_V$-dependent reddening law to compute synthetic extinction values.
We assume the Galactic average of $R_V = 3.1$ \citep{schlafly11} for calculating $R_{OF}$, and calculate $R_{RF}$ with the calibration-provided value for host galaxy $R_V$.
If parametrizing shape with $\Delta m_{15}$, the template term changes to $T_{RF}(t', \Delta m_{15})$ and the shape polynomial's $(s_{BV} - 1)$ terms change to $(\Delta m_{15} - 1.1)$.

The max\_model and max\_model2 also simultaneously fit light curves in multiple bandpasses, but relax the requirement that the photometry follows a well-characterized reddening law.
Like the previous models, the two max\_models fit for a global shape parameter and time of B-band maximum, but rather than fit for a distance modulus and host galaxy colour excess, these models fit for a peak apparent magnitude in each rest-frame bandpass ($m_{RF}$).
\begin{equation}
    \begin{split}
        m_{OF}(t) =& T_{RF}(t', s_{BV}) + m_{RF} + R_{OF}E(B-V)_\text{MW} \\
        & + K_{OF,RF}(T_{RF}(t', s_{BV}), z)
    \end{split}
    \label{eqn:max_model_fit}
\end{equation}
$m_{RF}$ is not necessarily equal to $m_{RF}(t=0)$ because the evolution of \snia{} SEDs produces peaks in different bandpasses at different times \citep[e.g.][]{kasen06b, krisciunas09, phillips12, burns14}.
While distance moduli are not fitting parameters, they can be calculated based on each apparent maximum using a Tripp-like formula, such as the one presented in Equation 4 of \citet{burns18}:
\begin{equation}
    m_X = P_0 + P_1(s_{BV} - 1) + P_2(s_{BV} - 1)^2 + \mu_\text{cos} + \beta_\text{max} (m_Y - m_Z)
    \label{eqn:max_model}
\end{equation}
where $m_X$, $m_Y$, and $m_Z$ are the peak apparent magnitudes determined by the max\_model fit in the bandpasses $X$, $Y$, and $Z$ (these arbitrary labels are not to be confused with the $Y$ or $Z$ bandpasses).
The difference between the max\_model and max\_model2 is that the latter allows for each bandpass to correspond to a unique time of $B$-band maximum light.

The Rv\_model is similar to the EBV\_model, in that it uses $\Delta m_{15}$ and models the luminosity-shape correlation as a linear function.
The primary difference is that the total-to-selective extinction of the host galaxy is a fitting parameter rather than a global constant taken from the calibration.
Additionally, rather than using the \citet{folatelli10} values for calibrating luminosity, shape, and colour, this model uses values from \citet{burns11}.

The color\_model infers the shape parameter (only $s_{BV}$), the time of $B$-band maximum, and the peak apparent $B$ magnitude, but also uses the difference between the observed colours and the intrinsic colours found in the 81 \sneia{} in \citet{burns14} to infer the host galaxy colour excess and $R_V$.
Unfortunately, this model requires observations in the rest-frame $B$-band, and as such we cannot evaluate this model.
It is possible to use cross-band k-corrections to infer a rest-frame $B$-band light curve, but doing so would increase our vulnerability to differences between the real and modelled SED.

Lastly, the SALT\_model and MLCS\_model are wrappers for running the SALT2 and MLCS2k2 fitters in the \texttt{SNooPy} framework.
Neither SALT2 \citep{guy07} nor MLCS2k2 support NIR bandpasses \citep{jha07}, so we do not consider these models for \hsf{}.

We decide to use EBV\_model2 and the max\_model because they support the use of $s_{BV}$ and quadratic luminosity-shape correlations, both of which are favored over their alternatives \citep{burns14}.
Furthermore, recent work has made use of both the EBV\_model2 \citep{phillips22, jones22, pierel22, peterson23} and the max\_model \citep{burns18, phillips22, uddin23, lu23}.
We do not use the color\_model because our observed bandpasses do not overlap with rest-frame $B$-band in the majority of the redshift range we cover.
We do not use the max\_model2 because our $J$-band light curves are sparse and often times insufficient for estimating the time of $B$-band maximum alone.

\subsubsection{Choice of Calibration}
The choice of ``calibration'' refers to the values parametrizing the correlation between luminosity, shape, and colour (e.g. $P_0$, $P_1$, $P_2$ and $\beta_\text{max}$ in Equation \ref{eqn:max_model}).
These values come from fits to samples of \sneia{} observed by CSP.
While the method of fitting varies ($\chi^2$ minimization in \citet{folatelli10} and Markov chain Monte Carlo (MCMC) methods in \citet{burns11, burns14, burns18}), differences in calibration values are primarily driven by variation in the samples used.
As an example, the first calibration from \citet{burns18} was produced from 137 \sneia{}, but there are alternative calibrations, one excluding \sneia{} with $s_{BV}$ values less than 0.5, one excluding those with $m_B - m_V$ pseudo-colours greater than 0.5 mag, and one excluding those that meet either criteria.
For our EBV\_model2 fits, we use the calibration from \citet{burns18} based on the full sample since the reduced $\chi^2$ values of the fits using the \citet{burns18} calibrations are typically lower than those using the \citet{folatelli10} calibrations (details in Appendix \ref{appendix:snpy_calibration}) and because there are \sneia{} in our sample that have $s_{BV}$ values less than 0.5 and $m_B-m_V$ colours greater than 0.5 mag.
For our max\_model fits we use Stan \citep{carpenter17, stan} to infer the nuisance parameters $P_0$, $P_1$, $P_2$, and $\beta_{max}$ using our photometry.
We omit the term correlating luminosity and host-galaxy mass to maintain consistency with EBV\_model2, which does not factor in galaxy mass.

\subsubsection{Choice of Reddening Law}
The final decision point is the choice of reddening law.
\texttt{SNooPy}'s default reddening law \citep[O94;][]{odonnell94} is a corrected version of the CCM89 reddening law \citep{cardelli89}.
It also natively supports the original, uncorrected version, as well as the reddening laws F99 \citep{fitzpatrick99} and FM07 \citet{fitzpatrick07}.
We have performed minor modifications to the \texttt{SNooPy} source code\footnote{Modified version available at \url{https://github.com/ado8/snpy}} to accommodate the reddening laws provided in the \texttt{dust\_extinction} package \citep{dust_extinction}.
After comparing fits produced with the O94, F99, and F19 \citep[also referred to as F20;][]{fitzpatrick19} reddening laws (details in Appendix \ref{appendix:snpy_calibration}) we find that the reduced $\chi^2$ values are typically lowest when using the F19 reddening law.
Thus we use the F19 reddening law for all \texttt{SNooPy} fits, which applies to both host galaxy and Galactic extinction in the EBV\_model2, but only Galactic extinction in the max\_model.
We assume the total-to-selective extinction parameter for Galactic dust is $R_V = 3.1$ and use the $R_V$ value defined in the \citet{burns18} calibration for host galaxy dust.
Galactic colour excess values comes from the SFD dust map \citep{schlegel98} with the 0.86 scaling factor described in \citet{schlafly11} (hereafter, the rescaled SFD dust map).

\subsubsection{Estimating Uncertainties}
\texttt{SNooPy} provides estimates of statistical uncertainty in all inferred parameters following either frequentist or Bayesian conventions.
Initial fits without priors produce statistical errors using the standard frequentist convention of inverting the Hessian matrix at the best-fitting parameters to produce a covariance matrix.\footnote{\url{https://users.obs.carnegiescience.edu/cburns/SNooPyDocs/html/fitting\_LM.html}}
When this matrix is singular, as can happen with undersampled light curves or for light curves of non-\sneia{}, the model becomes insensitive to one or more parameters and will not infer values for any of them.
After the initial fit, \texttt{SNooPy} offers an MCMC method which samples their posterior distributions with the package \texttt{emcee} \citep{emcee}.
The default priors are based on previous work with the CSP sample, but can be overwritten with arbitrary functions.

In addition to providing statistical errors, \texttt{SNooPy} provides an uncertainty floor for each parameter.
The floor in the distance modulus reflects the uncertainty in the various terms used to standardize \sneia{} luminosities.
These terms depend on the model used, but generally include filter-specific measurements of peak absolute magnitude and how that changes with $s_{BV}$.
Thus, the distance modulus accuracy has a systematic floor determined by the sample used to calibrate it and becomes less accurate as the shape factor deviates from its normal value.
The other floors have constant values derived from various analyses.
The uncertainty floor in $s_{BV}$ is 0.03, and comes from the dispersion around a quadratic fit of $s_{BV}$ to the SALT $x_1$ parameter (discussed in section \ref{sec:salt}) \citep{burns14}.
The host galaxy colour excess floor is 0.06 mag, coming from the intrinsic dispersion of the $m_B - m_V$ colours in the CSP sample after correcting for reddening. 
In the max\_model, the peak magnitudes in each bandpass are presented with uncertainty floors based on \citet{folatelli10}. 
Lastly, the time of B-band maximum is fixed to have an uncertainty floor of 0.34 days.
We define the uncertainty on each parameter estimate as the quadrature sum of the statistical uncertainty and the floor.

\subsubsection{K- and S-corrections}
\label{sec:K_S_corrections}
Observations of \sneia{} at significant redshift can lead to a mismatch between the observed and rest-frame spectral energy distribution (SEDs).
One could almost trivially account for this issue in spectral observations if the redshift is known (telluric corrections aside), but photometric observations require some knowledge of the underlying SED to determine what is shifted into and out of the effective bandpass.
The adjustments needed to compensate for the mismatches between observed and emitted SEDs are called ``K-corrections'' \citep{humason56, oke68}.

Similarly, variations in an optical system's transmission function leads to differences in instrumental magnitudes that depend on the SED observed.
\texttt{SNooPy} models are defined in the CSP photometric system, and using data from other bandpasses would introduce systematic errors in the parameter inferences.
The typical treatment for managing multiple filter sets is to observe a range of standard stars and perform linear fits of colour terms to transform one set to the other.
Using stellar standards produces equations capable of converting stellar observations between filter sets, but \sneia{} have non-stellar SEDs, and there are no perennially available standard \sneia{}.
The solution is to apply an ``S-correction'' \citep{burns11}.

\texttt{SNooPy} applies both of these corrections simultaneously by calculating a ``cross-band K-correction'' \citep{kim96} using the spectral library from \citet{hsiao07}, which combines $\sim 600$ heterogeneous spectra of $\sim 100$ \sneia{}.
Although the library covers a wide breadth, the available spectra cannot represent every kind of \snia{} at every possible epoch.
To account for levels of reddening and intrinsic colours not seen in the spectral library, \citet{hsiao07} describe a ``mangling'' process by which template spectra can be multiplied by a smoothly varying spline to match observed colours.
The statistical error on each K-correction and mangling varies between about 0.01 mag and 0.04 mag depending the amount of overlap between the redshifted rest-frame CSP bandpass and the observed bandpass.
Pairs with little overlap rely on extrapolation, and are more sensitive to the spectral template used \citep{hsiao07}, whereas a rest-frame bandpass that maps exactly on to an observed bandpass would be completely insensitive to the underlying spectrum.
The ATLAS $c$ and $o$ bandpasses are wider than those in the CSP photometric system, and so they necessarily belong to the former category.

\subsection{SALT}
\label{sec:salt}
SALT fits \sneia{} light-curves using a different approach \citep{guy05, guy07, guy10}.
Roughly speaking, where \texttt{SNooPy} attempts to fit observed light curves to well studied light curves, SALT attempts to fit observed light curves to a spectral time series.
This model is built from a term that describes the phase-independent effect of the colour law ($CL(\lambda)$) and two or more surfaces spanning flux, phase ($p$), and wavelength ($\lambda$), whose combinations describe the spectral flux  and evolution of all \sneia{}:
\begin{equation}
    \label{eqn:salt}
    F(p, \lambda) = x_0 [M_0(p, \lambda) + x_1 M_1(p, \lambda) + ... ] \times \exp{[c \times CL(\lambda)]},
\end{equation}
where $M_i$ is the $i$th surface, $x_i$ scales how much that surface contributes to the spectral flux, and $c$ scales the colour law \citep{guy07}.
The surfaces are empirically derived, with $M_0$ encapsulating the ``standard'' \snia{} spectral time series while the remaining surfaces describe all other modes of variation.
This means the surfaces themselves may not correlate exactly with the physical parameters of \sneia{}, but instead may be understood as principal components.
With that said, $x_1$ is often considered a shape factor like $s_{BV}$ or $\Delta m_{15}$ since light-curve shape seems to be the dominant mode of variation.
Each combination of $x$ terms defines a SED and evolution that can be further sculpted by $c$, the colour law, and redshift.
At any observational epoch, a filter set's transmission functions is used to make synthetic magnitudes which can be compared to real photometry.
Thus one can infer the most likely SALT parameters and their uncertainties given observations of a particular \snia{}.
These parameters provide a distance modulus ($\mu$) by the equation
\begin{equation}
    \label{eqn:salt_DM}
    \mu = m^*_B - M + \alpha x_1 - \beta c,
\end{equation}
where $m^*_B$ is the rest-frame Bessell $B$-band magnitude \citep{perlmutter97}, $M$ is the absolute magnitude of a \snia{} with $x_1=c=0$, and $\alpha$ and $\beta$ are standardization coefficients.
While $m^*_B$ can be approximated by $-2.5\log_{10}(x_0) + \text{const.}$, we calculate its value using synthetic photometry based on model parameters.

\citet{rubin20} suggested that \sneia{} luminosity variability may consist of three to five independent parameters.
Attempts to standardize \sneia{} luminosities using one or two parameters report an ``intrinsic scatter'' that cannot be explained by measurement error alone \citep[e.g.][]{scolnic18, brout22}.
\citet{rose20} explored the differences between two and seven-component fits using SNEMO \citep{saunders18}, and found that only CSP data had the SNR and coverage to constrain the additional parameters.
Put another way, a two-component fit with SALT compares to a seven-component fit with SNEMO for all but the most extensively covered light curves.
With that in mind, we use the two-component fits of SALT3-NIR \citep{pierel22}.
The only other SALT model that can process NIR light-curves is SALT2-Extended, but it was trained on optical data extrapolated to the NIR and is thus insensitive to correlations between SALT parameters and NIR light-curve properties \citep{pierel18}.
SALT3-NIR was jointly trained on the optical sample of 1083 \sneia{} from \citet{kenworthy21} and 166 \sneia{} with NIR data \citep{pierel22}.
We access the SALT3-NIR model through the Python package \texttt{SNCosmo} version 2.10.4 \citep{barbary22}, and utilize the convenience functions therein to account for Galactic extinction using the rescaled SFD dust map and the reddening law from \citet{fitzpatrick19} with $R_V=3.1$.
Notably, we use \texttt{SNCosmo} to calculate model fluxes given a set of \snia{} parameters, but do not use the built-in functions to estimate those parameters.
Instead, we use the fitting methodology of \citet{rubin23}, defining a $\chi^2$ function and using a downhill-simplex algorithm to iteratively identify the SALT parameters that minimize that function.

\subsubsection{Estimating Uncertainties}
The covariance matrices we obtain for each object's best-fitting SALT parameters (time of $B$-band maximum light, $x_0$, $x_1$, and $c$) reflect three sources of uncertainty.
Our NIR photometric methods produce correlation matrices, but we assume the measurements and errors from ATLAS, ASAS-SN, and ZTF are completely independent.
We incorporate the SALT3-NIR model uncertainties during our fitting process.
Lastly, we repeat each fit with slightly varied inputs to calculate derivatives between the fitting parameters and quantities like redshift, Galactic colour excess, and the photometric zero-point in each bandpass.


The error explicitly associated with K-corrections and S-corrections is ostensibly removed due to SALT's use of spectra when fitting.
However, if the intrinsic SED of a \snia{} differs from the form of Equation \ref{eqn:salt} truncated after $i=1$, the synthetic photometry will be inaccurate.
We assume these errors are encapsulated in the model uncertainties.

The distance modulus in Equation \ref{eqn:salt_DM} requires specifying the standardization coefficients $\alpha$ and $\beta$, which are typically calibrated empirically.
Fitting for $\alpha$ and $\beta$ by minimizing dispersion in the Hubble residuals introduces a form of Eddington bias due to uncertainties in $x_1$ and $c$.
We estimate the standardization coefficients using a Bayesian framework called UNITY \citep[Unified Nonlinear Inference for Type-Ia cosmologY;][]{rubin15, rubin23}.
UNITY assumes a Gaussian and skew normal distribution for the population distributions of the true value of each SN's $x_1$ and $c$ respectively, and uses flat hyperpriors for the means of each distribution and the log of their standard deviations.
This approach avoids Eddington bias, which would suppress both coefficients.
Although UNITY can model $\alpha$ and $\beta$ as broken-linear functions, we assume the coefficients are constants.
In Section \ref{sec:trend_w_redshift} we identify and discuss a systematic issue tied to this decision.

\section{Validating Data and Methodology}
\label{sec:validation}
In this section, we validate our data reduction and modelling techniques by partially reproducing the analysis of the DEHVILS survey \citep{peterson23} using our NIR photometry and fitting methodologies.
To evaluate the differences produced by these variations, we compare each inferred distance modulus ($\mu_\text{fit}$) and the theoretical distance modulus at its corresponding redshift in a fiducial cosmology ($\mu_\text{cos}$).
These Hubble residuals are calculated as
\begin{align}
    \Delta \mu &= \mu_\text{fit} - \mu_\text{cos} \\
    \mu_\text{cos} &= 5\log_{10}\Big{[}\Big{(}\frac{c z_\text{CMB}}{H_0}\Big{)}\Big{(}\frac{1+z_\text{hel}}{1+z_\text{CMB}}\Big{)}\Big{(}1+\frac{1-q_0}{2}z_\text{CMB}\Big{)}\Big{]} + 25,
\end{align}
where $H_0$ is the Hubble constant and $q_0$ is the cosmic deceleration parameter, which we take as $-0.53$ \citep{planck18}.
As stated in \citet{burns18}, the factor of $(1+z_\text{hel})/(1+z_\text{CMB})$ accounts for observational effects which should be corrected in a heliocentric rest-frame.
In each sample we define $H_0$ such that the inverse-variance weighted average of the Hubble residuals is 0 mag.

The dispersion in $\Delta \mu$ is typically characterized through RMS \citep[e.g.][]{blondin11, avelino19, foley21, jones22, pierel22, peterson23};
inverse-variance weighted RMS \citep[WRMS; e.g.][]{blondin11, avelino19, foley21},
or normalized median absolute deviation \citep[NMAD; e.g.][]{boone21, peterson23}.
\sneia{} analyses repeatedly find that measurement uncertainty alone cannot explain the observed dispersion, indicating that 
\sneia{} luminosities include some unmodelled variance commonly called intrinsic scatter \citep[$\sigma_\text{int}$; e.g.][]{blondin11, scolnic18, burns18}.

Lastly, we validate our treatment of max\_model parameters by using photometry from CSP-I DR3 \citet{krisciunas17} to re-derive the Tripp calibration parameters in Table 1 of \citet{burns18}.
We comment on the unique sensitivity of the max\_model to certain spectral subtypes of \sneia{}.

\subsection{Comparisons with DEHVILS}
\label{sec:compare_w_dehvils}
The DEHVILS survey collected data in tandem with \hsf{}, also using UKIRT's WFCAM to collect NIR observations of \sneia{} \citep{peterson23}.
Our programs differ in that DEHVILS collected photometry in the $Y$-, $J$-, and $H$-bands and pursued more observations (median 6 epochs per bandpass) for fewer SNe ($N=96$).
We shared $J$-band observations near peak to avoid redundancy, but reduced the data through independent photometric pipelines.
The DEHVILS analysis employs the following quality cuts: $|x_1| < 3$, $\sigma_{x_1} < 1$, $\sigma_{t_0} < 2$, $E(B-V)_\text{MW} < 0.2$ mag, and Type Ia LC fit probability $P_\text{fit} > 0.01$.
$\sigma_{x_1}$ and $\sigma_{t_0}$ refer to the uncertainty in the SALT parameter $x_1$ and the estimated time of maximum light, while $P_\text{fit}$ is defined in \texttt{SNANA} as the fraction of the $\chi^2$ distribution with $k$ degrees of freedom above a given $\chi^2$ value \citep{kessler09SNANA}:
\begin{equation}
    \label{eqn:fit_prob}
    P_\text{fit}(k, \chi^2) = \frac{1}{\Gamma(k/2)}\int_{\chi^2/2}^\infty t^{k/2-1}e^{-t}dt.
\end{equation}
Further, the target's host galaxy must have a spectroscopic redshift.
There are differences between the redshifts in the DEHVILS sample and the redshifts assembled following the methods described in Section \ref{sec:host z}.
In this section we use the published DEHVILS redshifts for a fairer comparison.
The sample analyzed in \citet{peterson23} comprises 47 of the 83 spectroscopically classified normal \sneia{} with DEHVILS photometry.
Using fitting parameters to define cuts means differences in fitting methods may lead to differences in the objects cut.
When we apply the same cuts using our implentation of the SALT3-NIR model we find that 56 of the optical-only (ATLAS $co$) fits pass all cuts, 47 fits using optical (ATLAS $co$) and DEHVILS-reduced NIR\footnote{Available at \url{https://github.com/erikpeterson23/DEHVILSDR1}} ($YJH$) photometry pass, and 30 DEHVILS-reduced NIR-only ($YJH$) fits pass.
Refitting the 83 \sneia{} assuming the reddening law from \citet{fitzpatrick99} for Milky Way extinction does not lead to any difference in the objects cut.

\subsubsection{Varying Sample Selection and Fitting Methodology}
\label{sec:vary_method}

The 47 \sneia{} analyzed in \citet{peterson23} do not exactly match the 47 that pass the same cuts in our analysis, indicating a difference between our methodologies.
Identifying the exact point of departure is of intrinsic interest, but more immediately concerning are the consequences of such a difference.
Our goal in this section is to compare the dispersion of Hubble residuals found in \citet{peterson23} to our values calculated with the same photometry but different methods.
We assume there are negligible differences in our ATLAS photometry and that there are no unstated quality cuts in the DEHVILS analysis.

We use DEHVILS photometric measurements for all NIR data and fit each of the normal \sneia{} in host galaxies with spectroscopic redshifts using \texttt{SNooPy}'s EBV\_model2, \texttt{SNooPy}'s max\_model, and SALT3-NIR using the bandpass combinations $co$, $coYJH$ and $YJH$ for all three fitters.
The DEHVILS cuts are based on the SALT fitting parameter $x_1$, which we approximate in the \texttt{SNooPy} fits as $0.65 < s_{BV} < 1.40$ using a conversion we empirically determine in Equation \ref{eqn:shape_comparison} of Section \ref{sec:outlier method 1}.
We also use this equation to convert the $\sigma_{x_1} < 1$ cut to $\sigma_{s_{BV}} < 0.125$.

The 83 normal \sneia{} are defined as such based on their spectroscopic classification as SN Ia on TNS, but after visually inspecting the light curves we believe there are nine non-normal \sneia{} in this sample.
We find four underluminous candidates (SNe 2020jsa, 2020rlj, 2020unl, and 2021mim), four SN 2006bt-like candidates (SNe 2020naj, 2020sme, 2020mbf, 2020tkp.), and one Ia-pec candidate (SN 2020kzn).
These targets are eliminated by the $P_\text{fit}$ cut in the DEHVILS analysis and are excluded from the analyses in this section based on their suspected classification.
These targets are not explicitly excluded in the main sample selection process described in Section \ref{sec:sample selection}, but all are removed by either quality cuts or the outlier detection methods described in Section \ref{sec:outlier detection} except for SN 2021mim.

The \texttt{SNooPy}-based fits produce higher $\chi^2$ values than the SALT3-NIR fits for targets that were succesfully fit and passed quality cuts (excluding the $P_\text{fit}$ cut).
\citet{andrae10} review the inherent problems with using $\chi^2$ values and degrees of freedom to assess model performance (especially non-linear models), but the $P_\text{fit}$ cut is a function of those parameters.
The median ratio between $\chi^2/\text{DoF}$ values from EBV\_model2 fits and SALT3-NIR fits using the $coYJH$ bandpasses is about 2.22.
However, when excluding model variance in both sets of $\chi^2$ calculations, the median ratio is 0.88.
This reversal when excluding model variance applies to fits using other bandpass combinations, with the ratio falling from 1.37 to 0.94 in the optical only fits, and from 2.31 to 0.8 in the NIR only fits.
This indicates the uncertainties in the SALT3-NIR model may be overestimated, at least compared to the \texttt{SNooPy} model uncertainties which may themselves be underestimated.
This finding is similar to that of \citet{taylor23} who compared SALT2 and SALT3 models trained on identical data and found that the SALT3 model showed multiple indicators of overestimated model-plus-data uncertainties.
However, this is inconsistent with \citet{peterson23} who performed a visual inspection of their fits and suspected the model uncertainties may be underestimated, leading to significant cuts due to fit probability.
Regardless of the reason for the different $\chi^2$ values found by different models, the application of a $P_\text{fit} > 0.01$ cut will produce imbalanced sample sizes cut at different quantiles of $P_\text{fit}$, which will skew the comparison of Hubble residual dispersions.
Instead, we define model-specific samples based on the 47 highest $P_\text{fit}$ values from each model's fits to the $coYJH$ bandpasses.
This is consistent with the DEHVILS analysis which analyzed the same 47 \sneia{} when fit with optical-only, optical and NIR, or NIR-only photometry.
The $\chi^2/\text{DoF}$ cut values are 1.31 for SALT3-NIR fits, 4.14 for EBV\_model2 fits, and 4.51 for max\_model fits.

With the differences in sample selection defined, we now describe the differences in fitting methodology.
By Equation \ref{eqn:max_model}, calculating distance moduli using the max\_model requires specifying a bandpass ($m_X$) and a colour ($m_Y - m_Z$), which makes comparisons between max\_model fits subject to systematic discrepancies when the bandpasses differ.
There is no bandpass and colour common to the bandpass combinations we examine, but we may still compare each implementation of the max\_model against the DEHVILS results.
For the $co$ combination, we use the $V$ bandpass and the $V - r$ colour; for $coYJH$, we use $J$ and $V - r$; and for $YJH$, we use $J$ and $Y - J$.
We calculate SALT-based distance moduli using $\alpha$ and $\beta$ parameters derived with UNITY \citep{rubin15}, except for the $YJH$ sample which encountered numerous problems during modelling and produced an anomalously low and noisy $\beta = 0.14 \pm 1.80$.
For this sample we calculate the $\alpha$ and $\beta$ values that minimize the standard deviation of the Hubble residuals.
The standardization coefficients for the $co$, $coYJH$, and $YJH$ samples are ($\alpha$, $\beta$) = (0.155, 3.3), (0.138, 3.702), and (0.111, 2.475) respectively.
For comparison, \citet{peterson23} used standardization coefficients of ($\alpha$, $\beta$) = (0.145, 2.359) and (0.075, 2.903) for the $co$ and $coYJH$ samples, with no standardization applied to the $YJH$ sample.
They characterize the dispersion in Hubble residuals using NMAD and standard deviation (STD), so we use the same statistics in this section.

Our methods noticeably differ in fitting one of the bandpass combinations.
In the DEHVILS analysis, the fit parameters $x_1$ and $c$ were held fixed at 0 for the NIR-only sample.
Our methodology does not hold these parameters fixed, and we found greater dispersion.
This is consistent with their finding that keeping $c$ constant while allowing $x_1$ to vary led to increased scatter.
For the other $co$ bandpass combination, we found dispersions in Hubble residuals roughly consistent with the DEHVILS values and errors presented in \citet{peterson23} and reproduced in Table \ref{tab:DEHVILS dispersion}.
Our NMAD values were lower and our STD values higher, implying our Hubble residuals are heavier-tailed than a Gaussian distribution.
This could be an effect of different sample selection, different treatment of ATLAS photometry, or different standardization coefficients.
For the $coYJH$ bandpass combination, our analysis with \texttt{SNooPy}'s EBV\_model2 is consistent with the DEHVILS values, but the other two models tend to produce higher dispersion values.
We note that in our SALT3-NIR analysis, if we use the $\alpha$ and $\beta$ values that minimize the standard deviation (0.100 and 3.052 respectively), we find a value of 0.162 mag and a NMAD of 0.124 mag, which is consistent with the DEHVILS values.
Our max\_model analysis is also not optimized against dispersion.
We use the $J$-band peak magnitude and $V-r$ pseudo-colour to infer distances because that is the methodology we apply to our own photometry, which does not include $Y$- or $H$-band observations.

The consistency between the dispersion values we measure and the values reported in \citet{peterson23} suggests that our methodology is comparable for fits when using optical data or optical and NIR data. 
Our methodology is inferior for fits using only NIR photometry, and max\_model fits using $coYJH$ photometry, indicating that we would need to adapt our methodology if we were to collect $Y$- and $H$-band data like the DEHVILS team and produce NIR-only samples.
The samples we produce using our own $J$-band data always include optical data.

Our samples are distinct from the one analysed in \citet{peterson23}.
However, the effects of a few mismatched SNe should be suppressed after bootstrap resampling the Hubble residuals.
As in the DEHVILS analysis, for each sample we perform 5,000 iterations of randomly choosing 47 residuals with replacement.
The dispersion values and uncertainties presented in Table \ref{tab:DEHVILS dispersion} are the averages and standard deviations of the values measured across the 5,000 iterations.

\begin{table}
    \begin{tabular}{ |r|c|c|c|c| }
        \hline
    Model & Filters & N & NMAD (mag) & STD (mag) \\ \hline
        DEHVILS & \textit{co} & 47 &
            $0.177(029)$ &
            $0.221(043)$ \\
        DEHVILS & \textit{coYJH} & 47 &
            $0.132(025)$ &
            $0.175(034)$ \\
        DEHVILS & \textit{YJH} & 47 &
            $0.139(026)$ &
            $0.172(027)$ \\ \hline
        EBV\_model2 & \textit{co} & 55 &
            $0.177(041)$ &
            $0.327(065)$ \\
        EBV\_model2 & \textit{coYJH} & 47 &
            $0.126(023)$ &
            $0.131(014)$ \\
        EBV\_model2 & \textit{YJH} & 50 &
            $0.152(025)$ &
            $0.165(022)$ \\ \hline
        max\_model & \textit{co} & 51 &
            $0.215(039)$ &
            $0.234(032)$ \\
        max\_model & \textit{coYJH} & 47 &
            $0.159(027)$ &
            $0.153(015)$ \\
        max\_model & \textit{YJH} & 47 &
            $0.181(034)$ &
            $0.182(023)$ \\ \hline
        SALT3-NIR & \textit{co} & 56 &
            $0.225(043)$ &
            $0.246(025)$ \\
        SALT3-NIR & \textit{coYJH} & 47 &
            $0.184(030)$ &
            $0.186(023)$ \\
        SALT3-NIR & \textit{YJH} & 30 &
            $0.164(036)$ &
            $0.161(020)$ \\ \hline
    \end{tabular}
    \caption{We highlight any differences due to methodology by using DEHVILS photometry and approximating their quality cuts.
    We did not replicate their findings when using only $YJH$ photometry, which is where our methodologies differ the most.
    The DEHVILS team fixed $x_1$ and $c$ to 0 for those fits and we allowed them to vary.
    For the $co$ bandpass combination, our methodology produced Hubble residual dispersions consistent with the values reported by DEHVILS.
    We find mixed results with the $coYJH$ combination, with our max\_model analysis producing larger dispersions, and our SALT3-NIR analysis producing consistent results only if we solve for the $\alpha$ and $\beta$ values that minimize dispersion in the Hubble residuals.
    }
    \label{tab:DEHVILS dispersion}
\end{table}

\begin{figure*}
    \centering
    \includegraphics[width=0.9\textwidth]{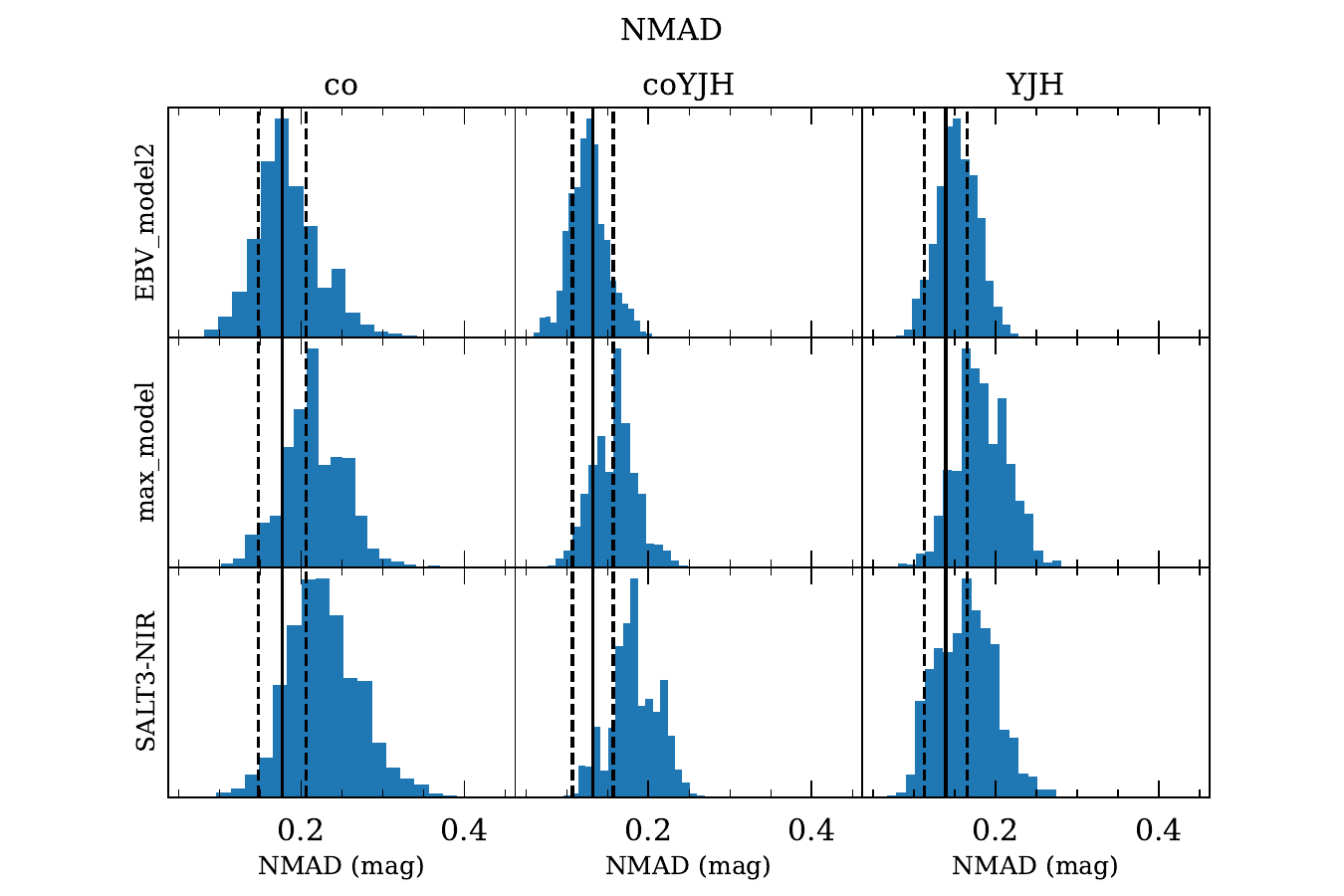}
    \caption{
    We show the distributions of NMAD in 5,000 bootstrap resamplings of each set of Hubble residuals.
    The columns correspond to the fitting model used the rows to the bandpass combination.
    The solid and dashed vertical lines show the values and uncertainties of the NMAD reported in the DEHVILS survey.
    Our methodology produces Hubble residuals with dispersions consistent with the values reported by the DEHVILS survey using \texttt{SNooPy}'s EBV\_model2, but not when using SALT3-NIR or \texttt{SNooPy}'s max\_model.}
    \label{fig:hsf_vs_dehvils_method_test}
\end{figure*}

\begin{figure*}
    \centering
    \includegraphics[width=0.9\textwidth]{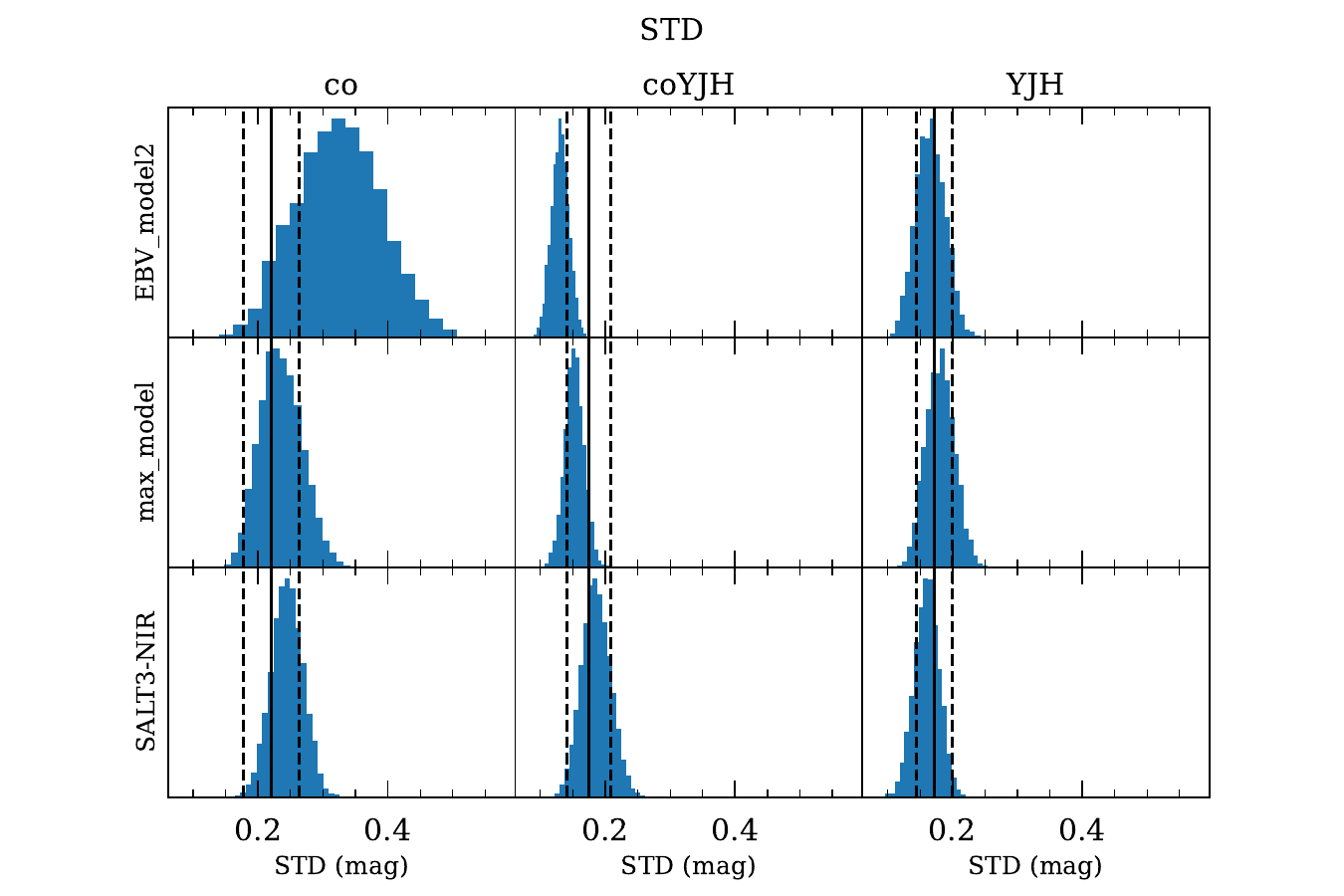}
    \caption{The same set of plots as Figure \ref{fig:hsf_vs_dehvils_method_test}, but showing standard deviation instead of NMAD.}
    \label{fig:hsf_vs_dehvils_method_test_STD}
\end{figure*}


\subsubsection{Varying Photometry}
\label{sec:vary_phot}

We repeat the comparative analysis of the previous section, this time isolating the effects of differing photometry.
We fit ATLAS and either our $J$-band data or that of the DEHVILS survey to create two sets of fits for each of our three models.
We apply the model-specific $\chi^2/\text{DoF}$ cuts based on the greater value between the fits using our photometry or that of DEHVILS.

Once more, we bootstrap resample the Hubble residuals to estimate the uncertainties in our dispersion measurements, but we include an additional set of statistics.
When varying methodology, we could only compare the distributions of our resampled dispersion measurements with the values reported in \citet{peterson23}, but in this analysis we can make pairwise comparisons between individual iterations of the resampling process.
For each iteration, we randomly choose \sneia{} with replacement, record the NMAD and STD of their Hubble residuals in our six samples, and additionally calculate the differences in dispersion between each model's sample using our $J$-band photometry and using DEHVILS photometry ($\Delta D = D_{\textrm{HSF}} - D_{\textrm{DEHVILS}}$ where $D$ is either NMAD or STD).
Thus, we not only produce distributions of NMAD and STD, but also distributions of $\Delta$NMAD and $\Delta$STD.

\begin{figure*}
    \centering
    \includegraphics[width=0.9\textwidth]{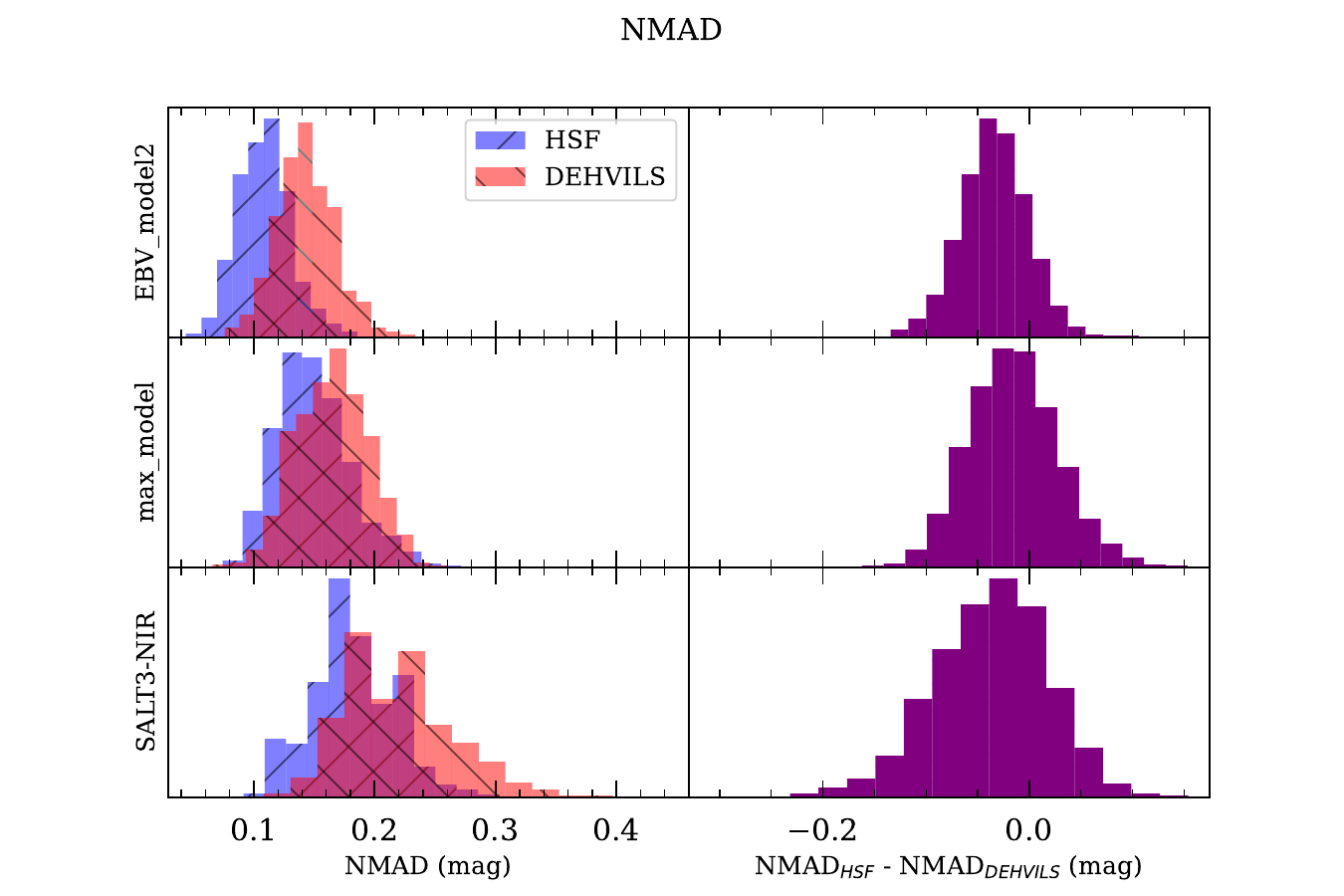}
    \caption{The blue and red histograms show the distributions of NMAD dispersion measurements after 5,000 iterations of bootstrap resampling Hubble residuals.
    The purple histograms show the distributions of differences in dispersion between the samples using HSF photometry and the samples using DEHVILS photometry in each iteration.
    Using our measurements instead of DEHVILS photometry may lead to a decrease in the dispersion of the Hubble residuals, but it is not statistically significant.
    }
    \label{fig:hsf_vs_dehvils_phot_test}
\end{figure*}

The averages and standard deviations of these values are presented in Table \ref{tab:var_phot} and the histograms of dispersions and differences are plotted in Figure \ref{fig:hsf_vs_dehvils_phot_test}.
None of the distributions indicate that using our photometry instead of DEHVILS photometry leads to increased dispersion measurements.
The averages are within one standard deviation of each other, and the differences within one standard deviation of no change in dispersion.

\begin{table}
    \begin{tabular}{ |r|c|c|c|c| }
        \hline
    Model & $J$ Data & N & NMAD (mag) & STD (mag) \\ \hline
        EBV\_model2 & HSF & 48 &
            $0.109(022)$ &
            $0.143(020)$ \\
        EBV\_model2 & DEHVILS & 48 &
            $0.142(024)$ &
            $0.152(017)$ \\ \hline
        max\_model & HSF & 50 &
            $0.144(032)$ &
            $0.196(029)$ \\
        max\_model & DEHVILS & 50 &
            $0.165(029)$ &
            $0.161(015)$ \\ \hline
        SALT3-NIR & HSF & 49 &
            $0.180(034)$ &
            $0.197(022)$ \\
        SALT3-NIR & DEHVILS & 49 &
            $0.218(044)$ &
            $0.227(021)$ \\ \hline
    \end{tabular}
    \caption{We use our methodology to calculate Hubble residuals using ATLAS photometry and either our $J$-band photometry or that of the DEHVILS survey.
    Bootstrap resampling these residuals 5,000 times shows the dispersion measurements are insensitive to any differences between our photometry.
    In \texttt{SNooPy}'s EBV\_model2 and max\_model and in SALT3-NIR, the change in dispersion is consistent with 0.
    }
    \label{tab:var_phot}
\end{table}

\begin{figure*}
    \centering
    \includegraphics[width=0.9\textwidth]{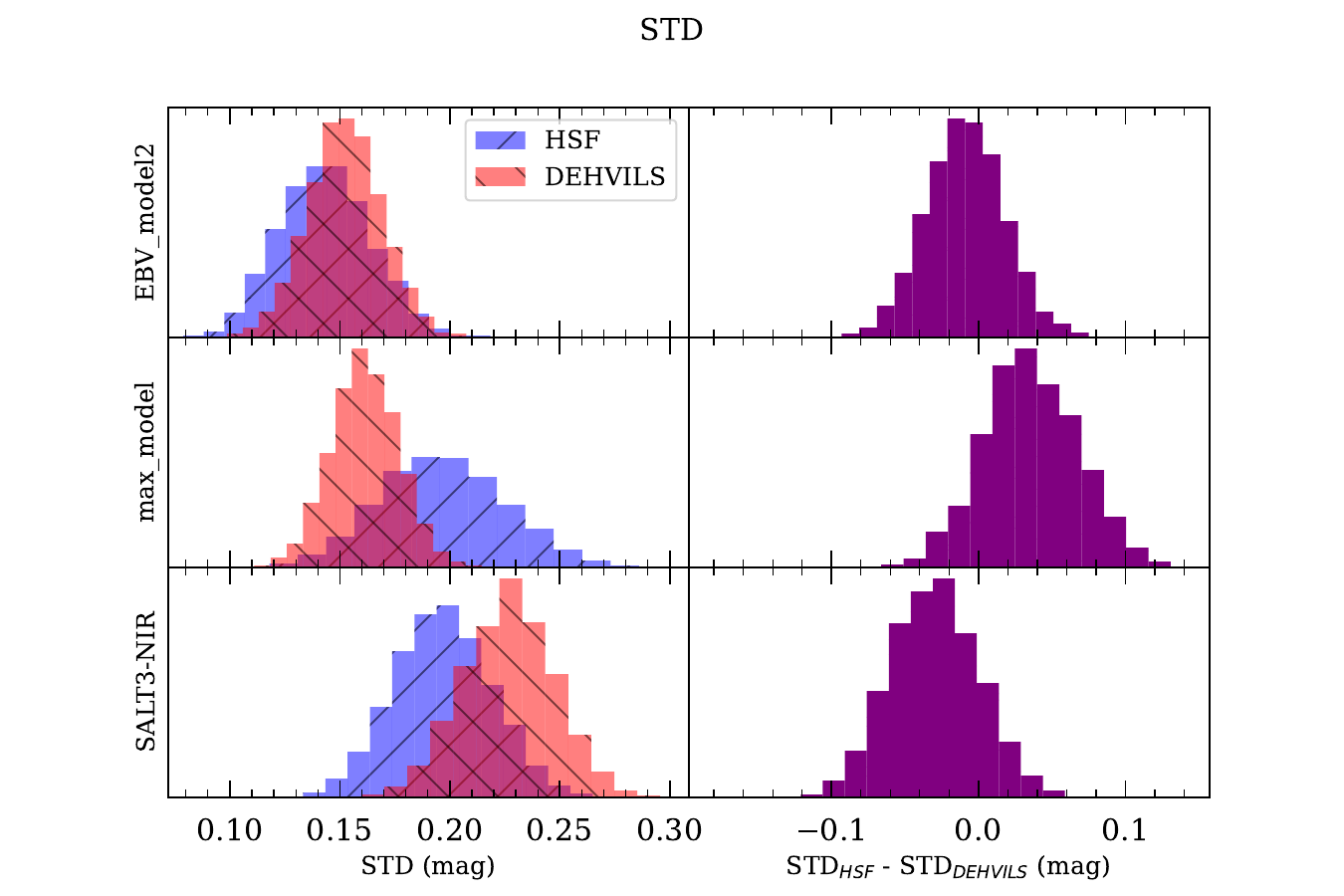}
    \caption{The same set of histograms as Figure \ref{fig:hsf_vs_dehvils_phot_test}, but applied to standard deviation rather than NMAD.
    Again, using our measurements instead of DEHVILS photometry does not lead to a statistically significant difference in the dispersion of the Hubble residuals.
    The closest case is the standard deviation in the max\_model, which increases by an average of $0.014 \pm 0.022$ mag.
    The more robust NMAD decreases by $0.015 \pm 0.038$ mag, indicating the increase in standard deviation is due to a few discrepant values rather than a systematically preferred set of photometry.
    }
    \label{fig:hsf_vs_dehvils_phot_test_std}
\end{figure*}

\subsection{Comparison with CSP}
\label{sec:comparison_w_csp}

The EBV\_model2 produces Hubble residuals with lower dispersion than those produced by either SALT3-NIR or the max\_model.
The greater dispersion in the max\_model was unexpected since the EBV\_model2 is calibrated to CSP observations of 36 SNe, whereas in this analysis we derived standardization coefficients for the max\_model using our observations of 47 SNe.

\subsubsection{Validating Tripp Calibration}
To test our derivation process, we used photometry from CSP-I DR3 \citep{krisciunas17} to solve for the calibration coefficients presented in Table 1 of \citet{burns18}.
We fit all CSP photometry with the \texttt{SNooPy} max\_model, parametrizing light-curve shape with $s_{BV}$.
We use the heliocentric redshifts provided in the data release rather than redshifts from HyperLEDA to focus on differences due to methodology.
Our Equation \ref{eqn:max_model} does not include a term for host-galaxy mass, but in order to match the CSP derivation methodology we reintroduce this term:
\begin{equation}
    \begin{split}
        m_X =& P_0 + P_1(s_{BV} - 1) + P_2(s_{BV} - 1)^2 + \mu_\text{cos} \\
         & + \beta_\text{max} (m_Y - m_Z) + \alpha_\text{M} \left(\log\left(M_\ast / M_\odot\right) - M_0\right)
    \end{split}
\end{equation}
Where $\alpha_\text{M}$ is the coefficient correlating magnitude and host-galaxy stellar mass ($M_\ast$) and $M_0$ is an arbitrary mass zero point, taken as $10^{11} M_\odot$.
We follow the methodology in Appendix B of \citet{burns18} for assembling host-galaxy stellar masses, primarily drawing from the 2MASS Extended Source Catalog \citep{jarrett00}, which we convert from $K$-band apparent magnitudes to stellar masses assuming a constant mass-to-light ratio.
\begin{equation}
    \log_{10}\left(M_\ast / M_\odot\right) = -0.4\left(m_K + \mu\right) + C
\end{equation}
where $\mu$ is the distance modulus and $C$ is a constant which CSP determined to be 1.04 dex by comparing masses from the 2MASS catalogue with mass estimates from \citet{neill09}.
We verify that this is the best-fitting value from a simple least-squares regression.
When there is no $K$-band magnitude available, we use estimates from \citet{neill09} and \citet{chang15} when possible, as \citet{burns18} did.

The coefficients in Equation \ref{eqn:max_model} derived in \citet{burns18} and re-derived with our methods are presented in Table \ref{tab:tripp_coefficients}.
The average deviation between the two sets of coefficients is 0.582 times the quadrature sum of the uncertainties.
Additionally, we derive a set of coefficients while not accounting for host-galaxy mass.
As expected, the average difference between this set and the original values is greater, albeit only slightly at 0.598 times the combined uncertainty.

\begin{table*}
    \begin{tabular}{ |r|c|c|c|c|c|c|c| }
        \hline
        Derivation & $P_0$ & $P_1$ & $P_2$ & $\beta_\text{max}$ & $\alpha_\text{gal}$ & $\sigma_\text{int}$ & $v_\text{pec}$ \\
        & (mag) & (mag) & (mag) & & (mag/dex) & (mag) & (km s$^{-1}$) \\ \hline
        CSP & -18.633(062) & -0.37(12) & 0.61(32) & 0.36(10) & -0.056(029) & 0.11 & 336 \\
        This Work & -18.626(028) & -0.407(128) & -0.021(344) & 0.292(096) & -0.044(032) & 0.083(033) & 384(57) \\
        This Work (No Masses) & -18.607(024) & -0.352(126) & 0.102(346) & 0.270(097) & N/A & 0.093(032) & 384(57) \\ \hline
    \end{tabular}
    \caption{
        We show some of the Tripp calibration coefficients presented in Table 1 of \citet{burns18} and our derivations using the same data with our methodology.
        Our values differ from the original values by an average of 0.582 times the combined uncertainty.
        When not accounting for host-galaxy masses, the average difference slightly increases to 0.598 times the combined uncertainty.
    }
    \label{tab:tripp_coefficients}
\end{table*}

We conclude that our methodology for calibrating the Tripp method is consistent with the method used in \citet{burns18}.
The difference in dispersion in Hubble residuals between the max\_model and EBV\_model2 seen in Section \ref{sec:vary_method} is not due to errors in determining the calibration coefficients.
Additionally, we do not find a significant difference in dispersion between the two models when examining the CSP data.
Using the max\_model, the Hubble residuals have a NMAD dispersion of 0.163 mag and a standard deviation of 0.233 mag, which is only marginally greater than the same values using EBV\_model2: 0.157 mag and 0.227 mag.

%

\section{Sample Selection}
\label{sec:sample selection}

We have NIR observations of 1,217 unique transients, but only about a quarter of those are presently useful for cosmology.
Our final sample is comprised of targets that pass three sets of cuts: one based on observational data, one based on fitting parameters, and one based on several outlier detection algorithms.
The number of targets discarded and remaining after each cut are presented in Tables \ref{tab:survey_wide_cuts} and \ref{tab:sample_specific_cuts}.

\subsection{First Cut: Observational Data}
The set of all our observed transients includes unclassified or misclassified non-\sneia{}, galaxies with photometric or unknown redshifts, and \sneia{} missing coverage near maximum light in one or more all-sky survey bandpasses.
In future work we intend to incorporate the unclassified transients that are photometrically consistent with \snia{} light curves, but for this paper, we do not include them in our analysis.
\citet{vincenzi23} describe the magnitude of biases in cosmological measurements when using photometrically classified samples and discuss various methods for mitigating them to sub-percent levels when estimating the dark energy equation of state parameter $w$.

Of the 1,217 observed transients, 668 have been spectroscopically classified as usable \sneia{}.
This number does not include \sneia{} subtypes that are unsuitable for distance inference using SALT3-NIR or \texttt{SNooPy}: 2002cx-like SNe \citep[sometimes called SNe Iax,][]{li03}, 2002ic-like SNe \citep[sometimes called SNe Ia-CSM,][]{hamuy03}, 2003fg-like SNe \citep[formerly called super-Chandrasekhar SNe or SNe Ia-SC,][]{howell06, hicken07, ashall21}, and generally peculiar \sneia{} (Ia-pec). 
This number does include several 2006bt-like candidates, which we discuss in Section \ref{sec:outlier detection}.

Spectroscopic host-galaxy redshifts are available or have been successfully measured for 603 of these 668 \sneia{}.
The remaining 65 include galaxies scheduled for spectroscopic observation, galaxies with spectral features manually judged to be too weak for accurate redshift determination, and galaxies with exceptionally low surface brightness, such that spectroscopic observation is prohibitively expensive.
We remove an additional 8 targets that have Galactic reddening greater than 0.3 mag according to \citet{schlafly11}.
As the last cut in this set, we remove targets with fewer than 5 optical and NIR observations, counting each quartet of ATLAS exposures as a single observation.
Of the remaining 595 \sneia{}, 76 are in galaxies for which we have unreduced spectroscopic observations, and 15 encountered errors during photometric analysis, leaving 504 \sneia{}.

\subsection{Second Cut: Fitting Parameters}
\label{sec:second cuts}
Removing targets based on fitting parameters necessarily requires successfully running each model's fitting procedure, which is not guaranteed for each possible permutation of input data.
Without sufficient phase coverage in photometry, the shape parameter of a \snia{} becomes underconstrained.
The same is true for insufficient wavelength coverage and the colour parameter or host-galaxy extinction.
These produce singular covariance matrices, indicating degeneracy in the fitting parameters.
Additionally, the models span finite combinations of phase and wavelength, making comparisons to some observations interpolative at best and often times extrapolative.
The fit is unsuccessful if all data in a given bandpass lie outside the model domain.
However, the phase of any observation is dependent on the estimated time of maximum light, which itself is a fitting parameter.
This means that the success of a fit is partially dependent on how the fitting parameters are initialized.
When a fit fails because one of the bandpasses has no data in a model's domain, we attempt to perform the same fit without data from the behaviour bandpass.
If that succeeds, we use those fitting parameters to initialize a new fit, reintroducing the excluded data.
Sometimes this leads to a successful fit using all available bandpasses, at other times a subset of available bandpasses, and occasionally the fit cannot be salvaged.
The success or failure of a fit acts as a cut.
We now define three distinct samples based on the three fitting models: SNPY\_EBV with 502 fits from \texttt{SNooPy}'s EBV\_model2, SNPY\_Max with 502 fits from \texttt{SNooPy}'s max\_model, and SALT with 503 fits from SALT3-NIR.

After fitting, we apply the following cuts.
In the SNPY samples we use quality cuts from \citet{jones22}, rejecting fits with shape factors outside the interval $0.6 < s_{BV} < 1.3$ (their ``loose'' cut) or with uncertainty $\sigma_{s_{BV}} > 0.2$, and for SNPY\_EBV, rejecting fits with host-galaxy $E(B-V)_\text{host} > 0.3$.
In the SNPY\_Max sample, the rest-frame bandpasses used for calculating distances depend on both the observed bandpasses and the redshift.
Since we infer distances using the $J$ band and the $V - r$ colour, we cut SNe from SNPY\_Max whenever the max\_model does not provide inferences for the maximum apparent magnitudes in those bandpasses.
While it is possible to force \texttt{SNooPy} to map to these bandpasses, the cross-band K-corrections required become much more sensitive to differences between the assumed and actual SED.
This acts as a cut based on redshift.
In the SALT sample we reject fits where $|x_1| > 3$, $\sigma_{x_1} > 1.5$, $|c| > 0.3$, or $\sigma_c > 0.2$
\citep{scolnic18, foley21, scolnic22}.
We use the temporal coverage cut from \citet{rubin23}, which is based on the calculated time of maximum light ($t_0$), the phase of the initial observation ($t_i$), and the phase of the final observation ($t_f$).
Given that $t_0$ can vary between the three samples, we apply this cut to each sample independently.
Adequately observed \sneia{} meet at least one of two sets of criteria.
The first set requires $t_i$ no more than 2 days after $t_0$, $t_f$ at least 8 days after $t_0$, and $t_f - t_i$ must span at least 10 days.
The second set allows for a later $t_i$, up to 6 days after $t_0$, as long as $t_f - t_i$ spans at least 15 days.
Lastly, we remove fits with reduced $\chi^2$ values above 4.14, 4.51, or 1.31 for fits in the SNPY\_EBV, SNPY\_Max, and SALT samples respectively.
These cut values come from the comparison to the DEHVILS sample in Section \ref{sec:compare_w_dehvils}.
This leaves our three samples with 363/502 objects in SNPY\_EBV, 330/502 in SNPY\_Max, and 368/503 in SALT.

\subsection{Third Cut: Outlier Detection}
\label{sec:outlier detection}
There are many vectors for outliers to appear in our sample: spectroscopic misclassification of core-collapse SNe, incorrectly assigned host-galaxy redshifts, errors in photometric reduction, or errors in fitting.
Even with `perfect' data and methods, an outlier could arise from anomalous astrophysical properties (e.g. an exotic progenitor system or detonation mechanism) or unclassified Type-Ia peculiarity.
In particular, 2006bt-like SNe are difficult to identify without $i$-band or NIR observations \citep{stritzinger11, phillips12}.
There are several objects in our sample that are classified as \sneia{} on TNS, but have NIR light-curves suggestive of 2006bt-like SNe: SN 2020naj, SN 2020tkp, SN 2020mbf, and SN 2020sme.
We employ two kinds of outlier detection methods.
The first compares inferred parameters for common targets between the samples, and the second is based on the mixture model of \citep{kunz07} as implemented through UNITY \citep{rubin15}.

\subsubsection{Divergent Model Inferences}
\label{sec:outlier method 1}

\begin{figure*}
    \centering
    \includegraphics[width=0.9\textwidth]{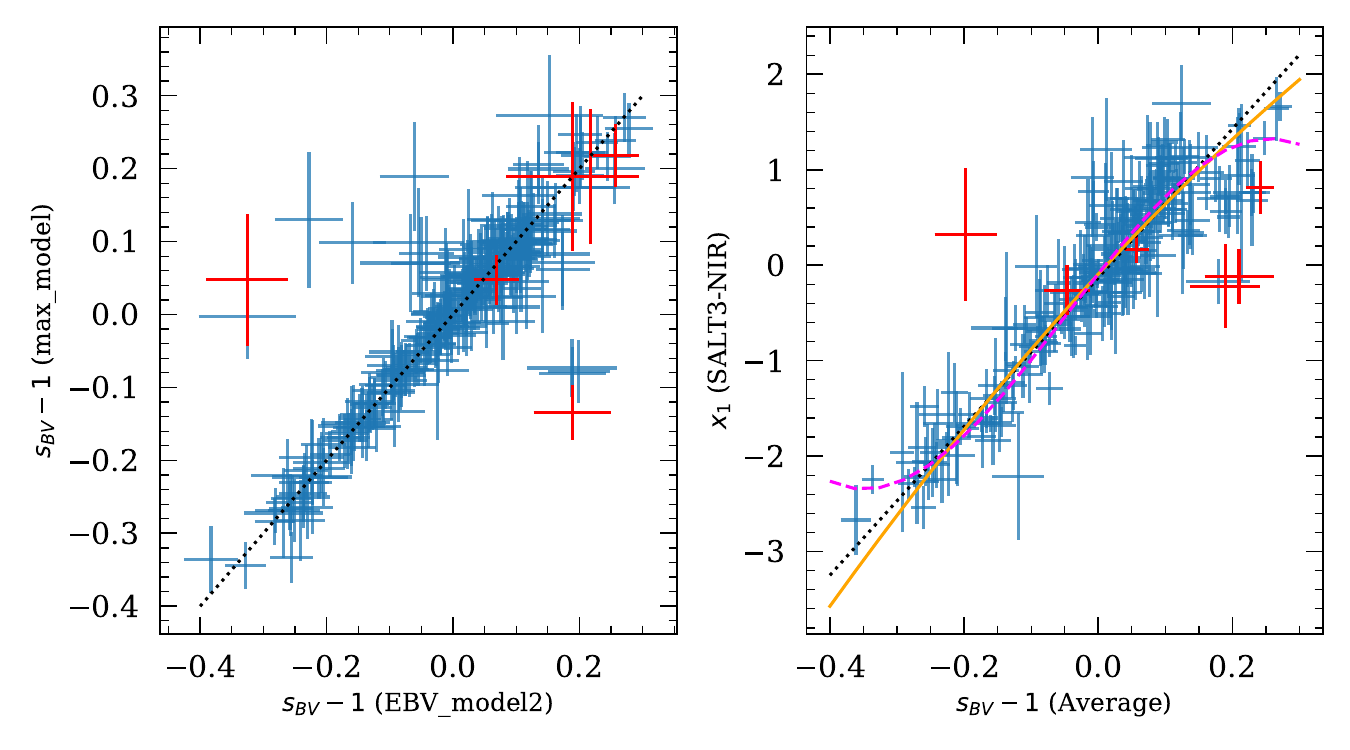}
    \caption{Left: The $s_{BV}$ values of each SN are inferred using the same data, and either \texttt{SNooPy}'s EBV\_model2 or max\_model.
    The one-to-one line is plotted in black.
    Right: Each SN's inverse-variance weighted average $s_{BV}$ value is compared to the $x_1$ value inferred by SALT3-NIR.
    We plot the linear (dotted black), quadratic (solid orange), and cubic (dashed magenta) polynomial fits determined through orthogonal distance regression.
    The Bayesian information criterion marginally favours the linear fit.
    In both plots, we colour outliers identified through divergent model inferences red.
    These outliers are ignored when calculating the parameter transformation equations.
    }
    \label{fig:shape_comparison}
\end{figure*}

\begin{figure}
    \centering
    \includegraphics[width=0.45\textwidth]{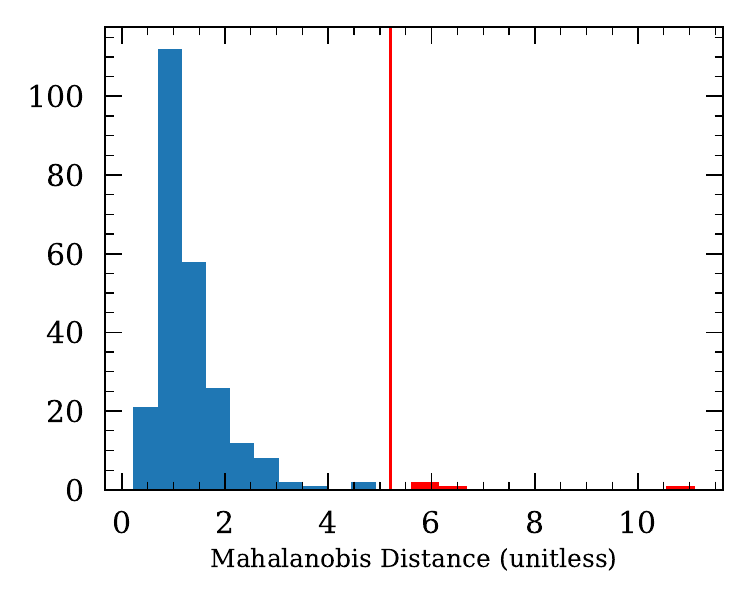}
    \caption{We identify outliers (red) based on disagreement between the three models inferences for a SN's time of maximum, shape, and colour.
    After transforming the \texttt{SNooPy} parameters $s_{BV}$, $E(B-V)_\text{host}$, and $m_V - m_J$ to $x_1$, $c$, and $c$, we calculate the standard deviations in each SN's three inferred values for $t_0$, $x_1$, and $c$.
    We then parametrize disagreement using the Mahalanobis distance between the standard deviations ($\sigma_{t_0,i}$, $\sigma_{x_1,i}$, and $\sigma_{c,i}$) and a distribution centred at the origin.
    When the three models produce consistent estimates the standard deviation is relatively low, but when they diverge the standard deviation increases.
    }
    \label{fig:mahalanobis_outliers}
\end{figure}

In a Bayesian framework, the physical parameters inferred by each fitting model should draw from the same posterior distribution of ``true'' physical parameters.
This common quantity allows for simple error detection in the 246 SNe common to all samples.
Where the estimates of the same parameter vary significantly, at least one model is likely to have converged on a local maximum in likelihood and is not reliable for inferring other parameters.
The top row of Figure \ref{fig:mahalanobis_outliers} shows the relationships between the corresponding fitting parameters in our samples.
The SNPY and SALT samples share a common definition for the time of maximum light, but differ in exactly how they quantify light-curve shape and colour.
\citet{burns18} described a linear transformation between the $x_1$ parameter in SALT2 and the $s_{BV}$ parameter in \texttt{SNooPy}.
We use orthogonal distance regression and find a slightly different relationship, potentially due to differences between SALT2 and SALT3-NIR.
After testing linear, quadratic, and cubic polynomial fits, the Bayesian information criterion favours a cubic relationship ($102.6$,~$103.5$,~$77.2$):
\begin{equation}
    \begin{split}
        x_1 &= -0.09(02) + 8.97(27)(s_{BV} - 1) \\
        & -4.73(92)(s_{BV} - 1)^2 \\
        & -34.35(04)(s_{BV} - 1)^3.
    \end{split}
    \label{eqn:shape_comparison}
\end{equation}
Here $s_{BV}$ is the average between the values inferred by \texttt{SNooPy}'s two models.
The relationship between $s_{BV}$ values from the two \texttt{SNooPy} models as well as the relationship between their average and the SALT $x_1$ parameter is shown in Figure \ref{fig:shape_comparison}.

The $c$ parameter in SALT represents both intrinsic colour variation in \sneia{} and reddening from dust, while the $E(B-V)_\text{host}$ fitting parameter in \texttt{SNooPy} is strictly concerned with the latter.
However, \citet{brout20} found that the correlation between intrinsic colour and luminosity may be weak, and that dust can provide the observed diversity of colours.
We test linear, quadratic, and cubic fits, and the Bayesian information criterion supports a linear fit ($-245.4$,~$-239.9$,~$-234.7$):
\begin{equation}
    c = -0.05(00) + 0.96(03)E(B-V)_\text{host} \text{ mag}^{-1}.
\end{equation}
Our colour information in the SNPY\_Max sample comes from the differences in apparent maxima.
To more effectively parametrize dust, we use the $m_V - m_J$ pseudo-colour.
We test the same polynomial fits, and find support for a cubic fit ($364.2$,~$369.4$,~$340.9$):
\begin{equation}
\begin{split}
    c =& 0.12(02) -0.17(10)(m_V-m_J) \text{ mag}^{-1} \\
      & -1.21(23)(m_V-m_J)^2 \text{ mag}^{-2} \\
      & -0.85(16)(m_V-m_J)^3 \text{ mag}^{-3}
\end{split}
\end{equation}

\begin{figure*}
    \centering
    \includegraphics[width=0.9\textwidth]{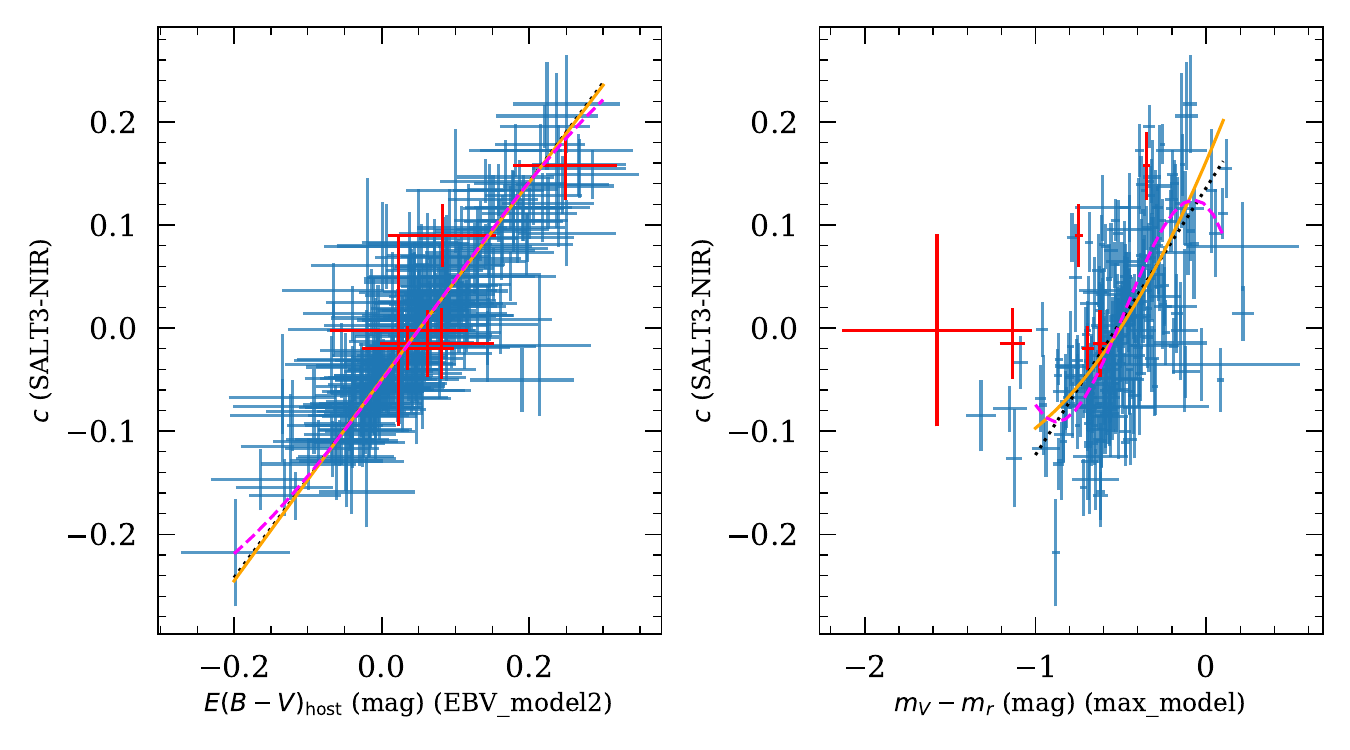}
    \caption{Left: The $E(B-V)_\text{host}$ values inferred by \texttt{SNooPy}'s EBV\_model2 are plotted against the $c$ values inferred by SALT3-NIR.
    As in Figure \ref{fig:shape_comparison}, we show the linear (dotted black), quadratic (solid orange), and cubic (dashed magenta) polynomial fits determined through orthogonal distance regression.
    The linear and quadratic relationships overlap.
    Right: For each SN, the pseudo-colour between the peak apparent magnitudes in $V$ and $J$ determined by \texttt{SNooPy}'s max\_model are plotted against $c$.
    Once again, we colour outliers identified through divergent model inferences red.
    These outliers are ignored when calculating the parameter transformation equations.
    }
    \label{fig:colour_comparison}
\end{figure*}

After converting the \texttt{SNooPy} parameters into SALT parameters, we can directly compare each model's inferences for each SN to find where they disagree.
We define $\sigma_{t_0,i}$, $\sigma_{x_1,i}$, and $\sigma_{c,i}$ as the standard deviation between the transformed fitting parameters of SN $i$ in the SNPY samples its parameters in the SALT sample.
We account for correlations between the differences by calculating the Mahalanobis distance between each point $m_i = (\sigma_{t_0,i}, \sigma_{x_1,i}, \sigma_{c,i}$) and a distribution $D$ centred at the origin with covariance matrix $\Sigma$ \citep{mahalanobis30}.
We approximate $\Sigma$ by bootstrap resampling the parameter differences 5,000 times, calculating each sample covariance $S$, and defining each element $\Sigma_{j,k}$ as the average of all sample elements $S_{j,k}$.
Each distance $d_i(m_i,D) = \sqrt{m_i\Sigma^{-1}m_i^T}$, and can be understood as the number of standard deviations between point $m_i$ and distribution $D$.
The quadrature sum of the standard deviations is a similar metric if all dimensions are normalized to have unit variance, but does not account for correlations.
Figure \ref{fig:mahalanobis_outliers} shows the histogram of distances.
There are 4 SNe with distances greater than 5 times the standard deviation in $d$, indicating significant disagreement between the models.
We recalculated the parameter transformation equations and Mahalanobis distances excluding these 4, and identified no additional outliers.
The equations and figures presented are the recalculated versions.
Disagreement alone leaves room for one or two of the models to have accurately fit the data, but while manual inspection often reveals which models fit the data well and which do not, we err on the side of caution by removing these 4 SNe from all three samples.

\subsubsection{Mixture-model Analysis}
The mixture model introduced as BEAMS \citep[Bayesian Estimate Applied to Multiple Species;][]{kunz07} posits that an imperfect \sneia{} survey will lead to measurements following the sum of multiple distributions.
Measurements of real \sneia{} should feature relatively low dispersion whereas measurements of survey contaminants will be more dispersed, and may have a different mean.
As implemented in UNITY, both populations are assumed to be Gaussian around a common mean, and the outlier population is assumed to have variances of one in $m^*_B$, $x_1$, and $c$ \citep{rubin15}.
UNITY's use of SALT parameters means in its present version it can only process the SALT sample.
Since the mixture-model framework is generalizable to arbitrary fitting parameters, future work could allow UNITY to process results from \texttt{SNooPy}.
At such a time comparing outliers between samples could indicate whether the \sneia{} is astrophysically exceptional, or whether one of the models is unreliable.
For now we are limited to examining SALT.

As UNITY sifts the data through its Bayesian hierarchical model, it produces a population level estimate of the fraction of outliers in the sample (with a prior of $\log{f^{outl}} \sim \mathcal{N}(-3, 0.5^2)$), and a pair of estimates for each object describing the likelihood it belongs to the normal or outlier population.
Our sample has an estimated outlier fraction of 0.012$\pm$0.004 and there are 2 SNe more likely to be outliers than a part of the normal population.
As before, we take the conservative approach of removing these objects from all samples.
Some of the 2 may have been eliminated from the SNPY samples by previous cuts, which is why the number cut at this stage may vary across the samples.

\subsection{Final Samples}
After all three sets of cuts the sizes of our samples are 357, 324, and 362 for the SNPY\_EBV, SNPY\_Max, and SALT samples respectively.
Tables \ref{tab:survey_wide_cuts} and \ref{tab:sample_specific_cuts} enumerate the effect of each cut.

\begin{table*}
    \begin{tabular}{lll}
        \hline
    \multicolumn{1}{|l|}{Cuts} &
        \multicolumn{1}{l|}{Number Cut} &
        \multicolumn{1}{l|}{Remaining Sample} \\ \hline
    \multicolumn{1}{|l|}{Spec. Classification} &
        \multicolumn{1}{l|}{327} &
        \multicolumn{1}{l|}{890} \\ \hline
    \multicolumn{1}{|l|}{\snia{}} &
        \multicolumn{1}{l|}{222} &
        \multicolumn{1}{l|}{668} \\ \hline
    \multicolumn{1}{|l|}{Spec. Redshift} &
        \multicolumn{1}{l|}{65} &
        \multicolumn{1}{l|}{603} \\ \hline
    \multicolumn{1}{|l|}{$E(B-V)_\text{MW}$ < 0.3} &
        \multicolumn{1}{l|}{8} &
        \multicolumn{1}{l|}{595} \\ \hline
    \multicolumn{1}{|l|}{$N_\text{obs} \geq 5$} &
        \multicolumn{1}{l|}{0} &
        \multicolumn{1}{l|}{595} \\ \hline
    \multicolumn{1}{|l|}{Successful Spectroscopic Reduction} &
        \multicolumn{1}{l|}{76} &
        \multicolumn{1}{l|}{519} \\ \hline
    \multicolumn{1}{|l|}{Successful Photometric Reduction} &
        \multicolumn{1}{l|}{15} &
        \multicolumn{1}{l|}{504} \\ \hline
    \end{tabular}
    \caption{Our first set of cuts are independent of the fitting model used and apply to all targets observed.}
    \label{tab:survey_wide_cuts}
\end{table*}

\begin{table*}
    \begin{tabular}{lllllll}
    \hline
    \multicolumn{1}{|l|}{Cuts} &
        \multicolumn{1}{l|}{SNPY\_EBV Cut} &
        \multicolumn{1}{l|}{Remaining} &
        \multicolumn{1}{l|}{SNPY\_Max Cut} &
        \multicolumn{1}{l|}{Remaining} &
        \multicolumn{1}{l|}{SALT Cut} &
        \multicolumn{1}{l|}{Remaining} \\ \hline
    \multicolumn{1}{|l|}{Successful Fit} &
        \multicolumn{1}{l|}{2} &
        \multicolumn{1}{l|}{502} &
        \multicolumn{1}{l|}{2} &
        \multicolumn{1}{l|}{502} &
        \multicolumn{1}{l|}{1} &
        \multicolumn{1}{l|}{503} \\ \hline
    \multicolumn{1}{|l|}{Rest frame $m_J$, $m_V$, and $m_r$} &
        \multicolumn{1}{l|}{\cellcolor[HTML]{C0C0C0}} &
        \multicolumn{1}{l|}{\cellcolor[HTML]{C0C0C0}} &
        \multicolumn{1}{l|}{100} &
        \multicolumn{1}{l|}{402} &
        \multicolumn{1}{l|}{\cellcolor[HTML]{C0C0C0}} &
        \multicolumn{1}{l|}{\cellcolor[HTML]{C0C0C0}} \\ \hline
    \multicolumn{1}{|l|}{0.6 < $s_{BV}$ < 1.3} &
        \multicolumn{1}{l|}{24} &
        \multicolumn{1}{l|}{478} &
        \multicolumn{1}{l|}{41} &
        \multicolumn{1}{l|}{361} &
        \multicolumn{1}{l|}{\cellcolor[HTML]{C0C0C0}} &
        \multicolumn{1}{l|}{\cellcolor[HTML]{C0C0C0}} \\ \hline
    \multicolumn{1}{|l|}{$\sigma_{s_{BV}}$ < 0.2} &
        \multicolumn{1}{l|}{2} &
        \multicolumn{1}{l|}{476} &
        \multicolumn{1}{l|}{2} &
        \multicolumn{1}{l|}{359} &
        \multicolumn{1}{l|}{\cellcolor[HTML]{C0C0C0}} &
        \multicolumn{1}{l|}{\cellcolor[HTML]{C0C0C0}} \\ \hline
    \multicolumn{1}{|l|}{$E(B-V)_\text{host}$ < 0.3 mag} &
        \multicolumn{1}{l|}{75} &
        \multicolumn{1}{l|}{401} &
        \multicolumn{1}{l|}{\cellcolor[HTML]{C0C0C0}} &
        \multicolumn{1}{l|}{\cellcolor[HTML]{C0C0C0}} &
        \multicolumn{1}{l|}{\cellcolor[HTML]{C0C0C0}} &
        \multicolumn{1}{l|}{\cellcolor[HTML]{C0C0C0}} \\ \hline
    \multicolumn{1}{|l|}{|$x_1$| < 3} &
        \multicolumn{1}{l|}{\cellcolor[HTML]{C0C0C0}} &
        \multicolumn{1}{l|}{\cellcolor[HTML]{C0C0C0}} &
        \multicolumn{1}{l|}{\cellcolor[HTML]{C0C0C0}} &
        \multicolumn{1}{l|}{\cellcolor[HTML]{C0C0C0}} &
        \multicolumn{1}{l|}{20} &
        \multicolumn{1}{l|}{483} \\ \hline
    \multicolumn{1}{|l|}{$\sigma_{x_1}$ < 1.5} &
        \multicolumn{1}{l|}{\cellcolor[HTML]{C0C0C0}} &
        \multicolumn{1}{l|}{\cellcolor[HTML]{C0C0C0}} &
        \multicolumn{1}{l|}{\cellcolor[HTML]{C0C0C0}} &
        \multicolumn{1}{l|}{\cellcolor[HTML]{C0C0C0}} &
        \multicolumn{1}{l|}{4} &
        \multicolumn{1}{l|}{479} \\ \hline
    \multicolumn{1}{|l|}{|c| < 0.3} &
        \multicolumn{1}{l|}{\cellcolor[HTML]{C0C0C0}} &
        \multicolumn{1}{l|}{\cellcolor[HTML]{C0C0C0}} &
        \multicolumn{1}{l|}{\cellcolor[HTML]{C0C0C0}} &
        \multicolumn{1}{l|}{\cellcolor[HTML]{C0C0C0}} &
        \multicolumn{1}{l|}{43} &
        \multicolumn{1}{l|}{436} \\ \hline
    \multicolumn{1}{|l|}{$\sigma_c$ < 0.2} &
        \multicolumn{1}{l|}{\cellcolor[HTML]{C0C0C0}} &
        \multicolumn{1}{l|}{\cellcolor[HTML]{C0C0C0}} &
        \multicolumn{1}{l|}{\cellcolor[HTML]{C0C0C0}} &
        \multicolumn{1}{l|}{\cellcolor[HTML]{C0C0C0}} &
        \multicolumn{1}{l|}{1} &
        \multicolumn{1}{l|}{435} \\ \hline
    \multicolumn{1}{|l|}{Phase Requirements} &
        \multicolumn{1}{l|}{0} &
        \multicolumn{1}{l|}{401} &
        \multicolumn{1}{l|}{1} &
        \multicolumn{1}{l|}{358} &
        \multicolumn{1}{l|}{2} &
        \multicolumn{1}{l|}{433} \\ \hline
    \multicolumn{1}{|l|}{Reduced $\chi^2$ < 4.14/4.51/1.31} &
        \multicolumn{1}{l|}{38} &
        \multicolumn{1}{l|}{363} &
        \multicolumn{1}{l|}{28} &
        \multicolumn{1}{l|}{330} &
        \multicolumn{1}{l|}{65} &
        \multicolumn{1}{l|}{368} \\ \hline
    \multicolumn{1}{|l|}{$d_{M} < 5 \sigma_{d_{M}}$} &
        \multicolumn{1}{l|}{4} &
        \multicolumn{1}{l|}{359} &
        \multicolumn{1}{l|}{4} &
        \multicolumn{1}{l|}{326} &
        \multicolumn{1}{l|}{4} &
        \multicolumn{1}{l|}{364} \\ \hline
    \multicolumn{1}{|l|}{UNITY Outlier} &
        \multicolumn{1}{l|}{2} &
        \multicolumn{1}{l|}{357} &
        \multicolumn{1}{l|}{2} &
        \multicolumn{1}{l|}{324} &
        \multicolumn{1}{l|}{2} &
        \multicolumn{1}{l|}{362} \\ \hline
    \end{tabular}
    \caption{The second set of cuts are based on the fitting model used, the SNPY\_EBV sample using \texttt{SNooPy}'s EBV\_model2, the SNPY\_Max sample using the max\_model, and the SALT sample using SALT3-NIR.
    We calculate the $\chi^2/\text{DoF}$ thresholds based on our comparison to the DEHVILS cut based on \texttt{SNANA}'s fit probability parameter.
    Our final two cuts are based on outlier detection.
    $d_M$ refers to the Mahalanobis distance described in Section \ref{sec:outlier method 1}.}
    \label{tab:sample_specific_cuts}
\end{table*}

\section{Results}
\label{sec:results}

\subsection{Hubble Diagrams}
We now present measurements of dispersion in the Hubble residuals of our three samples and their inferred intrinsic dispersions.
We do not list the value of $H_0$ used in each sample because it is not a direct result of the data as explained in Section \ref{sec:distances}.
To reiterate, $H_0$ is degenerate with the absolute magnitude of \sneia{} and we do not use alternative distance probes to estimate that magnitude.

\begin{table*}
    \begin{tabular}{ |r|c|c|c|c|c| }
        \hline
    Sample & N & RMS (mag) & WRMS (mag) & NMAD (mag) & $\sigma_{int}$ (mag) \\ \hline
        SNPY\_EBV & 357 & 0.165(010) & 0.152(008) & 0.123(010) & 0.121(011) \\
        SNPY\_Max & 324 & 0.245(024) & 0.214(028) & 0.164(011) & 0.212(028) \\
        SALT & 362 & 0.186(011) & 0.174(009) & 0.153(010) & 0.123(011) \\ \hline
    \end{tabular}
    \caption{We present various parametrizations of the dispersion in Hubble residuals in the SNPY\_EBV, SNPY\_Max, and SALT samples, as well as the intrinsic dispersion needed to reconcile the propagated uncertainties and measured dispersion.
    The samples overlap significantly, but they are not identical.
    For comparison purposes, we provide measurements of the common subset in Table \ref{tab:dispersion_common}.
    }
    \label{tab:dispersion}
\end{table*}

\begin{table*}
    \begin{tabular}{ |r|c|c|c|c|c| }
        \hline
    Sample & RMS (mag) & WRMS (mag) & NMAD (mag) & $\sigma_{int}$ (mag) \\ \hline
        SNPY\_EBV & 0.137(008) & 0.133(007) & 0.116(011) & 0.098(010) \\
        SNPY\_Max & 0.171(011) & 0.148(009) & 0.147(012) & 0.135(010) \\
        SALT & 0.146(007) & 0.150(008) & 0.142(012) & 0.103(010) \\ \hline
    \end{tabular}
    \caption{We present the values from Table \ref{tab:dispersion} derived from the intersection between the SNPY\_EBV, SNPY\_Max, and SALT samples.
    All samples are comprised of the same 240 objects.
    }
    \label{tab:dispersion_common}
\end{table*}

\begin{figure*}
    \includegraphics[width=0.9\textwidth]{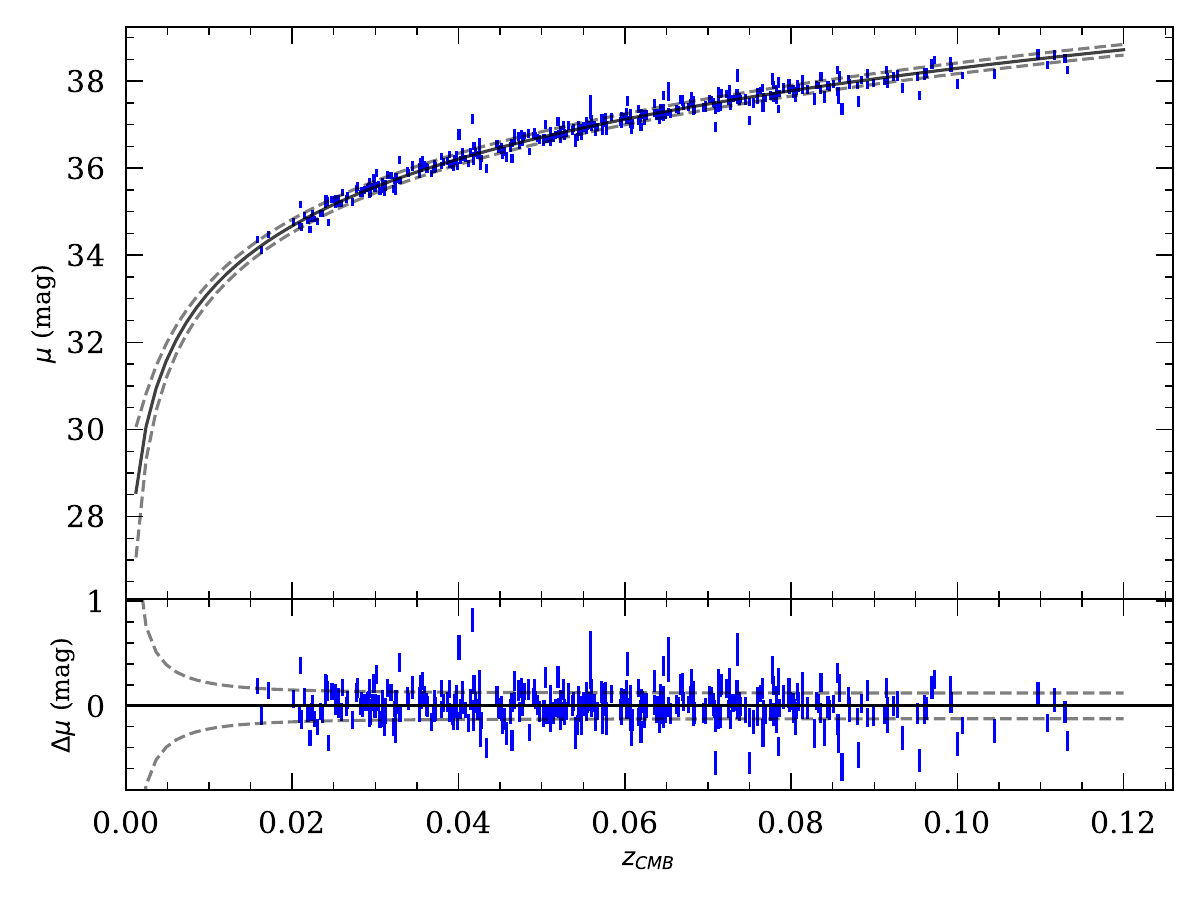}
    \caption{
        The top panel shows the Hubble diagram of the SNPY\_EBV sample with residuals plotted below.
        The value of $H_0$ is degenerate with the absolute magnitude of \sneia{}, amounting to a constant vertical offset.
        The solid black line shows the $\Lambda$CDM model that zeros the inverse-variance weighted residuals.
        The dashed lines show the combined uncertainty due to the sample's $\sigma_\text{int}$ and 250 km s$^{-1}$ of uncertainty in peculiar velocity converted to uncertainty in distance modulus via Equation \ref{eqn:redshift_uncertainty}.
    }
    \label{fig:EBVHD}
\end{figure*}

\begin{figure*}
    \includegraphics[width=0.9\textwidth]{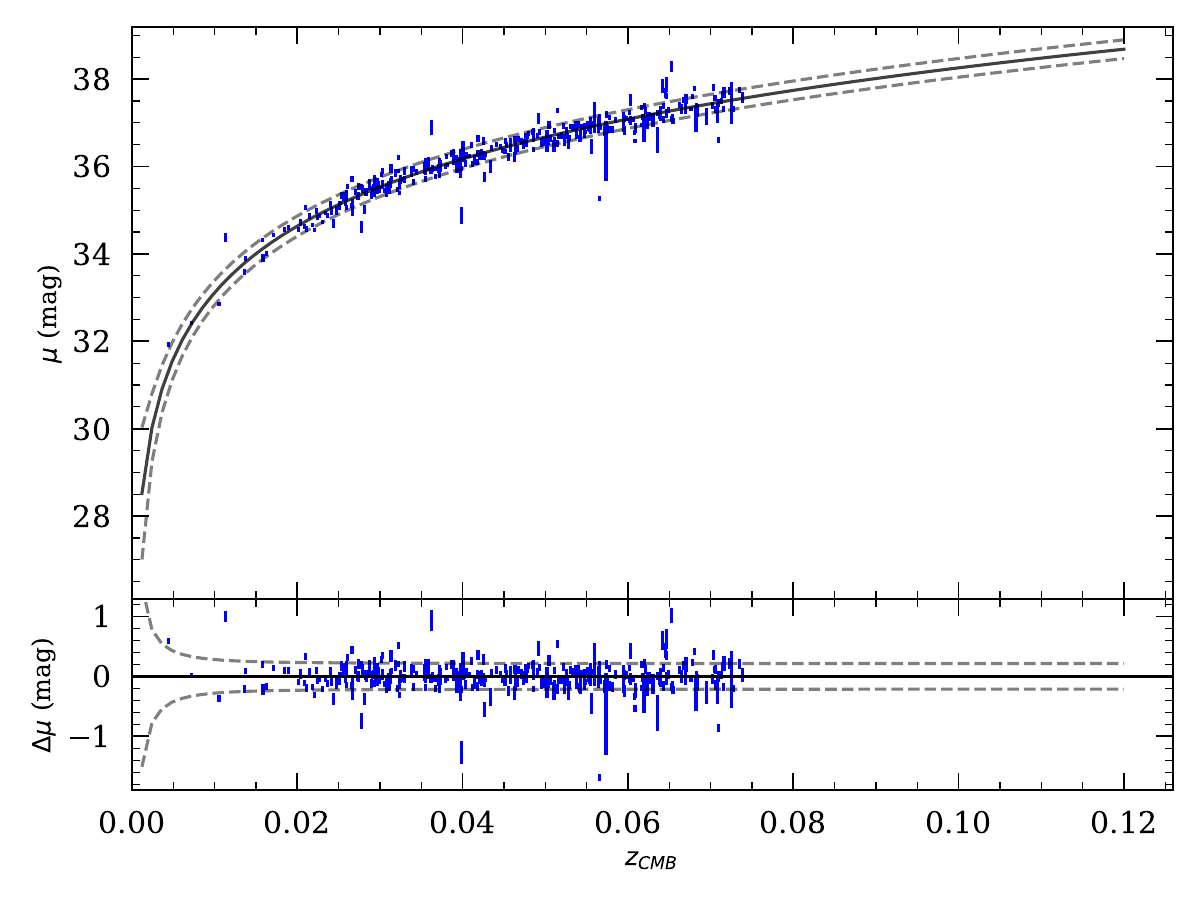}
    \caption{The same as Figure \ref{fig:EBVHD}, but using the SNPY\_Max sample.
    The limited redshift range is due to the cut requiring the observed filters to map to the CSP $V$ and $r$ filters in the rest frame.}
    \label{fig:MaxHD}
\end{figure*}

\begin{figure*}
    \includegraphics[width=0.9\textwidth]{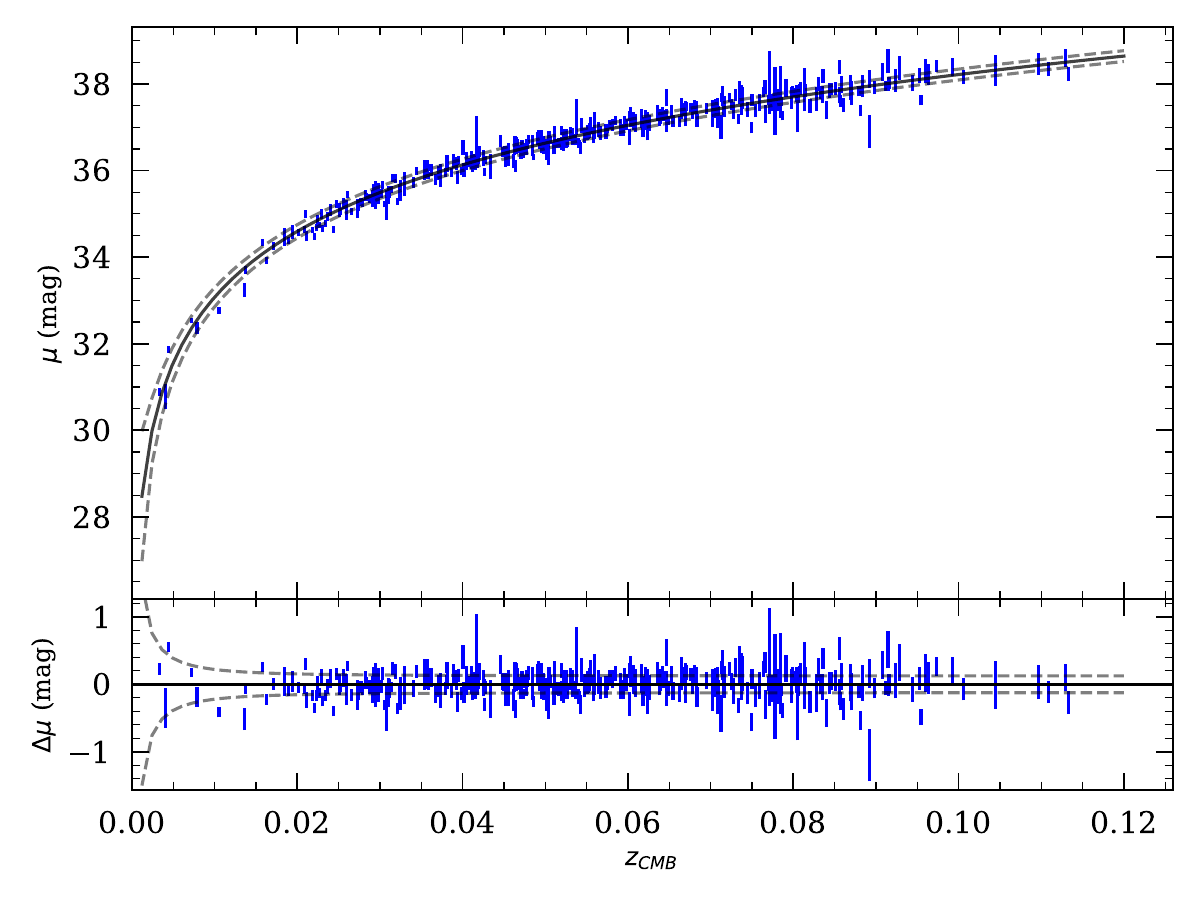}
    \caption{The same as Figure \ref{fig:EBVHD}, but using the SALT sample.}
    \label{fig:SALTHD}
\end{figure*}

The Hubble residuals in each sample are presented in Figures \ref{fig:EBVHD}--\ref{fig:SALTHD}.

\subsection{Trend with Redshift}
\label{sec:trend_w_redshift}
There is a trend between the Hubble residuals and redshift in the SNPY\_EBV and SALT samples.
We perform linear fits using the Bayesian approach detailed in \citet{jaynes99}, using flat priors in $\sin(\text{slope})$ and intercept, and a Jeffreys prior on scatter \citep{jeffreys46}.
\begin{align}
    \label{eqn:HR_redshift_trend}
    \Delta \mu_{\text{SNPY\_EBV}} =& -0.6(3) z_\text{CMB} \text{ mag} + 0.03(02) \text{ mag} \\
    \Delta \mu_{\text{SNPY\_Max}} =& 0.0(5) z_\text{CMB} \text{ mag} + 0.01(02) \text{ mag} \\
    \Delta \mu_{\text{SALT}} =& 0.4(4) z_\text{CMB} \text{ mag} -0.02(02) \text{ mag}
\end{align}
The Hubble residuals in the SNPY\_Max sample do not appear to trend with redshift.
The differing signs in the slopes of the SNPY\_EBV and SALT samples indicate that the issue is due to differences in the fitters rather than a real trend in the data or an issue in the estimation of $\mu_\text{cos}$.
A review of the \texttt{SNooPy} and \texttt{SNCosmo} code revealed no issue with the programmatic implementation of the methods in the literature \citep{burns11, burns14, guy05, guy07, kenworthy21, pierel22}.

\begin{figure}
    \includegraphics[width=0.45\textwidth]{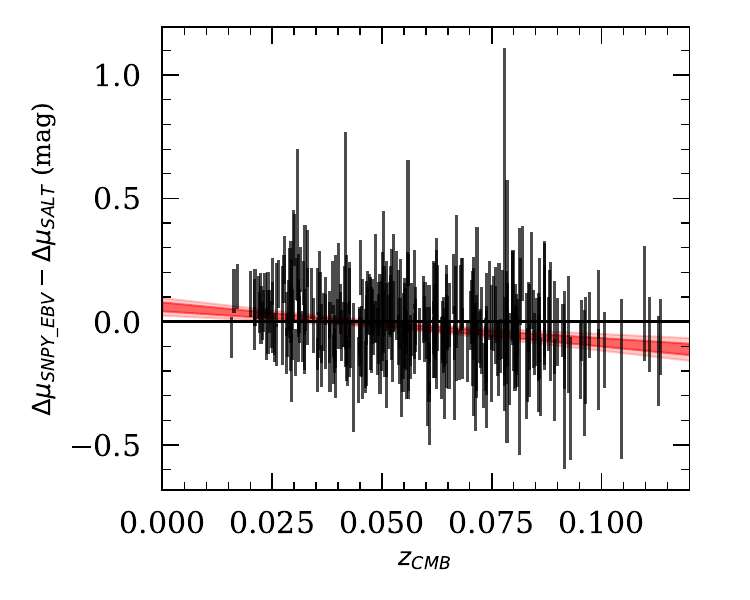}
    \caption{
        The difference between Hubble residuals from the SNPY\_EBV sample ($\Delta_{\text{SNPY\_EBV}}$) and the SALT sample ($\Delta_{\text{SALT}}$) is anti-correlated with redshift.
        We fit a linear trend to the data, finding a slope of -1.4(3) mag.
        The 1- and 2-$\sigma$ contours are in red and light-red.
    }
    \label{fig:residual_trend}
\end{figure}

The difference between SNPY\_EBV and SALT Hubble residuals is seen most clearly in Figure \ref{fig:residual_trend}.
Comparing residuals accounts for the zero-point offset in inferred $\mu$ in each sample and suppresses astrophysical properties that should affect both inferences equally, such as peculiar velocity or intrinsic variation in luminosity.
We calculate uncertainties for the differences using the Pearson correlation between the distance modulus uncertainties in each sample.

We tested the \texttt{SNooPy} EBV\_model2 and SALT3-NIR by fitting the $r$- and $i$-band photometry of a \snia{} at $z_\text{CMB} \approx 0.72$ using both programs.\footnote{SN 05D4ev from \citet{guy10}}
Using the $H_0$ and $M$ values from the SNPY\_EBV and SALT samples, we found the corresponding fitters produced residual distance moduli of $0.095 \pm 0.167$ and $-0.315 \pm 0.167$. 
This indicates that any trend between residual distance modulus and redshift does not continue at higher redshifts.

\begin{figure*}
    \includegraphics[width=0.9\textwidth]{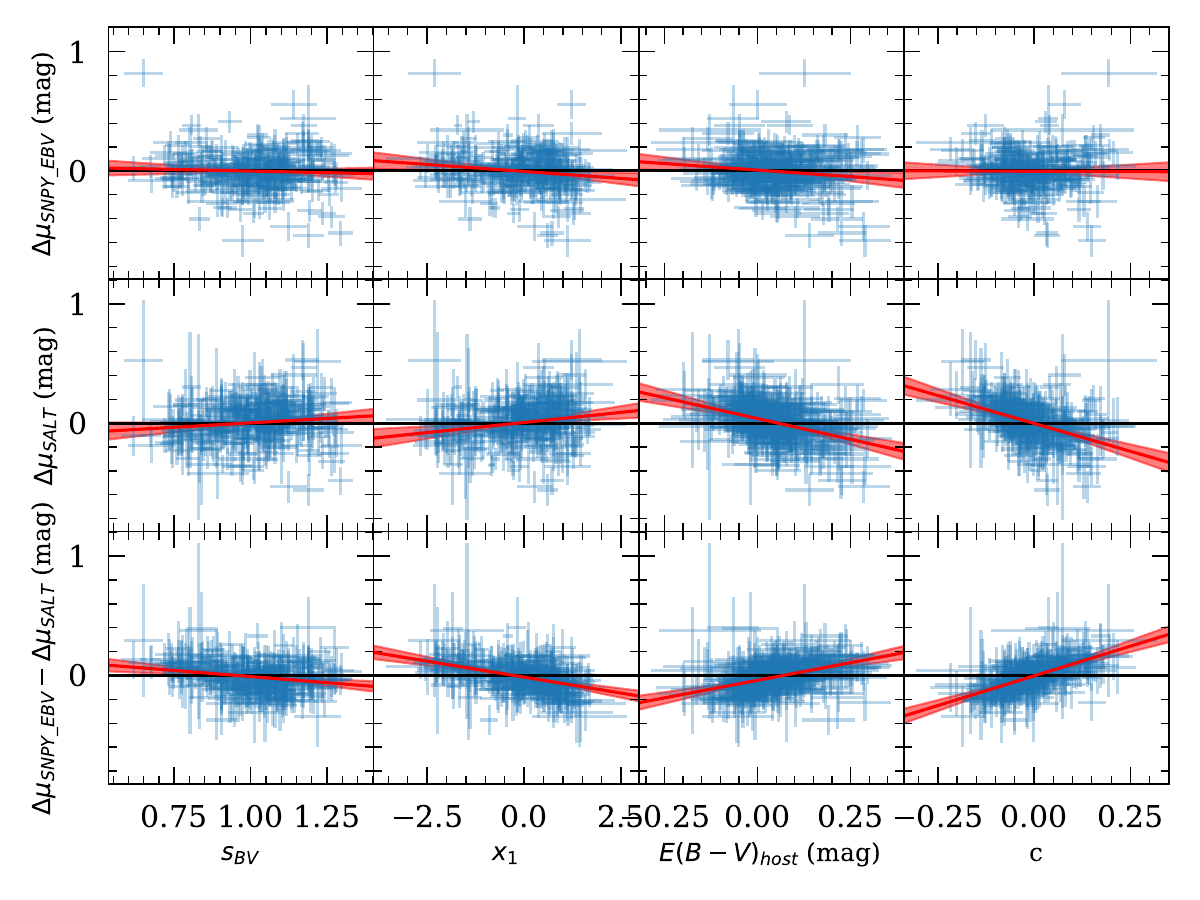}
    \caption{
        We compare the Hubble residuals in the SNPY\_EBV sample, the residuals in the SALT sample, and their differences to the \texttt{SNooPy} fitting parameters $s_{BV}$ and $E(B-V)_{\text{host}}$ and SALT fitting parameters $x_1$ and $c$.
        The best-fitting linear relation is plotted in red with a 95\% confidence interval in light red.
        The strong correlation with colour parameters implies the standardization coefficient $\beta$ is not calibrated correctly for our sample, and that the reddening law assumed in the EBV\_model2 may be inappropriate for our sample.
    }
    \label{fig:trend_vs_params}
\end{figure*}

The trend with redshift could be the result of differences in how the fitters account for shape or colour, which could both evolve with redshift due to selection effects.
We investigated whether the SNPY\_EBV residuals, SALT residuals, or their differences were correlated with the fitting parameters $s_{BV}$, $E(B-V)_{\text{host}}$, $x_1$, and $c$, plotting the results in Figure \ref{fig:trend_vs_params}.
Correlations imply the fitter is not properly accounting for the effect shape or colour has on the luminosity.
In SALT this would mean the standardization parameters $\alpha$ or $\beta$ are improperly calibrated.
In \texttt{SNooPy}'s EBV\_model2 a correlation between the Hubble residuals and $s_{BV}$ would imply there is a systematic difference between the light curves in our sample and the light curves used for interpolation.
A correlation with $E(B-V)_{\text{host}}$ would imply that the reddening law assumed in the EBV\_model2 does not fully capture the dust properties affecting our observations.

\begin{figure*}
    \includegraphics[width=0.9\textwidth]{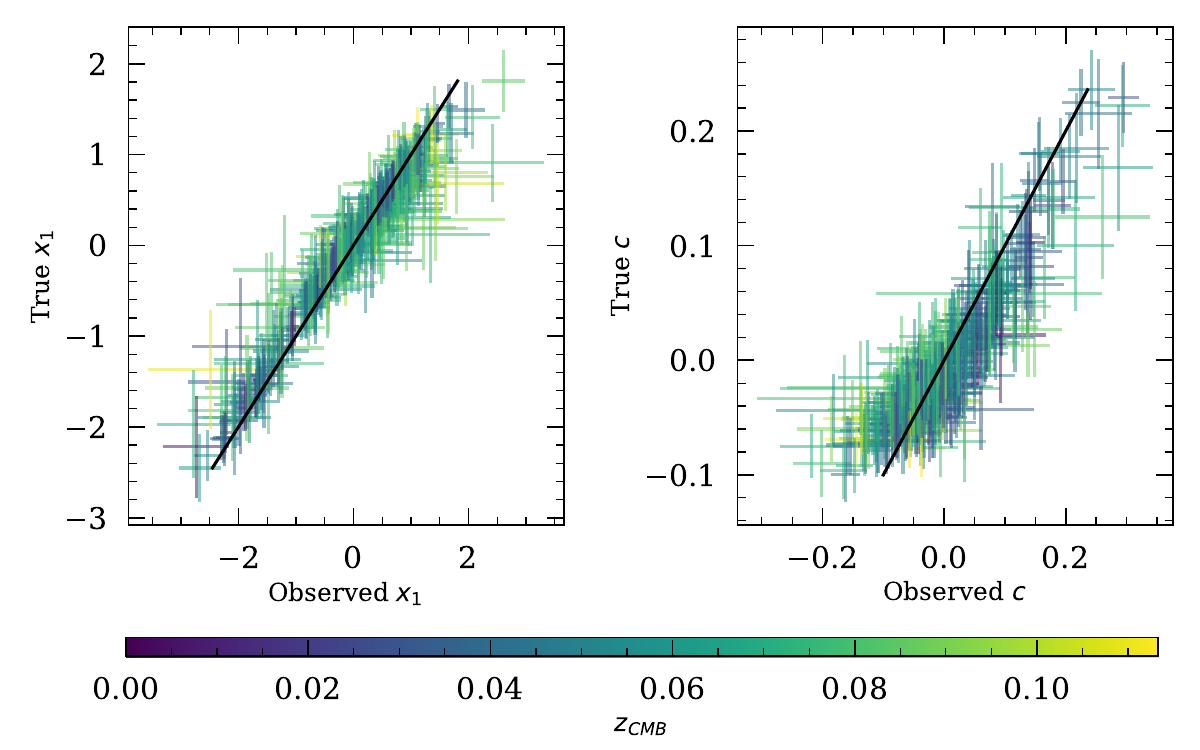}
    \caption{UNITY models the ``true'' values of $x_1$ and $c$ as latent variables in its hierarchical model. We compare these against the ``observed'' values that come from SALT3-NIR fits, with one-to-one correspondence lines plotted in black. The standardization coefficients from UNITY are calibrated against the true values, and will not minimize Hubble residuals when used with the observed values.}
    \label{fig:UNITY_true_vs_observed}
\end{figure*}

The residuals appear correlated with the colour parameters $E(B-V)_\text{host}$ and $c$, with the correlation most obvious in the SALT residuals and the residual differences.
This calls our $\beta$ coefficient into question, which comes from an analysis of the SALT sample performed in UNITY.
The coefficients are not inferred by minimizing dispersion in the Hubble residuals, but by maximizing a likelihood in a Bayesian hierarchical model.
UNITY models the ``true'' $x_1$ and $c$ parameters of a \snia{} as latent variables to account for Eddington bias.
The standardization coefficients operate on these ``true'' values rather than the ``observed'' values that come from a light curve fit.
We compare the ``true'' and ``observed'' $x_1$ and $c$ parameters in Figure \ref{fig:UNITY_true_vs_observed}.
Deviations from one-to-one correspondence come from both statistical error and Eddington bias, which manifests as ``observed'' parameters scattering away from 0.

\begin{figure}
    \includegraphics[width=0.45\textwidth]{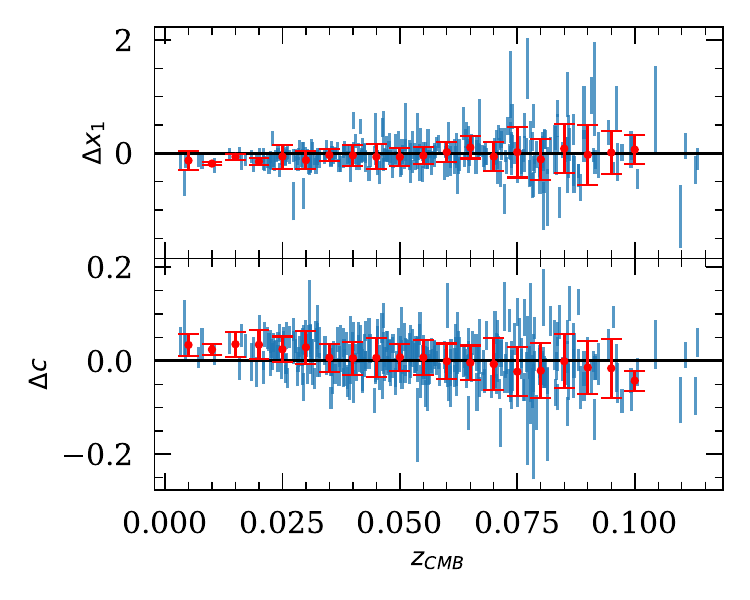}
    \caption{We parametrize the difference between the ``true'' and ``observed'' $x_1$ and $c$ parameters as $\Delta x_1$ and $\Delta c$. The errors in each difference are estimated using the sample Pearson correlation coefficient. The red markers and errorbars show the medians and standard deviations of differences in bins 0.005 wide in redshift space. The standard deviations generally increase with redshift until $z \sim 0.09$, where data is relatively sparse. $\Delta c$ appears offset from 0, especially at lower redshifts. This could imply that the hyperparameters UNITY uses to describe the distribution of $c$ may require more flexibility to accurately model our observations.}
    \label{fig:eddington_bias_vs_redshift}
\end{figure}

Figure \ref{fig:eddington_bias_vs_redshift} shows that the bias seems more prevalent at higher redshifts.
Parameter differences in bins at higher redshifts generally have larger standard deviations, with sample incompleteness heavily affecting bins beyond $z \sim 0.09$.
This could indicate that the uncertainties in $x_1$ or $c$ are underestimated in fits to noisier data, but verifying such a claim would require simulations beyond the scope of this paper (such as those in \citet{peterson24}).
Additionally, the median differences in $c$ appear non-zero, especially in lower redshift bins.
We have not identified a definitive cause for this behaviour, but speculate that our choice of hyperparameters in UNITY does not allow for the flexibility needed to model the distribution of $c$ over the parameter space spanned by our sample.
This could produce a correlation between SALT Hubble residuals and redshift or $c$ independent of $\beta$.

\begin{figure}
    \includegraphics[width=0.45\textwidth]{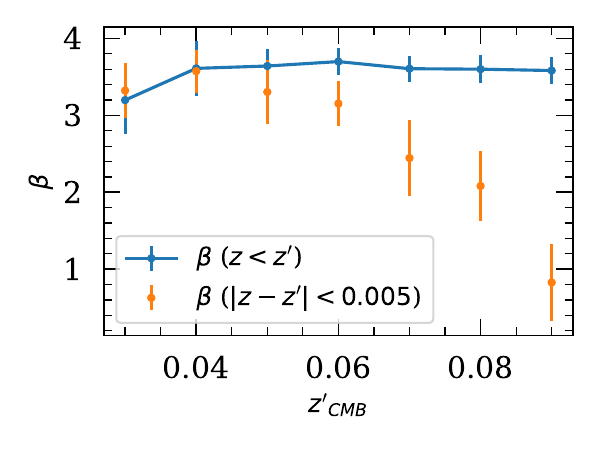}
    \caption{
        We analyse subsets of the SALT sample using UNITY and show the inferred $\beta$ values as a function of redshift cut value $z'_{CMB}$.
        The blue data show the results from subsets defined by a maximum redshift $z < z'$.
        $\beta$ is relatively stable as the sample expands to include higher redshift targets.
        The orange data are from redshift bins centred on $z'$ and 0.01 wide, such that $|z - z'| < 0.005$.
        SNe in higher redshift bins prefer lower $\beta$ values, which could be a result of Eddington bias becoming more significant at greater distances.
    }
    \label{fig:beta_vs_redshift}
\end{figure}

To ascertain the validity of using a single $\beta$ value over the entire redshift range, we analysed subsets of the SALT sample in UNITY.
We examined subsets consisting of targets within redshift bins of width 0.01, as well as a cumulative sum including all targets below a series of redshifts.
Figure \ref{fig:beta_vs_redshift} shows that the inferred $\beta$ decreases in higher redshift bins, but that it remains relatively stable in the cumulative case.
This suggests the hyperparameters describing the distribution of $c$ are robust against higher redshift \sneia{}, but that these SNe prefer a lower value for $\beta$.
This assumes the $c$ parameter follows a single distribution over the whole sample, rendering measurements that may suggest otherwise an effect of bias.
UNITY permits alternative parametrizations, such as a broken-linear form for $\beta$, but adding such complexity is beyond the scope of this paper.
Similarly, more sophisticated analyses of dust properties using \texttt{SNooPy} have been performed with its color\_model \citep[e.g.][]{burns18, johansson21} or by using more of the colour information in the max\_model \citep[e.g.][]{uddin20, uddin23}.
Implementing and evaluating these approaches will be necessary before using our data for robust cosmological analyses.

Thus, we do not find a satisfactory solution to eliminate the correlations between Hubble residuals and redshift or colour parameter.
Empirical corrections are possible using Equations \ref{eqn:HR_redshift_trend}, but such an approach is neither physically motivated nor statistically rigorous.
More detailed analyses are required to fully understand and rectify this issue.

\subsection{The Effect of NIR Photometry on Distance Measurements}
\label{sec:nir vs optical}

As mentioned in section \ref{sec:sneia_in_nir}, NIR photometry offers two key benefits when analyzing \sneia{}.
The effects of dust are suppressed and \sneia{} demonstrate less variable peak luminosities in the NIR.
We examine the benefits of NIR photometry by comparing fits using NIR and optical photometry to fits using only optical photometry.
We split our three samples into six: SNPY\_EBV\_OJ, SNPY\_Max\_OJ, and SALT\_OJ (the OJ samples), which include optical and $J$-band photometry and SNPY\_EBV\_O, SNPY\_Max\_O and SALT\_O (the O samples), which are their optical-only counterparts.
Unlike the SNPY\_EBV, SNPY\_Max, and SALT samples, which vary in size, target selection, and even bandpasses used to fit a given target, we enforce parity between the OJ and O samples.
To do this we prepare the O samples following the same methodology used to produce SNPY\_EBV, SNPY\_Max, and SALT except without $J$-band photometry.
The number of SNe discarded at each cut and the size of the final samples are listed in Table \ref{tab:cuts nir vs optical}.
The six samples are made of the SNe common to both the O and OJ samples.

\begin{figure}
    \centering
    \includegraphics[width=0.45\textwidth]{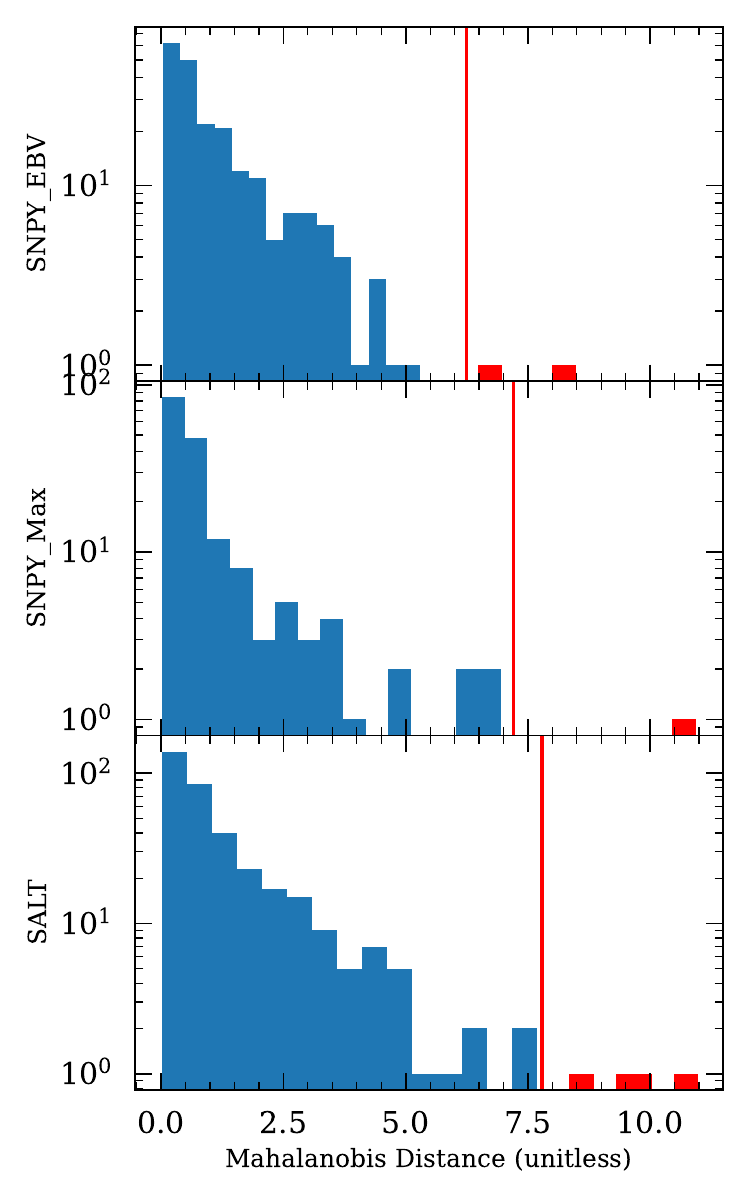}
    \caption{Our first outlier detection algorithm is based on agreement between models fit with and without $J$-band photometry as quantified by the Mahalanobis distance between parameter differences and the origin, representing a SN with identical estimates in the OJ and O samples.
    A greater distance indicates greater disagreement between fits, with significant disagreement indicating at least one of the models is unreliable.
    }
    \label{fig:mahalanobis distances}
\end{figure}

We modify the outlier detection method described in \ref{sec:outlier method 1} to highlight disagreement between the OJ and O samples rather than between SNPY\_EBV, SNPY\_Max, and SALT.
This precludes the need to transform \texttt{SNooPy} fitting parameters into SALT parameters.
The fitting parameters of SNe in the OJ and O samples produce difference vectors: $m = (\delta t_0, \delta s_{BV}, \delta E(B-V)_\text{host})$ for differences between SNPY\_EBV\_OJ and SNPY\_EBV\_O, $m = (\delta t_0, \delta s_{BV}, \delta(V - r))$ for differences between SNPY\_Max\_OJ and SNPY\_Max\_O, and $m = (\delta t_0, \delta m^*_B, \delta x_1, \delta c)$ for differences between SALT\_OJ and SALT\_O.
We use the Mahalanobis distance $d(m,D) = \sqrt{m\Sigma^{-1}m^T}$ to identify outliers, once again approximating the covariance matrix $\Sigma$ by bootstrap resampling the parameter differences 5,000 times and averaging the sample covariances.
The distributions of Mahalanobis distances for samples the three pairs of samples are given in Figure \ref{fig:mahalanobis distances}.

\begin{table*}
    \begin{tabular}{lllllll}
        \hline
    \multicolumn{1}{|l|}{Cuts} &
        \multicolumn{1}{l|}{SNPY\_EBV\_O Cut} &
        \multicolumn{1}{l|}{Remaining} &
        \multicolumn{1}{l|}{SNPY\_Max\_O Cut} &
        \multicolumn{1}{l|}{Remaining} &
        \multicolumn{1}{l|}{SALT\_O Cut} &
        \multicolumn{1}{l|}{Remaining} \\ \hline
    \multicolumn{1}{|l|}{Passed Sample Wide Cuts} &
        \multicolumn{1}{l|}{} &
        \multicolumn{1}{l|}{504} &
        \multicolumn{1}{l|}{} &
        \multicolumn{1}{l|}{504} &
        \multicolumn{1}{l|}{} &
        \multicolumn{1}{l|}{504} \\ \hline
    \multicolumn{1}{|l|}{Successful Fit} &
        \multicolumn{1}{l|}{2} &
        \multicolumn{1}{l|}{502} &
        \multicolumn{1}{l|}{2} &
        \multicolumn{1}{l|}{502} &
        \multicolumn{1}{l|}{1} &
        \multicolumn{1}{l|}{503} \\ \hline
    \multicolumn{1}{|l|}{Rest frame $m_V$ and $m_r$} &
        \multicolumn{1}{l|}{\cellcolor[HTML]{C0C0C0}} &
        \multicolumn{1}{l|}{\cellcolor[HTML]{C0C0C0}} &
        \multicolumn{1}{l|}{92} &
        \multicolumn{1}{l|}{410} &
        \multicolumn{1}{l|}{\cellcolor[HTML]{C0C0C0}} &
        \multicolumn{1}{l|}{\cellcolor[HTML]{C0C0C0}} \\ \hline
    \multicolumn{1}{|l|}{$0.6 < s_{BV} < 1.3$} &
        \multicolumn{1}{l|}{23} &
        \multicolumn{1}{l|}{479} &
        \multicolumn{1}{l|}{40} &
        \multicolumn{1}{l|}{370} &
        \multicolumn{1}{l|}{\cellcolor[HTML]{C0C0C0}} &
        \multicolumn{1}{l|}{\cellcolor[HTML]{C0C0C0}} \\ \hline
    \multicolumn{1}{|l|}{$\sigma_{s_{BV}} < 0.2$} &
        \multicolumn{1}{l|}{3} &
        \multicolumn{1}{l|}{476} &
        \multicolumn{1}{l|}{7} &
        \multicolumn{1}{l|}{363} &
        \multicolumn{1}{l|}{\cellcolor[HTML]{C0C0C0}} &
        \multicolumn{1}{l|}{\cellcolor[HTML]{C0C0C0}} \\ \hline
    \multicolumn{1}{|l|}{$E(B-V)_\text{host} < 0.3$ mag} &
        \multicolumn{1}{l|}{66} &
        \multicolumn{1}{l|}{410} &
        \multicolumn{1}{l|}{\cellcolor[HTML]{C0C0C0}} &
        \multicolumn{1}{l|}{\cellcolor[HTML]{C0C0C0}} &
        \multicolumn{1}{l|}{\cellcolor[HTML]{C0C0C0}} &
        \multicolumn{1}{l|}{\cellcolor[HTML]{C0C0C0}} \\ \hline
    \multicolumn{1}{|l|}{$|x_1| < 3$} &
        \multicolumn{1}{l|}{\cellcolor[HTML]{C0C0C0}} &
        \multicolumn{1}{l|}{\cellcolor[HTML]{C0C0C0}} &
        \multicolumn{1}{l|}{\cellcolor[HTML]{C0C0C0}} &
        \multicolumn{1}{l|}{\cellcolor[HTML]{C0C0C0}} &
        \multicolumn{1}{l|}{22} &
        \multicolumn{1}{l|}{481} \\ \hline
    \multicolumn{1}{|l|}{$\sigma_{x_1} < 1.5$} &
        \multicolumn{1}{l|}{\cellcolor[HTML]{C0C0C0}} &
        \multicolumn{1}{l|}{\cellcolor[HTML]{C0C0C0}} &
        \multicolumn{1}{l|}{\cellcolor[HTML]{C0C0C0}} &
        \multicolumn{1}{l|}{\cellcolor[HTML]{C0C0C0}} &
        \multicolumn{1}{l|}{5} &
        \multicolumn{1}{l|}{476} \\ \hline
    \multicolumn{1}{|l|}{$|c| < 0.3$} &
        \multicolumn{1}{l|}{\cellcolor[HTML]{C0C0C0}} &
        \multicolumn{1}{l|}{\cellcolor[HTML]{C0C0C0}} &
        \multicolumn{1}{l|}{\cellcolor[HTML]{C0C0C0}} &
        \multicolumn{1}{l|}{\cellcolor[HTML]{C0C0C0}} &
        \multicolumn{1}{l|}{38} &
        \multicolumn{1}{l|}{438} \\ \hline
    \multicolumn{1}{|l|}{$\sigma_c < 0.2$} &
        \multicolumn{1}{l|}{\cellcolor[HTML]{C0C0C0}} &
        \multicolumn{1}{l|}{\cellcolor[HTML]{C0C0C0}} &
        \multicolumn{1}{l|}{\cellcolor[HTML]{C0C0C0}} &
        \multicolumn{1}{l|}{\cellcolor[HTML]{C0C0C0}} &
        \multicolumn{1}{l|}{5} &
        \multicolumn{1}{l|}{433} \\ \hline
    \multicolumn{1}{|l|}{Phase Requirements} &
        \multicolumn{1}{l|}{1} &
        \multicolumn{1}{l|}{409} &
        \multicolumn{1}{l|}{1} &
        \multicolumn{1}{l|}{362} &
        \multicolumn{1}{l|}{2} &
        \multicolumn{1}{l|}{431} \\ \hline
    \multicolumn{1}{|l|}{Reduced $\chi^2$ < 4.14/4.51/1.31} &
        \multicolumn{1}{l|}{21} &
        \multicolumn{1}{l|}{388} &
        \multicolumn{1}{l|}{21} &
        \multicolumn{1}{l|}{341} &
        \multicolumn{1}{l|}{40} &
        \multicolumn{1}{l|}{391} \\ \hline
    \multicolumn{1}{|l|}{Also In $OJ$ Sample} &
        \multicolumn{1}{l|}{173} &
        \multicolumn{1}{l|}{215} &
        \multicolumn{1}{l|}{166} &
        \multicolumn{1}{l|}{175} &
        \multicolumn{1}{l|}{36} &
        \multicolumn{1}{l|}{355} \\ \hline
    \multicolumn{1}{|l|}{$d_{M} < 5 \sigma_{d_{M}}$} &
        \multicolumn{1}{l|}{2} &
        \multicolumn{1}{l|}{213} &
        \multicolumn{1}{l|}{1} &
        \multicolumn{1}{l|}{174} &
        \multicolumn{1}{l|}{5} &
        \multicolumn{1}{l|}{350} \\ \hline
    \multicolumn{1}{|l|}{UNITY Outlier} &
        \multicolumn{1}{l|}{1} &
        \multicolumn{1}{l|}{212} &
        \multicolumn{1}{l|}{0} &
        \multicolumn{1}{l|}{174} &
        \multicolumn{1}{l|}{2} &
        \multicolumn{1}{l|}{348} \\ \hline
    \end{tabular}
    \caption{Similar to Table \ref{tab:sample_specific_cuts}, we list the number of SNe discarded at each cut for our optical-only samples.
    We begin after the survey wide cuts of Table \ref{tab:survey_wide_cuts}, starting with the number of successful fits in \texttt{SNooPy}'s EBV\_model2 (SNPY\_EBV\_O), max\_model (SNPY\_Max\_O), and SALT3-NIR (SALT\_O).
    Our final two cuts are based on outlier detection.}
    \label{tab:cuts nir vs optical}
\end{table*}

\begin{table*}
    \begin{tabular}{ |r|c|c|c|c|c| }
        \hline
    Sample & RMS (mag) & WRMS (mag) & NMAD (mag) & $\sigma_{int}$ (mag) \\ \hline
        SNPY\_EBV\_OJ & 
            $0.166(012)$ &
            $0.154(010)$ &
            $0.120(011)$ &
            $0.122(014)$ \\
        SNPY\_EBV\_O & 
            $0.171(011)$ &
            $0.162(010)$ &
            $0.149(012)$ &
            $0.127(014)$ \\ \hline
        SNPY\_Max\_OJ & 
            $0.281(037)$ &
            $0.245(047)$ &
            $0.173(019)$ &
            $0.237(044)$ \\
        SNPY\_Max\_O & 
            $0.276(028)$ &
            $0.227(021)$ &
            $0.188(022)$ &
            $0.248(028)$ \\ \hline
        SALT\_OJ & 
            $0.170(008)$ &
            $0.171(009)$ &
            $0.146(010)$ &
            $0.122(011)$ \\
        SALT\_O & 
            $0.185(009)$ &
            $0.184(010)$ &
            $0.162(011)$ &
            $0.129(012)$ \\ \hline
    \end{tabular}
    \caption{Adding NIR photometry does not lead to statistically significant decreases in the various measures of dispersion.
    Each estimator is calculated after bootstrap resampling the Hubble residuals 5,000 times.
    The value is the average and the uncertainty is the standard deviation.
    }
    \label{tab:nir vs optical dispersion}
\end{table*}

\begin{figure*}
    \centering
    \includegraphics[width=0.9\textwidth]{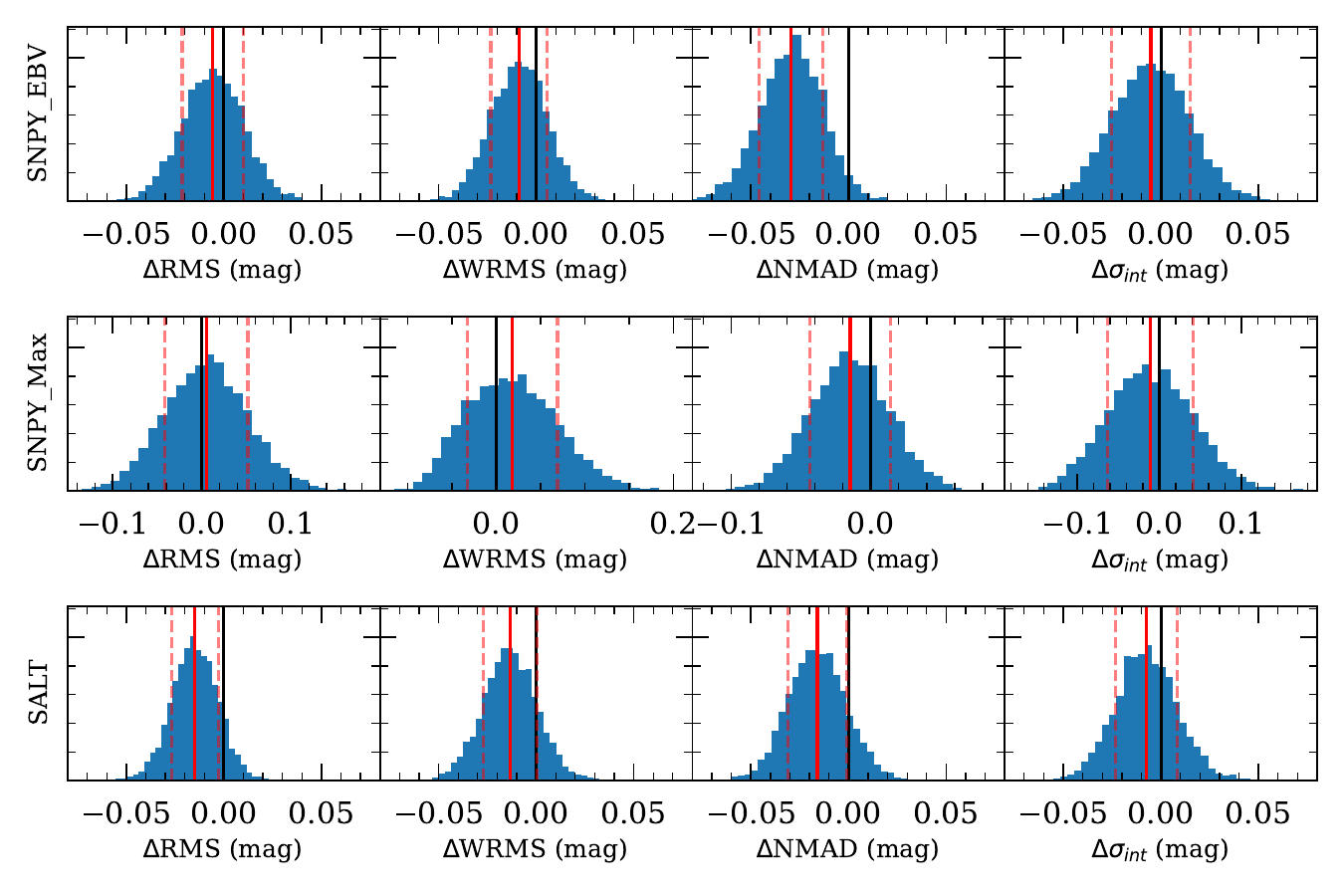}
    \caption{We recorded various measures of dispersion over 5,000 iterations of bootstrap resampling, and show the distributions of differences between the OJ and O samples with the averages given by the solid red lines.
    Including $J$-band photometry with the optical data typically leads to decreases in the three tested measures of dispersion in Hubble residuals and the inferred intrinsic dispersion, but those differences are usually within one standard deviation (red dashed lines) of 0 mag (solid black lines).
    }
    \label{fig:NIR_vs_optical_tests}
\end{figure*}

Measurements of dispersion in each sample's Hubble residuals are presented in Table \ref{tab:nir vs optical dispersion}.
We characterize the differences between the OJ and O samples with the bootstrapping method we used when varying photometry in Section \ref{sec:vary_phot}.
The values and uncertainties in Table \ref{tab:nir vs optical dispersion} are the averages and standard deviations of this process.
Histograms of the resampled dispersion differences are plotted in Figure \ref{fig:NIR_vs_optical_tests}.
The various dispersion estimators show a general decrease when adding $J$-band photometry to the O samples, but most of the differences are within one standard deviation of no change.
The exceptions are the NMAD in the SALT and SNPY\_EBV samples and the RMS in the SALT sample.
Interpreting these exceptions as indicators of decreased dispersion while ignoring the other measures is a classic case of the multiple comparisons problem.
To control the familywise error rate we use the sequentially rejective Bonferroni test \citep{holm79}.
None of the distributions are far enough from 0 to claim that including $J$-band photometry leads to statistically significant decreases in dispersion.
This does not imply other methodologies do not or cannot benefit from the $J$-band photometry, but that with our samples, cuts, and methods, we cannot definitively say NIR photometry leads to smaller Hubble residuals.

\section{Discussion}
The RMS of the Hubble residuals is 0.165 mag in the SNPY\_EBV sample ($N=357$), 0.245 mag in the SNPY\_Max sample ($N=324$), and 0.186 mag in the SALT sample ($N=362$).
Our result goes against a number of studies which support the use of NIR photometry in deriving distances to \sneia{}, but is not alone in finding relatively large dispersions.
\citet{stanishev18} combine optical and NIR light curves from numerous sources, including new observations, and find RMS values of $\sim$0.15 mag ($N\sim120$) while \citet{johansson21} did the same and found an RMS of 0.19 mag ($N=165$).
Notably, the sub-sample of 16 new \sneia{} presented in \citet{stanishev18} were only imaged once or twice in the NIR, and the RMS of their residuals is $\sim$0.2.
Sparsity may play a role in the greater dispersion, but \citet{mullerbravo22} found that the effect is relatively small, bringing an RMS of 0.166 mag to 0.180 mag ($N=36$) when removing all but one epoch from $J$-band light-curves.

One key difference between our work and those which find smaller dispersions is that our analysis does not force \sneia{} to be standard candles in the NIR.
It may be possible that variation in \snia{} NIR luminosity, if it does vary, is not parametrized by the correlations observed in the optical.
In our comparison with \citet{peterson23} we found that our fitting methods applied to their $YJH$ data resulted in large dispersion, whereas they fixed the shape and colour parameters in the NIR-only fits to be 0, removing any variation in luminosity between SNe, and measured lower dispersion than when using bandpass combinations including optical data from ATLAS.
This is similar to \citet{avelino19}, who treated \sneia{} as standard candles in the NIR and found that smaller Hubble residuals than those from optical-only fits using \texttt{SNooPy} or SALT2.
That said, not all studies favour this approach.
\citet{jones22} measured Hubble residual scatter over several analyses of 79 \sneia{} and measured a RMS of $\sim$0.17 mag using \texttt{SNooPy} to fit only NIR data, and $\sim$0.14 mag when including optical data with $R_V = 1.52$.
The optical and NIR RMS increased to $\sim$0.18 mag when using a Milky Way-like value of $R_V = 3.1$, emphasizing the importance of assumptions regarding dust.
Similarly, \citet{pierel22} examined the performance of SALT3-NIR, SALT3, and \texttt{SNooPy} over various bandpass combinations and model parametrizations.
They found a Hubble residual RMS of $\sim$0.12 mag ($N=24$) when using SALT3-NIR with optical and $YJH$ data, fitting for shape and colour, and a greater RMS of $\sim$0.13 mag for the same sample using only $YJH$ data and keeping the shape and colour parameters fixed at 0.

As explored in Section \ref{sec:trend_w_redshift}, there is a trend between redshift and the Hubble residuals in SNPY\_EBV and SALT.
Correcting this trend empirically will decrease the dispersion of the Hubble residuals, but such an a posteriori treatment invites bias.
There are several other obvious ways to decrease the measured dispersion.
One could calibrate $\alpha$ or $\beta$ by minimizing Hubble residuals, employ corrections by the redshift bin, or use cuts based on Hubble residuals such as Chauvanet's crietrion or $\sigma$-clipping.
There is ample motivation for using such techniques.
Our robust measure of dispersion, NMAD, is consistently lower than RMS and WRMS in all three samples, which suggests that there are \sneia{} in our samples could be considered outliers.
However, we choose to present our results as we found them to avoid contaminating them with ad hoc corrections.

The study of \sneia{} in the NIR has advanced as more data have become available, but there are still challenges that must be met to maximize the potential benefits.
At the moment it is unclear whether \sneia{} are standard candles in the NIR or simply require less standardization than in the optical.
Similarly, it is not clear if the shape-luminosity correlation observed in the optical is still the primary mode of variation in the NIR.
Answering these questions will require various kinds of data.
Spectral time series provide unique views into the physical mechanisms of \sneia{}, while also improving the accuracy of K-corrections.
High-cadence, multi-band observations like those pursued by the DEHVILS survey are vital for building standardization models.
The \hsf{} project provides a valuable test bed for \snia{} research through its unprecedented sample size.

\section{Conclusion}
This paper introduces the \hsf{} project, a peculiar velocity survey designed to obtain systematics-limited distances to \sneia{} while consuming minimal dedicated observational resources.
We review the observational components of our project: optical photometry from public all-sky surveys, NIR photometry from UKIRT, and optical spectroscopy from the UH 2.2 m and Subaru.
We validate our methods and data against external sources.
We use SDSS spectra to find that our redshift inferences are typically within 45 km s$^{-1}$ of the values in HyperLEDA.
The absolute wavelength calibration of our spectroscopic observations introduces minimal error, producing redshifts typically within 48 km s$^{-1}$ of their published values.
Using our methodology to fit data from our partner program DEHVILS, we found no increase in the dispersion of Hubble residuals when using only optical data, or using optical and NIR data with \texttt{SNooPy}'s EBV\_model2.
The increase in dispersion seen when using SALT3-NIR seems to come from our choice to calibrate the standardization coefficients with UNITY instead of only using the Hubble residuals.
Our independent photometric reductions of the same $J$-band observations are consistent, producing similar dispersions.
Given cuts on both \texttt{SNooPy} and SALT fitting parameters, our three final samples include 357, 324, and 362 SNe.
The RMS values of their Hubble residuals are 0.165, 0.245, and 0.186 mag.

The Nancy Grace Roman space telescope will obtain rest-frame NIR observations of \sneia{} within $z\sim0.7$ \citep{hounsell18, rose21}, necessitating the maturation of \snia{} cosmology in the NIR.
Thus far, the majority of publicly available NIR \snia{} light-curves have come from CSP-I \citep[N=123][]{krisciunas17}, CfAIR2 \citep[N=94][]{friedman15}, or recent work from our partner program DEHVILS \citep[N=96][]{peterson23}.
Data from CSP-II \citep[N=214][]{phillips19, hsiao19}, the SIRAH program \citep[N=24, HST-GO 15889; ][]{jha19}, and other exciting projects are expected in the near future.
Upon publication of this work, we will release NIR observations of 1,217 transients, including 668 spectroscopically classified \sneia{}, 437 of which are in at least one of our final samples, and 215 spectroscopic redshifts for \snia{} host-galaxies that have not been previously measured.
The NIR photometry of the \hsf{} project is the largest homogeneous collection of its kind in terms of unique \sneia{}.
This growing sample will provide increasing resolution into peculiar velocities as a function of position on the sky and redshift, permitting us to map the structure of dark matter.

\section{Data Availability}
The following data are available at \url{https://www.github.com/ado8/hsf_DR1}:
\begin{itemize}
    \item $J$-band light-curves of all observed targets regardless of spectroscopic classification
    \item Weighted cross-correlation results for all galaxies we observed with SNIFS or FOCAS
    \item Multiple sets of fitting parameters and uncertainties for all spectroscopically confirmed \sneia{}, with sets covering all combinations of fitting model (EBV\_model2, max\_model, SALT3-NIR) and data used (optical only or optical and NIR)
    \item Additional sets of fitting parameters for \sneia{} used in our comparisons with DEHVILS and CSP
\end{itemize}
The code used in our analysis can be found at \url{https://www.github.com/ado8/hsf_code}.
Data such as images and spectra may be available upon reasonable request.

\section{Acknowledgements}

This work, AD, and BS are supported by National Science Foundation grant AST-1911074.

AD and KSM are supported by the European Research Council (ERC) under the European Union’s Horizon 2020 research and innovation programme (Grant Agreement No. 101002652).

This publication makes use of data products from the Two Micron All Sky Survey, which is a joint project of the University of Massachusetts and the Infrared Processing and Analysis Center/California Institute of Technology, funded by the National Aeronautics and Space Administration and the National Science Foundation.

The Pan-STARR Surveys (PS1) and the PS1 public science archive have been made possible through contributions by the Institute for Astronomy, the University of Hawaii, the Pan-STARRS Project Office, the Max-Planck Society and its participating institutes, the Max Planck Institute for Astronomy, Heidelberg and the Max Planck Institute for Extraterrestrial Physics, Garching, The Johns Hopkins University, Durham University, the University of Edinburgh, the Queen's University Belfast, the Harvard-Smithsonian Center for Astrophysics, the Las Cumbres Observatory Global Telescope Network Incorporated, the National Central University of Taiwan, the Space Telescope Science Institute, the National Aeronautics and Space Administration under Grant No. NNX08AR22G issued through the Planetary Science Division of the NASA Science Mission Directorate, the National Science Foundation Grant No. AST-1238877, the University of Maryland, Eotvos Lorand University (ELTE), the Los Alamos National Laboratory, and the Gordon and Betty Moore Foundation.

This research has made use of NASA’s Astrophysics Data System.

We acknowledge the usage of the HyperLeda database (http://leda.univ-lyon1.fr).

This research has made use of the SIMBAD database, operated at CDS, Strasbourg, France

This research has made use of the NASA/IPAC Extragalactic Database (NED), which is funded by the National Aeronautics and Space Administration and operated by the California Institute of Technology.

UKIRT is owned by the University of Hawaii (UH) and operated by the UH Institute for Astronomy.
When (some of) the data reported here were obtained, the operations were enabled through the cooperation of the East Asian Observatory.

The ZTF forced-photometry service was funded under the Heising-Simons Foundation grant
\#12540303 (PI: Graham).

This work has made use of data from the European Space Agency (ESA) mission {\it Gaia} (\url{https://www.cosmos.esa.int/gaia}), processed by the {\it Gaia} Data Processing and Analysis Consortium (DPAC, \url{https://www.cosmos.esa.int/web/gaia/dpac/consortium}).
Funding for the DPAC has been provided by national institutions, in particular the institutions participating in the {\it Gaia} Multilateral Agreement.

We acknowledge ESA Gaia, DPAC and the Photometric Science Alerts Team (http://gsaweb.ast.cam.ac.uk/alerts).

Funding for the Sloan Digital Sky Survey (SDSS) has been provided by the Alfred P. Sloan Foundation, the Participating Institutions, the National Aeronautics and Space Administration, the National Science Foundation, the U.S. Department of Energy, the Japanese Monbukagakusho, and the Max Planck Society. The SDSS Web site is http://www.sdss.org/.

The SDSS is managed by the Astrophysical Research Consortium (ARC) for the Participating Institutions. The Participating Institutions are The University of Chicago, Fermilab, the Institute for Advanced Study, the Japan Participation Group, The Johns Hopkins University, the Korean Scientist Group, Los Alamos National Laboratory, the Max-Planck-Institute for Astronomy (MPIA), the Max-Planck-Institute for Astrophysics (MPA), New Mexico State University, University of Pittsburgh, University of Portsmouth, Princeton University, the United States Naval Observatory, and the University of Washington.

\bibliographystyle{mnras.bst}
\bibliography{bibliography.bib}

\appendix
\section{NIR Photometry and Late-time Observations}
\label{appendix:1D3_vs_2D}
Late-time observations are not always critical for accurate forward-modelled photometry.
When the surface-brightness profile of the host galaxy varies smoothly or has sharp features (e.g. galaxy nucleus, spiral arms) that do not overlap with the supernova, the forward-modelling code is able to cleanly separate the flux from the supernova and galaxy.
In these cases, a late-time observation does not provide new information, and the inferred photometry does not change.
However, there are supernova-galaxy configurations that critically dependent on late-time observations for accurate modelling.
This motivates the use of a mixture model to simultaneously infer the parameters of both populations.

For each SN, we create two ensembles of images; one with and one without late-time observations.
This provides two values for the magnitude of the SN at each observed epoch, $m_\text{ref}$ and $m_0$ respectively.
The differences $\Delta m = m_\text{ref} - m_0$ should be distributed about 0 mag if $m_\text{ref}$ and $m_0$ are normally distributed about the same mean.
We use Stan to infer the population means, standard deviations, and the mixing ratio.

Our priors are based on a crude analysis where we consider the subset $|\Delta m| \leq 0.5$ mag and $|\Delta m| > 0.5$ mag, where 0.5 was chosen arbitrarily.
The subset near 0 comprises 752 of our 832 observations.
Our prior on the mixing ratio is a Beta distribution with $\alpha$ = 3 and $\beta$ = 0.3 such that the mean expectation value $\frac{\alpha}{\alpha + \beta} \approx 752/832$.
The scale of $\alpha$ and $\beta$ was chosen to create a moderately informed prior.
Our priors on the population means (in magnitudes) are $\mu_\text{in} \sim \mathcal{N}(0 \text{ mag}, 0.1^2 \text{ mag}^2)$ for the tightly-dispersed population, and $\mu_\text{out} \sim \mathcal{U}(-\infty, \infty)$ for the late-time sensitive population.
Lastly, our priors on the standard deviations (in magnitudes) are $\sigma \sim \mathcal{N}(0 \text{ mag}, 2^2 \text{ mag}^2)$, with $0 < \sigma_\text{in} < \sigma_\text{out}$.

We fit a Gaussian mixture-model to the photometric differences using Stan \citep{carpenter17} and find 74.0 $\pm$ 2.3\% of the differences appear tightly-dispersed ($\Delta m \sim \mathcal{N}(0.01 \pm 0.004 \text{ mag}, (0.08 \pm 0.005 \text{ mag})^2)$), and the remaining 26.0\% vary much more dramatically ($\Delta m \sim \mathcal{N}(0.33 \pm 0.050 \text{ mag}, (0.68 \pm 0.037 \text{ mag})^2)$).
The fraction of targets reliant upon late-time observations for accurate photometry is vastly exaggerated in this analysis because the subsample comprises only targets manually determined to need late-time observations.
The critical information is the distribution of the tightly-dispersed population, which describes the effect late-time observations have on typical photometric measurements.

\begin{figure*}
    \centering
    \includegraphics[width=0.9\textwidth]{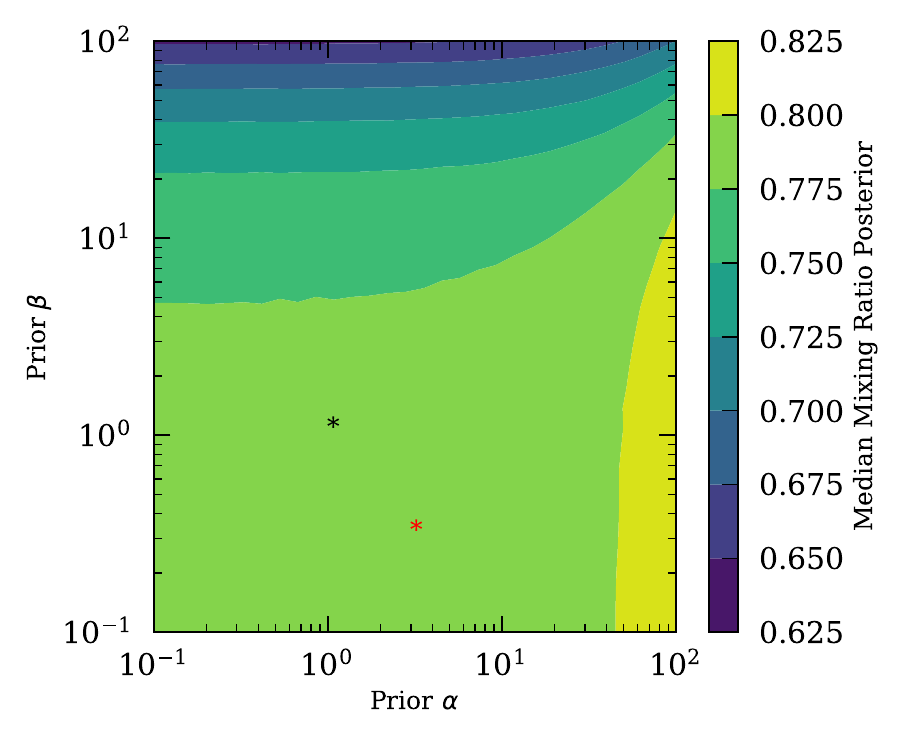}
    \caption{We resample our model using a grid of $\alpha$ and $\beta$ parameters for the prior Beta distribution of the mixing ratio and plot the median values of the mixing ratio posteriors.
    The set of values used in our analysis is marked with a red star, and the set that produces a flat prior is marked with a black star.
    The inferred mixing ratio is not sensitive to variations in the initial Beta distribution unless extremely strong priors are assumed.
    }
    \label{fig:sensitivity_analysis_mix_ratio}
\end{figure*}

We perform sensitivity analyses on the priors used for the mixing ratio and the population parameters of the tightly-dispersed group.
For testing the former, we tested prior Beta distributions parametrized by $\alpha$ and $\beta$ parameters drawn from a 30 by 30 grid spaced logarithmically between 0.1 and 100.
Figure \ref{fig:sensitivity_analysis_mix_ratio} shows that the recovered posterior estimate is not affected unless extreme values for $\alpha$ and $\beta$ are assumed, corresponding to an extremely strong prior.
More specifically, the recovered mixing ratio is within the joint uncertainty of the value inferred when using our the original priors ($\alpha = 3$, $\beta = 0.3$) unless $\alpha \approx 100$ while $\beta \lesssim 5$ or $\alpha \lesssim 3\beta - 50$ while $\beta \gtrsim 20$.
This implies that our inference of the mixing ratio is driven by data rather than the moderately informative prior we used.

\begin{figure*}
    \centering
    \includegraphics[width=0.9\textwidth]{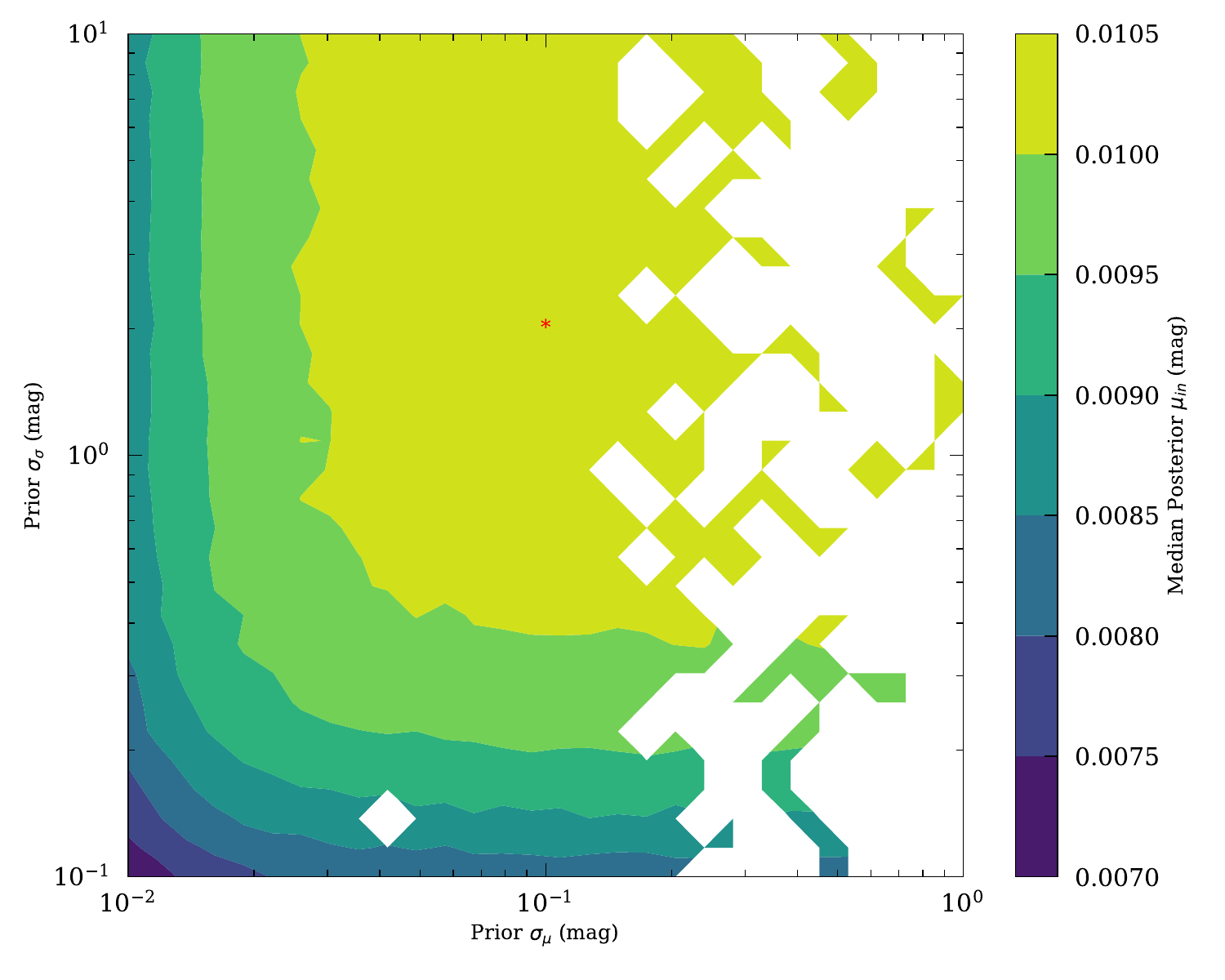}
    \caption{We resample our model using a grid of standard deviation values for the prior distributions of $\mu_{\text{in}}$ and $\sigma$, and plot the median values of the posterior $\mu_\text{in}$ samples.
    The set of values used in our analysis is marked with a red star.
    Increasing $\sigma_\mu$ leads to the convergence issues we observed when using a flat prior, so we mask the runs where the average Gelman-Rubin $\hat{r}$ value across the sampled parameters is greater than 1.05.
    }
    \label{fig:sensitivity_analysis_mu_in}
\end{figure*}

\begin{figure*}
    \centering
    \includegraphics[width=0.9\textwidth]{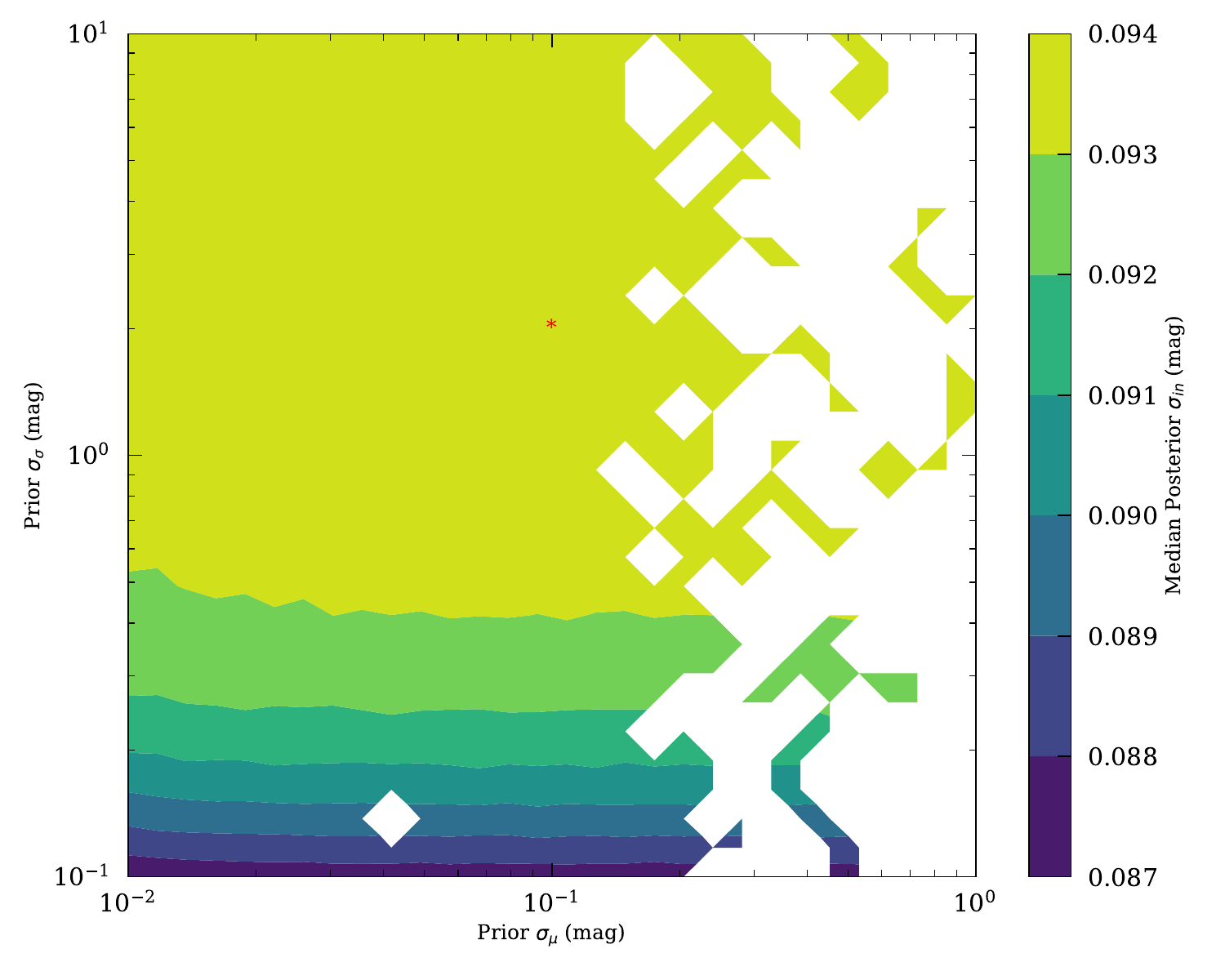}
    \caption{Similar to Figure \ref{fig:sensitivity_analysis_mu_in}, but plotting the median values of the posterior $\sigma_\text{in}$ samples rather than the $\mu_\text{in}$ samples.
    }
    \label{fig:sensitivity_analysis_sigma}
\end{figure*}

The priors on the population parameters of the tightly-dispersed group encode the assumption that the magnitude differences $\Delta m$ should be 0 mag if the late-time observations are not providing new information to break model degeneracies.
Alternatively, one could assume that there is a systematic offset that must be estimated, which could make a flat uniform prior more appropriate.
However, this leads to convergence issues when sampling our model with 7 chains and 5,000 steps using Stan's no-U-Turn Hamiltonian Monte Carlo sampler.
Instead, we examine the sensitivity of the posterior estimates to different levels of variance in the priors $\mu_\text{in} \sim \mathcal{N}(0 \text{ mag}, \sigma_{\mu}^2 \text{ mag}^2)$.
and $\sigma \sim \mathcal{N}(0 \text{ mag}, \sigma_\sigma^2 \text{ mag}^2)$ while maintaining $\sigma_\text{in} < \sigma_\text{out}$.
We sample our model using a 30 by 30 grid of $\sigma_\mu$ and $\sigma_\sigma$ values logarithmically spaced between 0.01 mag and 1 mag and between 0.1 mag and 10 mag respectively.
As $\sigma_{\mu}$ increases, the distribution approaches a flat prior, and we observe similar convergence issues as $\sigma_{\mu}$ approaches 1 mag.
We mask the runs where the average Gelman-Rubin $\widehat{R}$ value \citep{gelman92} across all parameters is greater than 1.05.
Figure \ref{fig:sensitivity_analysis_mu_in} and Figure \ref{fig:sensitivity_analysis_sigma} show the medians of the posterior $\mu_\text{in}$ and $\sigma_\text{in}$ samples respectively as a function of different values for the priors $\sigma_\mu$ and $\sigma_\sigma$.
Convergence issues aside, the inferred population parameters describing the tightly-dispersed group appear robust against variations in the priors.

A joint sensitivity analysis examining the effects of varying more than two scalar priors at a time is possible, but given the insensitivity demonstrated in the above analyses and the computational expense of increasing the dimensions of the problem, we deem that a joint analysis is not currently necessary.

\section{Host Galaxy Identification Algorithm} \label{appendix:DLR}
Before choosing to proceed with manual host galaxy identification, we investigated the Directional Light Radius method \citep[DLR;][]{sullivan06, sako18} which normalizes angular separation by the elliptical radius of a galaxy in the direction of the transient.
The morphological data came from the NASA/IPAC Extragalactic Database\footnote{\url{https://ned.ipac.caltech.edu/}} \citep[NED;][]{helou91, mazzarella07}, the Simbad Astronomical Database\footnote{\url{https://simbad.u-strasbg.fr/simbad/}} \citep[Simbad;][]{wenger00}, and the GLADE+ Galaxy Catalog\footnote{\url{https://glade.elte.hu/}} \citep[GLADE+;][]{dalya22}, which itself consolidates galaxies from the Gravitational Wave Galaxy Catalogue \citep{white11}, HyperLEDA\footnote{\url{http://leda.univ-lyon1.fr/}} \citep{makarov14}, and the 2MASS Extended Source Catalog \citep{jarrett00, skrutskie06}.
Unfortunately, the heterogeneity and sparsity of the available data presented several failure modes.
Firstly, the correct host galaxy could not be identified if it was not included in at least one of the aforementioned databases or catalogues.
Similarly, if a galaxy's morphological data was not available, the DLR method could not be applied.
Lastly, if multiple galaxies have categorically distinct morphological measurements, either using different filters or different metrics, the DLR method would have been biased.

\section{Alternative \texttt{SNooPy} Configurations}
\label{appendix:snpy_calibration}
In Section \ref{sec:snpy} we describe the decisions that affect the inference of distance moduli given a set of photometry.
Those decisions are:
\begin{itemize}
    \item Parametrizing shape with $s_{BV}$ instead of $\Delta m_{15}$.
    \item Performing fits with EBV\_model2 and the max\_model and not with the EBV\_model, max\_model2, Rv\_model, color\_model, SALT\_model, or MLCS\_model.
    \item Using the calibration from the full sample of \citet{burns18} to describe the correlation between SN parameters and absolute magnitude.
    \item Using the F19 reddening law.
\end{itemize}
In this section we present the quantitative analyses that led to our choices of shape parameter, calibration, and reddening law.

\subsection{Calibration}
The calibration used in the SNPY\_Max sample is defined based on our sample, but the calibration used in the SNPY\_EBV sample is selected from a list of available calibrations.
We reproduce fits of our SNPY\_EBV sample using the calibrations from Table \ref{tab:snpy_calibration_st_default} and \ref{tab:snpy_calibration_2018}.
We also reproduce our fits using $\Delta m_{15}$ to parameterize shape and the calibrations from Table \ref{tab:snpy_calibration_dm15_default}.
Each fit is performed assuming the default O94 reddening law \citep{odonnell94}.
In the EBV\_model2, the calibration values are $P_0$, $P_1$, $P_2$, and the host galaxy $R_V$, with a fit dispersion of $\sigma_\text{int}$ mag.

\begin{table*}
    \centering
    \begin{tabular}{|c|c|c|c|c|c|c|c|}
        \hline
        Calibration & Sample & Bandpass & $P_0$ & $P_1$ & $P_2$ & $R_V$ & $\sigma_\text{int}$ \\
        Number & Description & & (mag) & (mag) & (mag) & & (mag) \\
        \hline
        \multirow{9}{*}{0} & \multirow{9}{*}{\rotatebox[origin=c]{90}{{\parbox[c]{3cm}{\centering $u$-band \\ $s_{BV} > 0.5$}}}}
          & B & -19.310(025) & -0.675(110) & 3.415(387) & 1.518(082) & 0.072 \\
        & & V & -19.264(022) & -0.727(092) & 2.161(341) & 1.518(082) & 0.067 \\
        & & u & -18.945(035) & -1.077(163) & 4.066(533) & 1.518(082) & 0.148 \\
        & & g & -19.345(023) & -0.719(102) & 2.760(364) & 1.518(082) & 0.067 \\
        & & r & -19.146(023) & -0.619(094) & 1.968(347) & 1.518(082) & 0.076 \\
        & & i & -18.529(024) & -0.541(102) & 0.705(382) & 1.518(082) & 0.092 \\
        & & Y & -18.532(025) & -0.387(112) & 0.232(416) & 1.518(082) & 0.105 \\
        & & J & -18.646(026) & -0.719(126) & -0.714(462) & 1.518(082) & 0.117 \\
        & & H & -18.470(034) & -0.456(171) & -0.192(622) & 1.518(082) & 0.172 \\
        \hline
        \multirow{9}{*}{1} & \multirow{9}{*}{\rotatebox[origin=c]{90}{{\parbox[c]{3cm}{\centering $u$-band \\ $(m_B-m_V) < 0.3$}}}}
          & B & -19.317(025) & -0.655(108) & 3.500(378) & 1.746(180) & 0.071 \\
        & & V & -19.278(024) & -0.718(094) & 2.249(341) & 1.746(180) & 0.068 \\
        & & u & -18.972(031) & -1.028(143) & 4.416(418) & 1.746(180) & 0.124 \\
        & & g & -19.349(024) & -0.710(100) & 2.782(352) & 1.746(180) & 0.065 \\
        & & r & -19.162(025) & -0.613(095) & 2.049(351) & 1.746(180) & 0.077 \\
        & & i & -18.550(026) & -0.530(103) & 0.848(384) & 1.746(180) & 0.092 \\
        & & Y & -18.547(026) & -0.378(114) & 0.320(427) & 1.746(180) & 0.106 \\
        & & J & -18.665(028) & -0.697(129) & -0.538(475) & 1.746(180) & 0.119 \\
        & & H & -18.490(036) & -0.431(176) & -0.005(637) & 1.746(180) & 0.175 \\
        \hline
        \multirow{9}{*}{2} & \multirow{9}{*}{\rotatebox[origin=c]{90}{{\parbox[c]{3cm}{\centering$u$-band\\all objects}}}}
          & B & -19.325(022) & -0.676(103) & 3.804(262) & 1.531(081) & 0.069 \\
        & & V & -19.277(020) & -0.732(088) & 2.422(222) & 1.531(081) & 0.065 \\
        & & u & -18.969(032) & -1.123(142) & 4.742(253) & 1.531(081) & 0.153 \\
        & & g & -19.359(021) & -0.719(095) & 3.098(243) & 1.531(081) & 0.065 \\
        & & r & -19.154(020) & -0.637(089) & 2.048(221) & 1.531(081) & 0.074 \\
        & & i & -18.555(021) & -0.510(099) & 1.378(240) & 1.531(081) & 0.092 \\
        & & Y & -18.560(022) & -0.350(112) & 0.975(269) & 1.531(081) & 0.107 \\
        & & J & -18.686(026) & -0.639(143) & 0.460(313) & 1.531(081) & 0.140 \\
        & & H & -18.499(030) & -0.416(165) & 0.637(354) & 1.531(081) & 0.168 \\
        \hline
        \multirow{8}{*}{3} & \multirow{8}{*}{\rotatebox[origin=c]{90}{{\parbox[c]{3cm}{\centering No $u$-band\\$s_{BV} > 0.5$}}}}
          & B & -19.271(024) & -0.753(116) & 2.928(411) & 1.699(089) & 0.078 \\
        & & V & -19.246(021) & -0.791(093) & 1.867(343) & 1.699(089) & 0.066 \\
        & & g & -19.315(022) & -0.785(105) & 2.369(375) & 1.699(089) & 0.070 \\
        & & r & -19.134(021) & -0.678(094) & 1.728(346) & 1.699(089) & 0.075 \\
        & & i & -18.518(023) & -0.599(100) & 0.476(374) & 1.699(089) & 0.090 \\
        & & Y & -18.528(023) & -0.415(108) & 0.123(398) & 1.699(089) & 0.102 \\
        & & J & -18.638(025) & -0.743(122) & -0.827(445) & 1.699(089) & 0.112 \\
        & & H & -18.462(032) & -0.513(168) & -0.374(606) & 1.699(089) & 0.169 \\
        \hline
        \multirow{8}{*}{4} & \multirow{8}{*}{\rotatebox[origin=c]{90}{{\parbox[c]{3cm}{\centering No $u$-band\\$(m_B-m_V) < 0.3$}}}}
          & B & -19.276(025) & -0.730(116) & 3.053(412) & 1.716(202) & 0.078 \\
        & & V & -19.247(022) & -0.780(095) & 1.909(351) & 1.716(202) & 0.068 \\
        & & g & -19.311(023) & -0.782(106) & 2.363(383) & 1.716(202) & 0.070 \\
        & & r & -19.134(023) & -0.672(095) & 1.744(354) & 1.716(202) & 0.076 \\
        & & i & -18.524(024) & -0.589(101) & 0.557(381) & 1.716(202) & 0.092 \\
        & & Y & -18.529(025) & -0.409(111) & 0.146(418) & 1.716(202) & 0.104 \\
        & & J & -18.646(026) & -0.728(126) & -0.720(465) & 1.716(202) & 0.115 \\
        & & H & -18.477(035) & -0.489(174) & -0.188(644) & 1.716(202) & 0.175 \\
        \hline
        \multirow{8}{*}{5} & \multirow{8}{*}{\rotatebox[origin=c]{90}{{\parbox[c]{3cm}{\centering No $u$-band\\all objects}}}}
          & B & -19.304(022) & -0.682(113) & 3.916(321) & 1.729(089) & 0.077 \\
        & & V & -19.270(019) & -0.751(092) & 2.460(254) & 1.729(089) & 0.065 \\
        & & g & -19.344(021) & -0.727(102) & 3.166(292) & 1.729(089) & 0.067 \\
        & & r & -19.154(019) & -0.655(092) & 2.155(238) & 1.729(089) & 0.074 \\
        & & i & -18.553(020) & -0.536(099) & 1.409(238) & 1.729(089) & 0.089 \\
        & & Y & -18.561(021) & -0.360(107) & 1.024(244) & 1.729(089) & 0.103 \\
        & & J & -18.687(025) & -0.633(139) & 0.639(296) & 1.729(089) & 0.139 \\
        & & H & -18.495(028) & -0.456(161) & 0.594(331) & 1.729(089) & 0.164 \\
        \hline
    \end{tabular}
    \caption{The calibrations available for the EBV\_model2 in \texttt{SNooPy} version 2.6.0.
    All bandpasses are from the natural CSP photometric system.
    The shape factor used to calculate these values is $s_{BV}$.
    }
    \label{tab:snpy_calibration_st_default}
\end{table*}

\begin{table*}
    \centering
    \begin{tabular}{|c|c|c|c|c|c|c|c|}
        \hline
        Calibration & Sample & Bandpass & $P_0$ & $P_1$ & $P_2$ & $R_V$ & $\sigma_\text{int}$ \\
        Number & Description & & (mag) & (mag) & (mag) & & (mag) \\
        \hline
        \multirow{9}{*}{6} & \multirow{9}{*}{\rotatebox[origin=c]{90}{{\parbox[c]{3cm}{\centering Full sample}}}}
          & B & -19.182(062) & -0.89(11) & -0.02(30) & 1.65(08) & 0.13 \\
        & & V & -19.181(061) & -0.89(11) & -0.02(30) & 1.65(08) & 0.13 \\
        & & u & -18.818(097) & -1.28(17) & 0.32(44) & 1.13(52) & 0.22 \\
        & & g & -19.229(084) & -0.90(11) & -0.13(31) & 1.57(09) & 0.13 \\
        & & r & -19.099(059) & -0.74(10) & 0.38(27) & 1.78(08) & 0.12 \\
        & & i & -18.523(059) & -0.48(10) & 0.41(27) & 1.85(09) & 0.12 \\
        & & Y & -18.517(077) & -0.07(11) & 1.19(30) & 1.34(21) & 0.12 \\
        & & J & -18.633(062) & -0.37(12) & 0.61(32) & 1.27(36) & 0.11 \\
        & & H & -18.431(062) & -0.05(12) & 1.18(31) & 1.28(57) & 0.11 \\
        \hline
        \multirow{9}{*}{0} & \multirow{9}{*}{\rotatebox[origin=c]{90}{{\parbox[c]{3cm}{\centering $(m_B-m_V) < 0.5$}}}}
          & B & -19.161(062) & -0.94(11) & -0.36(43) & 1.54(14) & 0.13 \\
        & & V & -19.161(061) & -0.94(11) & -0.37(44) & 1.54(14) & 0.13 \\
        & & u & -18.793(095) & -1.35(18) & -0.47(64) & 1.12(51) & 0.21 \\
        & & g & -19.206(082) & -0.97(11) & -0.57(43) & 1.48(14) & 0.13 \\
        & & r & -19.081(060) & -0.77(10) & 0.12(41) & 1.67(13) & 0.13 \\
        & & i & -18.501(060) & -0.52(10) & -0.10(41) & 1.79(17) & 0.13 \\
        & & Y & -18.497(076) & -0.10(11) & 0.34(41) & 1.69(35) & 0.12 \\
        & & J & -18.601(062) & -0.43(11) & -0.42(45) & 1.51(58) & 0.11 \\
        & & H & -18.400(062) & -0.10(12) & 0.17(47) & 1.33(85) & 0.11 \\
        \hline
        \multirow{9}{*}{0} & \multirow{9}{*}{\rotatebox[origin=c]{90}{{\parbox[c]{3cm}{\centering $s_{BV} > 0.5$}}}}
          & B & -19.159(062) & -0.93(12) & -0.61(43) & 1.64(09) & 0.13 \\
        & & V & -19.159(061) & -0.94(11) & -0.62(43) & 1.64(09) & 0.13 \\
        & & u & -18.790(097) & -1.32(18) & -0.35(70) & 1.10(45) & 0.22 \\
        & & g & -19.204(084) & -0.96(12) & -0.80(43) & 1.56(09) & 0.13 \\
        & & r & -19.081(060) & -0.77(11) & -0.05(39) & 1.76(08) & 0.12 \\
        & & i & -18.499(059) & -0.52(10) & -0.21(38) & 1.82(10) & 0.12 \\
        & & Y & -18.480(076) & -0.11(11) & 0.32(42) & 1.18(22) & 0.11 \\
        & & J & -18.593(060) & -0.44(12) & -0.35(45) & 1.02(36) & 0.11 \\
        & & H & -18.394(061) & -0.10(12) & 0.13(47) & 0.82(52) & 0.11 \\
        \hline
        \multirow{9}{*}{0} & \multirow{9}{*}{\rotatebox[origin=c]{90}{{\parbox[c]{3cm}{\centering $s_{BV} > 0.5$\\$(m_B-m_V) < 0.5$}}}}
          & B & -19.162(061) & -0.94(11) & -0.30(46) & 1.55(14) & 0.13 \\
        & & V & -19.163(061) & -0.94(11) & -0.31(46) & 1.55(14) & 0.13 \\
        & & u & -18.796(095) & -1.35(17) & -0.42(69) & 1.12(51) & 0.21 \\
        & & g & -19.207(083) & -0.96(11) & -0.53(46) & 1.48(15) & 0.13 \\
        & & r & -19.083(060) & -0.77(10) & 0.17(42) & 1.68(13) & 0.13 \\
        & & i & -18.501(061) & -0.52(10) & -0.10(43) & 1.78(17) & 0.13 \\
        & & Y & -18.489(075) & -0.10(10) & 0.15(42) & 1.59(35) & 0.12 \\
        & & J & -18.598(063) & -0.43(12) & -0.48(47) & 1.48(57) & 0.11 \\
        & & H & -18.395(061) & -0.11(12) & 0.03(48) & 1.24(84) & 0.11 \\
        \hline
    \end{tabular}
    \caption{Similar to Table \ref{tab:snpy_calibration_st_default}, except the values are those presented in Table 1 of \citet{burns18}.
    }
    \label{tab:snpy_calibration_2018}
\end{table*}

\begin{table*}
    \centering
    \begin{tabular}{|c|c|c|c|c|c|c|c|}
        \hline
        Calibration & Sample& Bandpass & $P_0$ & $P_1$ & $P_2$ & $R_V$ & $\sigma_\text{int}$ \\
        Number & Description & & (mag) & (mag) & (mag) & & (mag) \\
        \hline
        \multirow{9}{*}{10} & \multirow{9}{*}{\rotatebox[origin=c]{90}{{\parbox[c]{3cm}{\centering $u$-band \\ $s_{BV} > 0.5$}}}}
          & B & -19.360(030) & 0.433(090) & 2.356(293) & 1.533(084) & 0.076 \\
        & & V & -19.282(027) & 0.540(080) & 1.253(260) & 1.533(084) & 0.070 \\
        & & u & -18.979(042) & 0.751(136) & 2.526(433) & 1.533(084) & 0.150 \\
        & & g & -19.380(028) & 0.473(084) & 1.958(275) & 1.533(084) & 0.069 \\
        & & r & -19.171(027) & 0.489(081) & 1.187(262) & 1.533(084) & 0.077 \\
        & & i & -18.548(027) & 0.379(088) & 0.383(278) & 1.533(084) & 0.093 \\
        & & Y & -18.549(027) & 0.082(091) & 0.372(282) & 1.533(084) & 0.095 \\
        & & J & -18.662(028) & 0.175(100) & 0.152(311) & 1.533(084) & 0.107 \\
        & & H & -18.475(037) & 0.143(136) & 0.035(418) & 1.533(084) & 0.165 \\
        \hline
        \multirow{9}{*}{11} & \multirow{9}{*}{\rotatebox[origin=c]{90}{{\parbox[c]{3cm}{\centering $u$-band \\ $(m_B-m_V) < 0.3$}}}}
          & B & -19.369(030) & 0.419(089) & 2.425(297) & 1.589(155) & 0.076 \\
        & & V & -19.292(028) & 0.528(080) & 1.322(265) & 1.589(155) & 0.070 \\
        & & u & -19.016(039) & 0.717(120) & 2.788(393) & 1.589(155) & 0.126 \\
        & & g & -19.384(029) & 0.464(082) & 1.992(274) & 1.589(155) & 0.066 \\
        & & r & -19.182(029) & 0.478(081) & 1.256(266) & 1.589(155) & 0.079 \\
        & & i & -18.568(029) & 0.362(088) & 0.511(283) & 1.589(155) & 0.093 \\
        & & Y & -18.560(028) & 0.070(091) & 0.446(285) & 1.589(155) & 0.094 \\
        & & J & -18.679(029) & 0.156(100) & 0.274(312) & 1.589(155) & 0.108 \\
        & & H & -18.494(038) & 0.122(136) & 0.176(424) & 1.589(155) & 0.165 \\
        \hline
        \multirow{9}{*}{12} & \multirow{9}{*}{\rotatebox[origin=c]{90}{{\parbox[c]{3cm}{\centering$u$-band\\all objects}}}}
          & B & -19.394(029) & 0.370(089) & 2.820(267) & 1.593(084) & 0.074 \\
        & & V & -19.320(025) & 0.472(077) & 1.736(214) & 1.593(084) & 0.067 \\
        & & u & -19.056(042) & 0.579(144) & 3.793(373) & 1.593(084) & 0.173 \\
        & & g & -19.409(027) & 0.424(082) & 2.306(243) & 1.593(084) & 0.067 \\
        & & r & -19.205(024) & 0.430(078) & 1.591(204) & 1.593(084) & 0.075 \\
        & & i & -18.598(025) & 0.279(085) & 1.070(207) & 1.593(084) & 0.093 \\
        & & Y & -18.594(024) & -0.010(088) & 1.013(207) & 1.593(084) & 0.095 \\
        & & J & -18.707(027) & 0.087(103) & 0.825(228) & 1.593(084) & 0.123 \\
        & & H & -18.509(032) & 0.066(128) & 0.559(275) & 1.593(084) & 0.161 \\
        \hline
        \multirow{8}{*}{13} & \multirow{8}{*}{\rotatebox[origin=c]{90}{{\parbox[c]{3cm}{\centering No $u$-band\\$s_{BV} > 0.5$}}}}
          & B & -19.281(030) & 0.516(095) & 1.768(306) & 1.727(097) & 0.086 \\
        & & V & -19.235(026) & 0.613(081) & 0.840(259) & 1.727(097) & 0.071 \\
        & & g & -19.314(028) & 0.547(087) & 1.454(282) & 1.727(097) & 0.074 \\
        & & r & -19.134(026) & 0.558(081) & 0.833(258) & 1.727(097) & 0.077 \\
        & & i & -18.516(026) & 0.445(088) & 0.064(276) & 1.727(097) & 0.092 \\
        & & Y & -18.529(026) & 0.123(090) & 0.180(277) & 1.727(097) & 0.095 \\
        & & J & -18.643(027) & 0.202(098) & -0.000(301) & 1.727(097) & 0.107 \\
        & & H & -18.453(036) & 0.211(136) & -0.212(418) & 1.727(097) & 0.168 \\
        \hline
        \multirow{8}{*}{14} & \multirow{8}{*}{\rotatebox[origin=c]{90}{{\parbox[c]{3cm}{\centering No $u$-band\\$(m_B-m_V) < 0.3$}}}}
          & B & -19.287(031) & 0.512(096) & 1.824(314) & 1.544(181) & 0.087 \\
        & & V & -19.228(027) & 0.610(081) & 0.840(265) & 1.544(181) & 0.073 \\
        & & g & -19.308(028) & 0.554(088) & 1.430(286) & 1.544(181) & 0.075 \\
        & & r & -19.125(027) & 0.552(081) & 0.826(262) & 1.544(181) & 0.078 \\
        & & i & -18.516(028) & 0.435(087) & 0.112(278) & 1.544(181) & 0.092 \\
        & & Y & -18.523(027) & 0.121(090) & 0.173(282) & 1.544(181) & 0.095 \\
        & & J & -18.644(028) & 0.205(098) & 0.016(307) & 1.544(181) & 0.108 \\
        & & H & -18.465(038) & 0.207(138) & -0.129(434) & 1.544(181) & 0.169 \\
        \hline
        \multirow{8}{*}{15} & \multirow{8}{*}{\rotatebox[origin=c]{90}{{\parbox[c]{3cm}{\centering No $u$-band\\all objects}}}}
          & B & -19.335(028) & 0.410(092) & 2.559(273) & 1.824(096) & 0.084 \\
        & & V & -19.288(023) & 0.515(076) & 1.549(209) & 1.824(096) & 0.067 \\
        & & g & -19.360(025) & 0.463(083) & 2.065(247) & 1.824(096) & 0.070 \\
        & & r & -19.182(022) & 0.469(076) & 1.447(190) & 1.824(096) & 0.074 \\
        & & i & -18.582(022) & 0.308(082) & 1.013(184) & 1.824(096) & 0.093 \\
        & & Y & -18.586(022) & 0.002(084) & 1.017(178) & 1.824(096) & 0.096 \\
        & & J & -18.704(025) & 0.070(100) & 0.961(203) & 1.824(096) & 0.125 \\
        & & H & -18.512(031) & 0.072(128) & 0.737(250) & 1.824(096) & 0.171 \\
        \hline
    \end{tabular}
    \caption{Similar to Table \ref{tab:snpy_calibration_st_default}, except the shape parameter used is $\Delta m_{15}$.}
    \label{tab:snpy_calibration_dm15_default}
\end{table*}

We compare the $\chi^2$ values of the resultant fits to determine which calibration to use.
All $\chi^2$ values are calculated as the square of the data-model residual divided by the quadrature sum of the uncertainties in the data and the model.
Table \ref{tab:snpy_calibration_chi2} lists the median $\chi^2/\text{DoF}$ in the SNPY\_EBV sample fit using the listed calibrations.
The table also includes the median $\chi_\text{bp}^2 / N_\text{bp}$ for each bandpass, where $\chi_\text{bp}^2$ is the sum of $\chi^2$ values in that bandpass and $N_\text{bp}$ is the corresponding number of photometric epochs.
These are not reduced $\chi^2$ values because the four fitting parameters are not removed from the DoF.
As such, $\chi^2/\text{DoF}$ is not an average of the bandpass specific values $\sum_\text{bp} \chi_\text{bp}^2 / N_\text{bp}$, but rather $\sum_\text{bp} \chi_\text{bp}^2 / (\sum_\text{bp} N_\text{bp} - 4)$.

\begin{table*}
    \centering
    \begin{tabular}{|c|c|c|c|c|c|c|}
        \hline
        Calibration & $\chi^2$/DoF & $\chi_g^2 / N_g$ & $\chi_c^2 / N_c$ & $\chi_r^2 / N_r$ & $\chi_o^2 / N_o$ & $\chi_J^2 / N_J$ \\
        \hline
        0 & 1.051 & 0.491 & 1.021 & 0.782 & 0.876 & 1.234 \\
        1 & 1.036 & 0.491 & 1.013 & 0.760 & 0.866 & 1.189 \\
        2 & 1.045 & 0.471 & 0.980 & 0.777 & 0.858 & 1.233 \\
        3 & 1.015 & 0.489 & 1.010 & 0.753 & 0.867 & 1.126 \\
        4 & 1.012 & 0.491 & 1.013 & 0.746 & 0.865 & 1.148 \\
        5 & 1.011 & 0.470 & 0.978 & 0.731 & 0.848 & 1.108 \\
        \hline
        6 & 0.955 & 0.484 & 1.105 & 0.636 & 0.846 & 1.052 \\
        7 & 0.953 & 0.467 & 0.963 & 0.667 & 0.866 & 1.002 \\
        8 & 0.937 & 0.460 & 0.974 & 0.637 & 0.867 & 1.032 \\
        9 & 0.947 & 0.458 & 0.963 & 0.657 & 0.874 & 0.997 \\
        \hline
        10 & 1.009 & 0.570 & 1.115 & 0.670 & 0.892 & 1.181 \\
        11 & 1.002 & 0.576 & 1.102 & 0.657 & 0.891 & 1.157 \\
        12 & 0.991 & 0.559 & 1.115 & 0.648 & 0.891 & 1.142 \\
        13 & 0.996 & 0.548 & 1.085 & 0.592 & 0.887 & 1.123 \\
        14 & 1.008 & 0.567 & 1.108 & 0.613 & 0.888 & 1.162 \\
        15 & 0.968 & 0.555 & 1.101 & 0.588 & 0.871 & 1.115 \\
        \hline
    \end{tabular}
    \caption{
        Various $\chi^2$ metrics are presented for the SNPY\_EBV sample fit with the 16 calibrations listed in Tables \ref{tab:snpy_calibration_st_default}, \ref{tab:snpy_calibration_2018}, and \ref{tab:snpy_calibration_dm15_default}.
        After the $\chi^2$/DoF column, each column lists the sum of $\chi^2$ values in the subscripted bandpass divided by the corresponding number of photometric epochs.
        Each listed value is the median across all \sneia{} in the SNPY\_EBV sample fit with the calibration in the first column.
        The $c$- and $o$-bandpasses are from ATLAS and the $g$- and $r$-bandpasses are from ZTF.
    }
    \label{tab:snpy_calibration_chi2}
\end{table*}

The $\chi^2/\text{DoF}$ values are consistently lowest in the calibrations sourced from Table 1 of \citet{burns18}.
Calibration 8, which was calculated without \sneia{} with $s_{BV} < 0.5$, has the lowest $\chi^2/\text{DoF}$ of all.
However, given that our uncut sample contains \sneia{} with $s_{BV} < 0.5$ we choose to use calibration 6 for all EBV\_model2 fits.

The ZTF $g$- and $r$-bandpasses and the ATLAS $o$-bandpass have $\chi_\text{bp}^2 / N_\text{bp}$ values below 1 across all calibrations, suggesting that the uncertainties in the photometry or in the model may be overestimated.
Our decision to combine photometry from each ATLAS quad into a single measurement based on a weighted median (see Section \ref{sec:atlas}) could produce such an overestimate, but it is not clear why $\chi_c^2 / N_c$ would be consistently larger than $\chi_o^2 / N_o$.

\subsection{Reddening Law}

A thorough review of the effects of dust \citep[e.g.][]{mccall04, gontcharov16}, even limited to studies of \sneia{} \citep[e.g.][]{brout23}, is beyond the scope of this work, but we will review a few definitions to contextualize the present analysis and our decision.
Extinction is parametrized as a function of wavelength, where the observed flux at wavelength $\lambda$ is decreased by $A(\lambda)$ mag due to dust.
The extinction curve $A(\lambda)$ is roughly inversely proportional to wavelength, meaning the intrinsic colour of an object is reddened.
This reddening, or colour excess, is conventionally defined as the differential or selective extinction between the Johnson-Cousins $B$- and $V$-bands ($E(B-V) = A(B) - A(V)$).
The total-to-selective extinction parameter $R$ is defined as the ratio between the total extinction at a given wavelength and the colour excess ($R_\lambda = A(\lambda)/E(B-V)$).
Both the total and selective extinction are linearly proportional to the amount of dust along the line-of-sight, which leaves $R$ constant across different amounts of interposing dust.
However, the scattering cross-section of dust varies with the shape and size of the dust grains, producing diverse extinction curves and $R$ values.
\citet{cardelli89} found that the stellar samples from \citet{fitzpatrick86, fitzpatrick88} permitted normalized extinction curves from the ultraviolet (UV) to the NIR that depend on only one parameter chosen to be $R_V$.
This is an example of a ``reddening law'' or ``extinction law'' which is a function that uses $R_V$ \citep[or additional parameters, e.g.][]{gordon16} to infer $R$ at a given wavelength or integrated across a given bandpass.

The definition of \texttt{SNooPy}'s EBV\_model2 (Equation \ref{eqn:snpy_ebv_model2}) involves two terms that describe an $R$ value multiplied by an estimate of colour excess; one to account for for Galactic extinction and one for host galaxy extinction.
The rescaled SFD dust map provides estimates of Galactic colour excess while the colour excesses of the host galaxies are inferred during the fitting process.
The inference of colour excess is largely degenerate with the inference of $R$ values, so the model requires the assumption of a reddening law and $R_V$ values for the Milky Way and the host galaxies.

We reproduce the fits of SNPY\_EBV sample using three reddening laws: O94 \citep{odonnell94}, F99 \citep{fitzpatrick99}, and F19 \citep{fitzpatrick19}.
These specific reddening laws are chosen for the following reasons.
O94 is the default reddening law used in \texttt{SNooPy} and in the derivation of the \citet{folatelli10} calibrations.
The analysis of \citet{schlafly11} found that the reddening measured in SDSS stellar spectra agreed with the rescaled SFD dust map better when using the F99 reddening law than when using the O94 law.
The F19 reddening law presents several improvements over the F99 law.
The foundational data used to derive the F19 law are spectrophotometric, which allows for normalization based on a single wavelength (4,400 and 5,500 \AA) rather than broadband photometry ($B$- and $V$-bands).
Additionally, the new data set spans the gap between the UV and optical regimes with homogeneous coverage whereas other reddening laws extrapolate or interpolate between qualitatively different data sets to cover this gap.

The three reddening laws we examine were defined using data spanning specific ranges in wavelength and $R_V$.
O94 is based on data spanning wavelengths between about 3,030 and 9,090 \AA and $R_V$ values between 2.85 and 5.6.
F99 uses spectra from the International Ultraviolet Explorer \citep[IUE;][]{boggess78a, boggess78b} and photometry in the Johnson broadband $UBVRIJHKLM$ system and intermediate-band Str\"{o}mgren $uvby$ system, effectively spanning wavelengths between 1,150 \AA and 6 $\mu m$.
The fit assumes that $A(\lambda)$ approaches 0 as wavelength approaches infinity, but the author cautions that the curve ``should be treated as very approximate'' beyond 6 $\mu m$.
The $R_V$ values of the data range between about 2 and 6.
The F19 law uses spectra spanning 1,150 to 10,000 \AA and 2MASS photometry in the $JHK$-bandpasses which extends the red end to about $2.2 \mu m$.
This fit also assumes that $A(\lambda)$ approaches 0 as wavelength approaches infinity.
The $R_V$ values of the data span a slightly smaller range than the data used to define the F99 law, spanning about 2.5 to 6.
The $J$-band data used in our project is redder than the data used to calculate the O94 law, and the host galaxy $R_V$ values in calibration 6 (1.1 to 1.9) are all below the minimum $R_V$ values used to define the O94, F99, and F19 laws.
The low $R_V$ values are likely due to the conflation of intrinsic \snia{} colour and the effects of host galaxy extinction in the EBV\_model2.
We edit the allowed wavelength and $R_V$ ranges in the \texttt{dust\_extinction} package \citep{dust_extinction} to allow for the extrapolations we require.

\begin{table*}
    \centering
    \begin{tabular}{|c|c|c|c|c|c|c|}
        \hline
        Reddening Law & $\chi^2/\text{DoF}$ & $\chi_g^2 / N_g$ & $\chi_c^2 / N_c$ & $\chi_r^2 / N_r$ & $\chi_o^2 / N_o$ & $\chi_J^2 / N_J$ \\
        \hline
        O94 & 0.955 & 0.484 & 1.105 & 0.636 & 0.846 & 1.052 \\
        F99 & 0.947 & 0.464 & 1.139 & 0.609 & 0.844 & 1.135 \\
        F19 & 0.961 & 0.493 & 1.089 & 0.633 & 0.837 & 1.004 \\
        \hline
    \end{tabular}
    \caption{
        Similar to Table \ref{tab:snpy_calibration_chi2}, but presenting various $\chi^2$ metrics for the O94, F99, and F19 reddening laws rather than the calibrations.
    }
    \label{tab:snpy_redlaw}
\end{table*}

The $\chi^2/\text{DoF}$ values are similar across the three reddening laws, which implies that the choice of reddening law does not significantly impact the accuracy of the EBV\_model2.
The fits using the F99 law have the lowest $\chi^2/\text{DoF}$ value, but this is not the case over all bandpasses.
The F99 law produces the lowest $\chi_\text{bp}^2/N_\text{bp}$ values for the ZTF $g$- and $r$-bands, while simultaneously producing the highest values the ATLAS $c$-band and the $J$-band.
The ATLAS, ZTF, and \hsf{} observing strategies produce more epochs of photometry in the former two bandpasses than the latter two, which accounts for the lower $\chi^2/\text{DoF}$ in F99 via weighting.
However, we do not presently understand why the $\chi_\text{bp}^2 / N_\text{bp}$ values are so much lower than 1 for the $gro$-bands, and note that the F19 law produces the lowest values for the $coJ$-bands.
Even though the median $\chi^2/\text{DoF}$ value is highest in the fits assuming the F19 law, it is still indicative of a good set of fits.
Thus, we choose to adopt the F19 reddening law for our EBV\_model2 fits.

\section{All-Sky Survey Independence}
\label{appendix:photometric_independence}
When multiple observers record photometric time series of a single source, the correlation between the resultant light curves is based primarily on the time-evolution of the astrophysical source, but is also affected by correlated observational or instrumental effects.
For example, the orbital motion of the Earth Doppler shifts the SED of any observed target, leading to slight annual correlations for non-flat SEDs.
Unmodelled variability in reference stars used by multiple surveys could lead to common errors in zero-point calibration.
We assume independence between the ATLAS, ASAS-SN, and ZTF photometry in the sense that we consider any correlated observational or instrumental effects as insignificant.

To justify this assumption, we analyze forced photometry of CALSPEC stars in the footprint of all surveys and fainter than 15 mag in $V$ to avoid saturation.
To account for proper motion, we access ASAS-SN lightcurves from the ASAS-SN Sky Patrol\footnote{\url{http://asas-sn.ifa.hawaii.edu/skypatrol/}} \citep{shappee14, hart23} and ZTF lightcurves from the ZTF DR 21 archive \citep{masci19} hosted at the NASA/IPAC Infrared Science Archive\footnote{\url{https://irsa.ipac.caltech.edu/Missions/ztf.html}}.
The CALSPEC targets in NGC 6681 are excluded due to crowding.
The list of CALSPEC stars and their synthetic magnitudes in the bandpasses of the three surveys are presented in Table \ref{tab:calspec}
CALSPEC stars demonstrate minimal stellar variability \citep{rubin22}, which we use to exclude astrophysical time-evolution as a source of correlation between light curves.
As mentioned in Section \ref{sec:atlas}, we combine ATLAS data from the same nights with a weighted median.

\begin{table*}
    \centering
    \begin{tabular}{|c|c|c|c|c|c|c|}
        \hline
        Name & ATLAS $c$ & ATLAS $o$ & ASAS-SN $g$ & ZTF $g$ & ZTF $r$ \\
         & (mag) & (mag) & (mag) & (mag) & (mag) \\
        \hline
        C26202 & 16.55 & 16.32 & 16.73 & 16.69 & 16.34 \\
        \hline
        HS2027+0651 & 16.56 & 16.96 & 16.39 & 16.44 & 16.89 \\
        \hline
        NGC2506-G31	& 17.99 & 17.66 & 18.24 & 18.18 & 17.69 \\
        \hline
        SDSS132811 & 17.05 & 17.33 & 17.01 & 16.99 & 17.28 \\
        \hline
        SDSSJ151421 & 15.81 & 16.23 & 15.66 & 15.68 & 16.16 \\
        \hline
        SF1615+001A & 16.82 & 16.48 & 17.07 & 17.01 & 16.52 \\
        \hline
        SNAP-2 & 16.29 & 15.98 & 16.53 & 16.47 & 16.01 \\
        \hline
        VB8 & 17.04 & 14.2 & 17.85 & 17.57 & 15.58 \\
        \hline
        WD0947+857 & 15.66 & 16.14 & 15.47 & 15.51 & 16.06 \\
        \hline
        WD1026+453 & 16.03 & 16.5 & 15.85 & 15.88 & 16.43 \\
        \hline
        WD1657+343 & 16.35 & 16.83 & 16.15 & 16.2 & 16.75 \\
        \hline
    \end{tabular}
    \caption{This table lists the CALSPEC stars used in our analysis, which were selected to be fainter than 16 mag in $V$ to avoid saturation issues. We do not include the stars in NGC 6681 due to crowding. The columns list the synthetic magnitudes in the five bandpasses examined.}
    \label{tab:calspec}
\end{table*}

For the 10 bandpass pairs possible using ATLAS $c$, ATLAS $o$, ASAS-SN $g$, ZTF $g$, and ZTF $r$, we identify observations where a given star was observed in both bandpasses within 12 hours.
This makes our analysis sensitive to correlated effects on characteristic timescales greater than half a day, but insensitive to effects that vary on shorter timescales.
We calculate observed-synthetic magnitude residuals and normalize by the recorded uncertainties to produce z-scores.
We do not include pairs where either observation is in the bottom or top 5\% of z-scores for that bandpass and star.
We concatenate the rest of the z-score pairs into equal length sets for both bandpasses.
Table \ref{tab:calspec_correlation} shows the calculated Pearson correlation coefficients between those sets.

\begin{table*}
    \centering
    \begin{tabular}{|c|c|c|c|c|c|c|}
        \hline
        Bandpass 1 & Bandpass 2 & $N$ & $r$ & CI-95\% & p-value \\ \hline
        ATLAS $c$ & ATLAS $o$ & 53 & 0.011 & (-0.26, 0.28) & 0.940\\ \hline
        ATLAS $c$ & ASAS-SN $g$ & 345 & -0.057 & (-0.16, 0.05) & 0.291\\ \hline
        ATLAS $c$ & ZTF $g$ & 235 & -0.086 & (-0.21, 0.04) & 0.188\\ \hline
        ATLAS $c$ & ZTF $r$ & 244 & -0.031 & (-0.16, 0.09) & 0.628\\ \hline
        ATLAS $o$ & ASAS-SN $g$ & 922 & -0.012 & (-0.08, 0.05) & 0.719\\ \hline
        ATLAS $o$ & ZTF $g$ & 629 & 0.020 & (-0.06, 0.1) & 0.613\\ \hline
        ATLAS $o$ & ZTF $r$ & 686 & 0.035 & (-0.04, 0.11) & 0.359\\ \hline
        ASAS-SN $g$ & ZTF $g$ & 804 & 0.030 & (-0.04, 0.1) & 0.396\\ \hline
        ASAS-SN $g$ & ZTF $r$ & 808 & -0.012 & (-0.08, 0.06) & 0.741\\ \hline
        ZTF $g$ & ZTF $r$ & 1669 & 0.053 & (0.01, 0.1) & 0.030\\ \hline
    \end{tabular}
    \caption{We present correlation measurements between the 10 bandpass pairs between the five survey bandpasses. For each pair we assemble all $N$ observations of common targets performed on the same date and calculate the Pearson correlation coefficient $r$. We present the 95\% confidence intervals (CI-95\%) and p-values, finding that all combinations but ZTF $g$ and ZTF $r$ are consistent with no correlation and are not significant at the $p < 0.05$ level.
    }
    \label{tab:calspec_correlation}
\end{table*}

All bandpass pairs besides ZTF $g$ and ZTF $r$ are consistent with a correlation coefficient of 0 at the 95\% level.
The correlation between the two ZTF bandpasses implies there is at least one observational or instrumental effect that applies to both sets of observations, but the magnitude of such a correlation is small at about 0.05.
This is to be expected since both sets of observations come from the Palomar 48 inch Schmidt telescope.
Perhaps more surprising is that the ATLAS $c$ and $o$ data do not appear correlated.
This could be due to the distribution of observations across multiple sites (Haleakala and Maunaloa in Hawai`i, El Sauce Observatory in Chile, Sutherland Observing Station in South Africa), or due to the low number of observations in both bandpasses.

\label{lastpage}
\end{document}